\documentclass[%
reprint,
 amsmath,amssymb,
 aps,
 prl,
 physics,
]{revtex4-1}

\usepackage{preamble}

\begin{document}

\author{Kohei Fukai}
\email{k.fukai@issp.u-tokyo.ac.jp }
 \affiliation{The Institute for Solid State Physics, The University of Tokyo, Kashiwa, Chiba 277-8581, Japan}

\date{\today}

\title{All Local Conserved Quantities of the One-Dimensional Hubbard Model}
\begin{abstract}
We present the exact expression for all local conserved quantities of the one-dimensional Hubbard model. 
We identify the operator basis constructing the local charges and find that nontrivial coefficients appear in the higher-order charges.
We derive the recursion equation for these coefficients, and some of them are explicitly given. 
There are no other local charges independent of those we obtained.
\end{abstract}

\maketitle

\textit{Introduction.---}
Quantum integrability and local conservation laws are two sides of the same coin.
Quantum integrable systems are exactly solvable many-body systems by the Bethe Ansatz~\cite{Bethe1931} and have an extensive number of local conserved quantities $\{Q_k\}_{k\geq2}$, which is the foundation of their solvability.
Recently, quantum integrable systems are becoming an arena for the studies of nonequilibrium quantum dynamics, inspired by their experimental realization with ultracold atoms~\cite{Kinoshita2006,RevModPhys.80.885,RevModPhys.83.863,Langen_2016}, where $Q_k$ play a crucial role:
their existence leads to the absence of thermalization~\cite{PhysRevLett.106.140405,PhysRevLett.124.040603,PhysRevResearch.2.033403} and the conjectured long-time steady-state is the generalized Gibbs ensemble~\cite{PhysRevLett.98.050405,doi:10.1126/science.1257026,Vidmar2016}, involving all local (and also quasi-local) conserved quantities as well as Hamiltonian~\cite{pozsgay2013,Wouters2014,PhysRevLett.113.117203,Ilievski2015,PhysRevLett.115.157201}.
The large-scale nonequilibrium behavior is described by Generalized hydrodynamics~\cite{PhysRevX.6.041065,Bertini2016}, which is based on the local continuity equations of $Q_k$.
In quantum inverse scattering methods~\cite{korepin_bogoliubov_izergin_1993,Baxter1982}, the existence of the local conserved quantities is understood from the commutativity of the transfer matrices $\bck{T(\lambda), T(\mu)}=0$: $Q_k$ is obtained from the expansion of $\ln T(\lambda)$ in terms of the spectral parameter $\lambda$, and usually, the leading term $Q_2=H$ is Hamiltonian itself.
Another way to calculate $Q_k$ is the use of the boost operator $B$~\cite{1982JETP...55..306T,10.1143/PTP.69.431,THACKER1986348} if it exists: local charges can be calculated recursively by $\bck{B,Q_k}=Q_{k+1}$.

Although a procedure to generate the local conserved quantities $Q_k$ has been known, it is still practically difficult to obtain their expressions.
This difficulty lies not only in the expensive computational cost for higher-order charges but also in finding a general pattern in the huge amounts of data that emerge out of this calculation~\cite{GRABOWSKI1995299}.
This problem has been investigated particularly for the spin-1/2 XYZ chain~\cite{Sutherland1970,PhysRevLett.26.832,PhysRevLett.26.834,BAXTER1972193,BAXTER1972323,BAXTER19731,BAXTER197325,BAXTER197348} and the one-dimensional Hubbard model~\cite{PhysRevLett.20.1445,Shastry1986,PhysRevLett.56.1529,olmedilla1987yang,Olmedilla1988,shastry1988BF01022987,Martins1998,essler2005one}.
The former case is now deeply understood: the explicit expressions for the isotropic XXX case are obtained independently in~\cite{anshelevich1980first} and~\cite{doi:10.1142/S0217732394002057}. 
For the general XYZ case, Grabowski and Mathieu found the structure of $Q_k$ and derived the recursion relations to construct $Q_k$ using boost operator~\cite{GRABOWSKI1995299}, and recently, its explicit expression was obtained by Nozawa and the author~\cite{Nozawa2020} using the doubling-product notation~\cite{Shiraishi2019}, and for the XXZ case, independently obtained by Nienhuis and Huijgen using the Temperley-Lieb algebra~\cite{Nienhuis2021}.

On the other hand, for the one-dimensional Hubbard model, the structure of the local conserved quantities $Q_k$ remains a mystery.
The problem is that there was no recursive way to construct them~\cite{wadachinote}, unlike the XYZ case, because of the absence of the boost operator~\cite{GRABOWSKI1995299, boostnote}.
This comes from the fact that the Hubbard model is not Lorentz invariant due to the separation of spin and charge excitations with different velocities~\cite{PhysRevB.42.10553,PhysRevLett.86.5096,10.21468/SciPostPhys.9.3.040}.
The first three non-trivial charges have been found before:
$Q_3$~\cite{Shastry1986,shastry1988BF01022987}, $Q_4$~\cite{Grosse1989,Zhou1990}, and $Q_5$~\cite{GRABOWSKI1995299}.
From these expressions, Grabowski and Mathieu conjectured that $Q_k$ is constructed of products of local conserved densities of the spin-$1/2$ XX chain~\cite{GRABOWSKI1995299}. 
However, what kind of products of the XX charges are allowed in $Q_k$ was unknown.

In this Letter, we reveal the structure of the local conserved quantities $Q_k$ in the one-dimensional Hubbard model and present their exact expressions.
We prove $Q_k$ is a linear combination of \textit{connected diagrams}, a notation for the particular kind of products of the XX charges.
With this notation, we find the expressions of the higher-order charges $Q_{k\geq 6}$, and nontrivial coefficients appear there.
We derive the recursion equation for these coefficients of the connected diagrams in $Q_k$.
There are no other local charges independent of our $Q_k$, and any local charges are written in the linear combination of $Q_k$.

\textit{Hamiltonian and notations.---}
The Hamiltonian of the one-dimensional Hubbard model is
\begin{align}
    H
    =&
    -
    2
    \sum_{j=1}^L
    \sum_{\sigma=\uparrow,\downarrow}
    \paren{
    c_{j,\sigma}^\dag c_{j+1,\sigma}+\text{h.c.}
    }
    \nonumber\\
    &\hspace{3em}
    +4U 
    \sum_{j=1}^L
    \paren{n_{j,\uparrow}-\frac{1}{2}}
    \paren{n_{j,\downarrow}-\frac{1}{2}}
    ,
    \label{Hamiltonian}
\end{align}
where the periodic boundary condition is imposed and $n_{j\sigma}\equiv c_{j,\sigma}^\dag c_{j,\sigma}$ and $U$ is the coupling constant.

We denote the $k$-th local conserved quantity in terms of the polynomial of $U$ by:
\begin{equation}
    Q_k
    =
    \sum_{j=0}^{j_{f}}
    U^j
    Q^j_k
    ,
\end{equation}
where $j_f=k-1~(k-2)$ for even~(odd) $k$ and $Q_k^{j}$ is independent of $U$.
$Q_k$ is a linear combination of operators that act on at most $k$ adjacent sites.
We determine $Q_k^j$ to satisfy $\bck{Q_k,H}=0$.

We introduce some notations to represent $Q_k$.
We define a \textit{unit} of $+$~type starting from $j$-th site by:
\vspace{-0.75em}
\begin{equation}
\label{landdef}
\begin{tikzpicture}[baseline=0.25*\rd,scale=1]
\osdotu{0}{2}{3}
\osu{3}{4}
\ubrace{0}{4}{ $n$}
\end{tikzpicture}
_\sigma (j)
\coloneqq
2
\paren{
c_{j,\sigma} c_{j+n,\sigma}^{\dagger}
+(-1)^n
c_{j,\sigma}^{\dagger} c_{j+n,\sigma}
}
,
\end{equation}
where $n(>0)$ is the \textit{length} of the unit.
We define the zero-length unit by   
$
\begin{tikzpicture}[baseline=0.25*\rd,scale=1,]
    \zu{0}
\end{tikzpicture}
\hspace{0.1em}_\sigma (j)
\coloneqq
2n_{j,\sigma}-1
$, and define its type as $-$.
A unit of $-$ type with non-zero length is defined by
$
\begin{tikzpicture}[baseline=0.25*\rd,scale=1,]
    \osu{0}{1}
    \dottedlineup{1}{2}
    \osu{2}{3}
    \zu{3}
\end{tikzpicture}
\hspace{0.1em}_\sigma (j)
(
=
\begin{tikzpicture}[baseline=0.25*\rd,scale=1,]
    \osu{0}{1}
    \dottedlineup{1}{2}
    \osu{2}{3}
    \zu{0}
\end{tikzpicture}
\hspace{0.1em}_\sigma (j)
)
\coloneqq 
\begin{tikzpicture}[baseline=0.25*\rd,scale=1,]
    \zu{0}
\end{tikzpicture}
\hspace{0.1em}_\sigma (j)
\times
\begin{tikzpicture}[baseline=0.25*\rd,scale=1,]
    \osu{0}{1}
    \dottedlineup{1}{2}
    \osu{2}{3}
\end{tikzpicture}
\hspace{0.1em}_\sigma (j)
$~\cite{doublingnote}.

A \textit{diagram} represents a product of units, denoted by $\dv(i)=\prod_{\alpha=1}^{l_\dv} \psi_{\sigma_\alpha}^{t_\alpha, n_\alpha}(j_{\alpha, i})$ where $\psi_{\sigma}^{t, n}(j)$ is the unit starting from $j$-th site with type $t$, length $n$, and spin $\sigma$, and $j_{\alpha, i} \equiv i+ j_\alpha$, $j_1=0$, $j_{\alpha}\leq j_{\alpha+1}$.
$j_\alpha$ and $\sigma_\alpha$ satisfy if $\sigma_\alpha = \sigma_\beta (\alpha<\beta)$, then $j_\alpha^\prime < j_\beta $ ($j^\prime_{\alpha} \equiv j_{\alpha}+n_{\alpha}$) and if $j_{\alpha} = j_{\alpha+1}$, then $\sigma_\alpha = \uparrow, \sigma_{\alpha+1} = \downarrow$.
$l_\dv$ is the number of units in $\dv(i)$.
Note that units in a diagram mutually commute.
A diagram $\dv(i)$ has a graphical representation by a two-row sequence: $\psi_{\sigma_\alpha}^{t_\alpha, n_\alpha}(j_{\alpha,i})$ is placed on the upper~(lower) row for $\sigma_\alpha = \uparrow(\downarrow)$, with $j_\alpha$ columns being on its left.
Positions without a unit are denoted by~~\Ifig.
For example, the diagram $\dv(i)=\psi_{\uparrow}^{-,2}(i)\psi_{\downarrow}^{+,3}(i+1)\psi_{\uparrow}^{-,1}(i+4)$ is represented as:
\begin{equation}
\label{exampleof61diagram}
\hspace{-0.35em}
\begin{tikzpicture}[baseline=-0.5*\rd]
    \oszu{0}{2}
    \Isu{2}{4}
    \oszu{4}{5}
    \Isd{0}{1}
    \osd{1}{4}
    \Isd{4}{5}
\end{tikzpicture}
(i)
=
\begin{tikzpicture}[baseline=0.25*\rd]
    \oszu{0}{2}
\end{tikzpicture}
_{\hspace{0.1em}\uparrow}(i)
\!
\times
\begin{tikzpicture}[baseline=0.25*\rd]
    \osu{1}{4}
\end{tikzpicture}
_\downarrow
(i+1)
\!
\times
\begin{tikzpicture}[baseline=0.25*\rd]
    \oszu{4}{5}
\end{tikzpicture}
_{\hspace{0.1em}\uparrow}
(i+4)
.
\end{equation}
Note that units on the same row are separated by~\Ifig.
The interaction term is written as
$
\interaction(j)
=
\begin{tikzpicture}[baseline=0.5*\rd]
    \zu{0}
\end{tikzpicture}
_{\uparrow} (j)
\times
\begin{tikzpicture}[baseline=0.5*\rd]
    \zu{0}
\end{tikzpicture} 
_{\downarrow} (j)
$.
A diagram without a site index denotes the site translation summation, $\dv \coloneqq \sum_{i=1}^L\dv(i)$.

We define some integers for a diagram $\dv$.
First, we define $p_i$ by $\{p_1,\ldots,p_{2l_\dv}\} = \{j_1,\ldots,j_{l_\dv}, j^\prime_1,\ldots,j^\prime_{l_\dv}\}$ where $p_{i}\leq p_{i+1}$.
Then, we define \textit{support} by $s_\dv \coloneqq j_{l_\dv}+n_{l_\dv}+1$, \textit{double} by $d_\dv \coloneqq \sum_{i=1}^{l_\dv-1} (p_{2i+1} - p_{2i})$, \textit{gap-number} by $g_\dv \coloneqq \sum_{i=1}^{l_\dv-1} \delta_i(p_{2i+1} - p_{2i})$ where $\delta_i = 1 $ for $p_{2i+1} \in \{j_1,\ldots,j_{l_\dv}\}$ and otherwise $0$.
$s_\dv-1$ corresponds to the total number of columns in the two-row graphical representation of $\dv$.
We refer to a column 
$
\begin{tikzpicture}[baseline=-0.5*\rd]
    \osu{0}{1}
    \osd{0}{1}
\end{tikzpicture}
\paren{
\begin{tikzpicture}[baseline=-0.5*\rd]
    \gap{0}{1}
\end{tikzpicture}
}
$ as \textit{overlap}~(\textit{gap}).
$g_\dv$~($d_\dv$) corresponds to the total number of columns of gap~(gap and overlap).
$(s,d)$-diagram is a diagram satisfying $s_\dv = s$ and $d_\dv = d$.
We note that $s_\dv>d_\dv\geq g_\dv$. 
For the diagram of~\eqref{exampleof61diagram}, the integers are $(s_\dv,d_\dv,g_\dv,l_\dv)=(6,1,0,3)$.

\begin{figure}[t]
    \centering
    \includegraphics[clip]{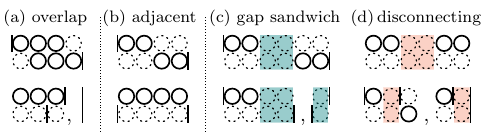}
    \caption{
        Examples of connections of units~(a)--(c), and the non-connected diagrams~(d).
        Units on the upper and lower rows of the diagrams in~(a)--(c) are connected.
        The gaps in (c~(d)) are indicated by the teal-~(orange-)~shaded area.
        The diagram on the bottom right in~(d) does not satisfy condition~(ii), while the others do not satisfy~(i).
        }
    \label{connectionfig}
\end{figure}

Two units in a diagram $\dv$, indexed by $\alpha$ and $\beta~(\alpha<\beta)$, are \textit{connected} if $\sigma_\alpha \neq \sigma_\beta$ and for any $\gamma (\neq \alpha, \beta)$, either of $j_{\gamma}^\prime < j_{\alpha}^\prime$ or $j_\beta < j_\gamma$ holds.
This condition can be categorized into three cases, as illustrated in Fig.~\ref{connectionfig}~(a)--(c).
A \textit{connected diagram} is a diagram satisfying (i)~for any two units in a diagram, indexed by $\alpha$ and $\beta$, there exists a sequence of indices of unit $\paren{\alpha=\gamma_0, \gamma_1,\ldots,\gamma_N,\gamma_{N+1}=\beta}$ where the $\gamma_i$-th and $\gamma_{i+1}$-th units are connected, and (ii)~the type of $\alpha$-th unit is $t_\alpha = (-)^{C_\alpha}$ where $C_\alpha$ is the number of units connected with it.
The diagrams in Fig.~\ref{connectionfig}~(a)--(c) and~\eqref{exampleof61diagram} are connected diagrams, and examples of non-connected diagrams are given in Fig.~\ref{connectionfig}~(d).
We found that $Q_k$ is a linear combination of connected diagrams.

\textit{Structure of $Q_k$.---}
We show the explicit form of lower-order charges previously found in terms of connected diagrams:
$
Q_2=H=
    \begin{tikzpicture}[baseline=-0.5ex,scale=1]
    \osu{0}{1}
    \Isd{0}{1}
    \end{tikzpicture}
    +
    \begin{tikzpicture}[baseline=-0.5ex,scale=1]
    \osd{0}{1}
    \Isu{0}{1}
    \end{tikzpicture}
    +
    U
    \begin{tikzpicture}[baseline=-0.5ex,scale=1]
    \zu{0}
    \zd{0}
    \end{tikzpicture}
$ 
and 
$
Q_3
    =
    \begin{tikzpicture}[baseline=-0.5*\rd]
        \osu{0}{2}
        \Isd{0}{2}
    \end{tikzpicture}
    +
    U
    \paren{
    \begin{tikzpicture}[baseline=-0.5*\rd]
        \oszu{0}{1}
        \Isd{0}{1}
        \zd{1}
    \end{tikzpicture}
    +
    \begin{tikzpicture}[baseline=-0.5*\rd]
        \osu{0}{1}
        \zu{0}
        \Isd{0}{1}
        \zd{0}
    \end{tikzpicture}
    }
    +
    \updownarrow
$, where $\updownarrow$ represents the diagrams with the upper and lower rows reversed, excluding those that remain invariant under this operation.
$Q_4$ and $Q_5$ is written as:
\begin{align}
Q_4
&
=
\begin{tikzpicture}[baseline=-0.5*\rd]
    \osu{0}{3}
    \Isd{0}{3}
\end{tikzpicture}
+
U
\left(
\ 
\begin{tikzpicture}[baseline=-0.5*\rd]
    \oszu{0}{2}
    \Isd{0}{2}
    \zd{0}
\end{tikzpicture}
+
\begin{tikzpicture}[baseline=-0.5*\rd]
    \oszu{0}{2}
    \Isd{0}{2}
    \zd{1}
\end{tikzpicture}
+
\begin{tikzpicture}[baseline=-0.5*\rd]
    \oszu{0}{2}
    \Isd{0}{2}
    \zd{2}
\end{tikzpicture}
+
\begin{tikzpicture}[baseline=-0.5*\rd]
    \oszu{0}{1}
    \Isu{1}{2}
    \oszd{1}{2}
    \Isd{0}{1}
\end{tikzpicture}
\right.
\nonumber\\
&
\quad\quad
\left.
+
\begin{tikzpicture}[baseline=-0.5*\rd]
    \osboth{0}{1}
    \zboth{1}
\end{tikzpicture}
    -
\begin{tikzpicture}[baseline=-0.5*\rd]
    \zu{0}
    \gap{0}{1}
    \zd{1}
\end{tikzpicture}
-
\begin{tikzpicture}[baseline=-0.5*\rd]
    \zboth{0}
\end{tikzpicture}
\ 
\right)
+
U^2
\begin{tikzpicture}[baseline=-0.5*\rd]
    \osu{0}{1}
    \Isd{0}{1}
    \zd{0}
    \zd{1}
\end{tikzpicture}
-
U^3
\begin{tikzpicture}[baseline=-0.5*\rd]
    \zboth{0}
\end{tikzpicture}
\
+
\updownarrow
,
\\
Q_5
& 
=
\begin{tikzpicture}[baseline=-0.5*\rd]
    \osu{0}{4}
    \Isd{0}{4}
\end{tikzpicture}
+
U(12 \text{ terms})
+
U^2
\left(\ \begin{tikzpicture}[baseline=-0.5*\rd]
\osu{0}{2}\zd{0}\zd{1}\Isd{0}{1}\Isd{1}{2}
\end{tikzpicture}+\begin{tikzpicture}[baseline=-0.5*\rd]
\osu{0}{2}\zd{0}\zd{2}\Isd{0}{2}
\end{tikzpicture}
\right.
\nonumber\\
&
\hspace{2em}
\left.
+\begin{tikzpicture}[baseline=-0.5*\rd]
\osu{0}{2}\zd{1}\zd{2}\Isd{1}{2}\Isd{0}{1}
\end{tikzpicture}
+\begin{tikzpicture}[baseline=-0.5*\rd]
\zu{0}\oszu{1}{2}\Isu{0}{1}\osd{0}{1}\Isd{1}{2}
\end{tikzpicture}+\begin{tikzpicture}[baseline=-0.5*\rd]
\oszu{0}{1}\zu{2}\Isu{1}{2}\osd{1}{2}\Isd{0}{1}
\end{tikzpicture}\right)
-U^3\left(\begin{tikzpicture}[baseline=-0.5*\rd]
\oszu{0}{1}\zd{0}\Isd{0}{1}
\end{tikzpicture}+\begin{tikzpicture}[baseline=-0.5*\rd]
\oszu{0}{1}\zd{1}\Isd{0}{1}
\end{tikzpicture}\right)
\
+
\updownarrow
,
\end{align}
where we omit the 12 terms of $Q_5^1$.
We newly obtained the explicit forms of higher-order $Q_k$ for $k\geq 6$ and found nontrivial coefficients that is not $\pm1$ appear, for example: 
\begin{align}
    Q_6^{j=3}
    &=
    \begin{tikzpicture}[baseline=-0.5*\rd]
\oszu{0}{2}\zd{0}\zd{1}\zd{2}\Isd{0}{1}\Isd{1}{2}
\end{tikzpicture}
    -\begin{tikzpicture}[baseline=-0.5*\rd]
\oszu{0}{2}\zd{0}\Isd{0}{2}
\end{tikzpicture}-\begin{tikzpicture}[baseline=-0.5*\rd]
\oszu{0}{2}\zd{1}\Isd{0}{1}\Isd{1}{2}
\end{tikzpicture}-\begin{tikzpicture}[baseline=-0.5*\rd]
\oszu{0}{2}\zd{2}\Isd{0}{2}
\end{tikzpicture}
-\begin{tikzpicture}[baseline=-0.5*\rd]
\oszu{0}{1}\oszd{1}{2}\Isd{0}{1}\Isu{1}{2}
\end{tikzpicture}+\begin{tikzpicture}[baseline=-0.5*\rd]
\zu{0}\osu{1}{2}\Isu{0}{1}\osd{0}{1}\zd{2}\Isd{1}{2}
\end{tikzpicture}
\nonumber\\
&\hspace{2em}
-2\begin{tikzpicture}[baseline=-0.5*\rd]
\oszu{0}{1}\oszd{0}{1}
\end{tikzpicture}+2\begin{tikzpicture}[baseline=-0.5*\rd]
\zu{0}\zd{1}\Isd{0}{1}\Isu{0}{1}
\end{tikzpicture}+5\begin{tikzpicture}[baseline=-0.5*\rd]
\zu{0}\zd{0}
\end{tikzpicture}
\
+
\updownarrow
.
\end{align}
We give the explicit forms of $Q_6, Q_7, Q_8$  as examples of higher-order $Q_k$ in~\cite{supp}.

\begin{figure}[b]
    \centering
    \includegraphics[clip, width=0.875\linewidth]{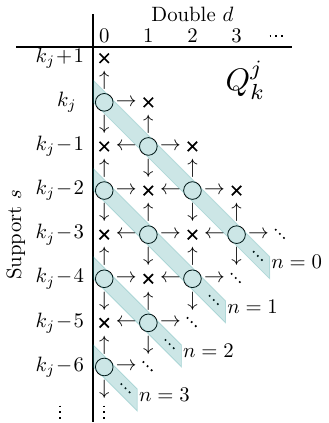}
    \caption{
    Structure of $Q_k^j$.
    $k_j\equiv k-j$.
    Circles at $(s,d)$ represent $(s,d)$-connected diagrams in $Q_k^j$ ($s>d$).
    The commutator of diagrams in the circle at $(s,d)$ with $H_0$ generates the diagrams in the crosses at $(s\pm 1,d)$ and $(s,d\pm 1)$, indicated by the solid arrow tip.
    }
    \label{Qkjfig}
\end{figure}

\begin{figure}[b]
    \centering
    \includegraphics[width=\linewidth]{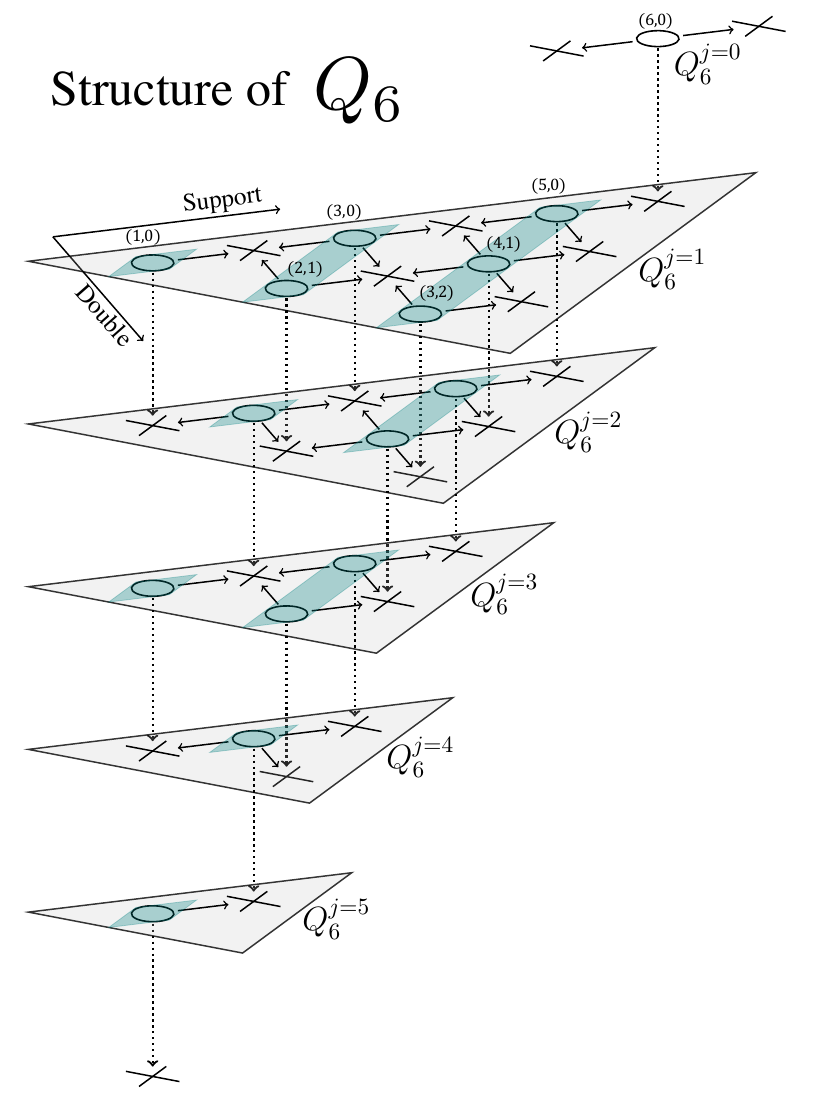}
    \caption{
    Structure of $Q_{k}$ for $k=6$.
    Each plane represents the structure of $Q_{k}^j$ in Fig.~\ref{Qkjfig}.
    The commutator of diagrams in the circle at $(s,d)$ in $Q_k^{j-1}$ with $H_\mathrm{int}$ generates diagrams in the cross at  $(s,d)$ in $Q_k^{j}$, indicated by the vertical dotted arrow tip.
    The diagrams generated in the crosses are to be canceled.
    }
    \label{Q63dfig}
\end{figure}

The diagrams in $Q_k^j$ are classified as $(s,d)$-connected diagrams as shown in Fig.~\ref{Qkjfig},~\ref{Q63dfig}, where circles represent $(s,d)$-connected diagrams in $Q^j_k$, and crosses represent diagrams generated by the commutator with the Hamiltonian $H=H_0+UH_1$, where $H_0=\begin{tikzpicture}[baseline=-0.5*\rd]\osu{0}{1}\Isd{0}{1}\end{tikzpicture}+\updownarrow$ and $H_1=\begin{tikzpicture}[baseline=-0.5*\rd]\zboth{0}\end{tikzpicture}\hspace{0.1em}$. 
The solid arrow in Fig.~\ref{Qkjfig},~\ref{Q63dfig} indicates the commutator with $H_0$. The vertical dotted arrow in Fig.~\ref{Q63dfig} indicates the commutator with $H_1$.
The diagrams at the crosses are to be canceled for the conservation law.

We give an example of commutators of units with $H_0$ and $H_1$~\cite{commutator_note}:
$
\bck{
\begin{tikzpicture}[baseline=-0.5*\rd]
\oszu{0}{1}
\Isd{0}{1}
\zd{0}
\end{tikzpicture}
,H_0
}
=
\begin{tikzpicture}[baseline=-0.5*\rd]
\oszu{0}{2}
\Isd{0}{2}
\zd{0}
\end{tikzpicture}
+
2
\begin{tikzpicture}[baseline=-0.5*\rd]
\oszu{0}{0}
\Isd{0}{0}
\zd{0}
\end{tikzpicture}
-
\begin{tikzpicture}[baseline=-0.5*\rd]
\oszu{-1}{1}
\Isd{-1}{1}
\zd{0}
\end{tikzpicture}
-
2
\begin{tikzpicture}[baseline=-0.5*\rd]
\zu{1}
\Isu{0}{1}
\Isd{0}{1}
\zd{0}
\end{tikzpicture}
+
\begin{tikzpicture}[baseline=-0.5*\rd]
\oszu{0}{1}
\oszd{0}{1}
\end{tikzpicture}
-
\begin{tikzpicture}[baseline=-0.5*\rd]
\Isu{-1}{0}
\oszu{0}{1}
\oszd{-1}{0}
\Isd{0}{1}
\end{tikzpicture}
$
and
$
\bck{
\begin{tikzpicture}[baseline=-0.5*\rd]
\oszu{0}{1}
\Isd{0}{1}
\zd{0}
\end{tikzpicture}
,H_1
}
=
\begin{tikzpicture}[baseline=-0.5*\rd]
\osu{0}{1}
\Isd{0}{1}
\zd{0}
\zd{1}
\end{tikzpicture}
-
\begin{tikzpicture}[baseline=-0.5*\rd]
\osu{0}{1}
\Isd{0}{1}
\end{tikzpicture}
$.
The commutator of connected diagrams and $H$ also generates a non-connected diagram; the details are given in~\cite{supp}.
We can construct $Q_k^j$ recursively by calculating the cancellation at the crosses in Fig.~\ref{Q63dfig} from top to bottom.

\textit{Exact expressions.---}
We define \textit{list} of diagram $\dv$, $\lambdabold_\dv=\bce{\lambda_1,\lambda_2,\ldots,\lambda_{l_\dv}}$ by  $\lambda_\alpha = p_{2\alpha}-p_{2\alpha-1}-\eta_\alpha$ where $\eta_\alpha = 0$ for $\alpha \in \{1, l_\dv\}$ and otherwise $\eta_\alpha = 1$. 
$\lambda_i$ represents the length of a sequence of \textit{coast} which we define as consecutive columns of 
$
\begin{tikzpicture}[baseline=-0.5*\rd]
    \osu{0}{2}
    \Isd{0}{2}
    \dottedlinemiddle{-1}{0}
    \dottedlinemiddle{2}{3}
\end{tikzpicture}
$
or 
$
\begin{tikzpicture}[baseline=-0.5*\rd]
    \osd{0}{2}
    \Isu{0}{2}
    \dottedlinemiddle{-1}{0}
    \dottedlinemiddle{2}{3}
\end{tikzpicture}
$
, as illustrated below:
\begin{equation}
    \dv=
    \begin{tikzpicture}[baseline=-0.5ex,scale=1,]
    \encloseboth{15.05}{16.95}{teal}{ fill=teal, opacity=0.35}
    \oszu{0}{3}
    \Isu{3}{5}
    \oszu{5}{15}
    \Isu{15}{23}
    \zu{20}
    \Isd{0}{2}
    \osd{2}{6}
    \Isd{6}{10}
    \oszd{10}{12}
    \Isd{12}{17}
    \osd{17}{23}
    \spacelengthabove{3.1}{4.9}{$\lambda_2\!+\!1$}{2.5}
    \spacelengthabove{17}{19.8}{$\lambda_5\!+\!1$}{2.5}
    \spacelengthabove{20.2}{23}{$\lambda_6$}{2.5}
    \spacelengthbelow{0}{1.9}{$\lambda_1$}{-2.5}
    \spacelengthbelow{6.1}{9.9}{$\lambda_3\!+\!1$}{-2.5}
    \spacelengthbelow{12.1}{14.9}{\hspace{0.em}$\lambda_4\!+\!1$}{-2.5}
    \end{tikzpicture}
\nonumber
\
,
\end{equation}
where $\lambdabold_\dv=\{2,1,3,2,2,3\}$,  and $\dv$ is the  $(24,6)$-connected diagram and $g_\dv=2$ and  $l_\dv=6$.  
The lengths of the coasts are indicated by the arrows, and the gap is indicated by the teal-shaded area.

We show the exact expression of $Q_k$.
$Q_k^0$ is the $(k,0)$-diagram:
$
Q_k^0
=
\begin{tikzpicture}[baseline=-0.5*\rd]
\osu{0}{1}
\dottedlineup{1}{2}
\osu{2}{3}
\Isd{0}{1}
\dottedlinedown{1}{2}
\Isd{2}{3}
\ubrace{0}{3}{{\footnotesize$ k-1$}}
\end{tikzpicture}
+
\updownarrow
$.
Note that $Q_k^0$ is the local charge for the $U=0$ case.
For $Q^j_k(j\geq 1)$, we obtain the following result.
\begin{theorem}
For $j\geq 1$, 
\begin{align}
    Q_k^j
    =
    \!\!
    {\sum_{\substack{0\leq n+d <  \lceil{\frac{k-j}{2} \rceil}, \\ n,d \geq 0 }}}
    \sum_{m=0}^{\lfloor{\frac{j-1}{2}}\rfloor}
    \sum_{g=0}^{d}
    (-1)^{n+m+g}
    \smashoperator{
    \sum_{\Psi\in \mathcal{S}_{n,d,g}^{k,j,m}}
    }
    C^{j,m}_{n,d}(\boldsymbol{\lambda}_{\Psi})
    \Psi
    ,
    \label{nonzerolinearcombination}
\end{align}
where $\connectedset_{n,d,g}^{k,j,m}$ is the set of  $(k-j-2n-d,d)$-connected diagrams $\dv$ with $l_\dv = j+1-2m$ and $g_\dv = g$. 
$C^{j,m}_{n,d}\paren{\lambda_1\ldots\lambda_{l}}\in \mathbb{Z}_{>0}$  is invariant under the permutation of $\lambda_i (2\leq i\leq l-1)$, and the exchange of $\lambda_1$ and $\lambda_{l}$.
\label{thm:qkj}
\end{theorem}
We note that the freedom to add $Q_{k'<k}$ to $Q_k$ is fixed by the above choice of $Q_k^0$ and the constraint of $l_\dv\geq 2$ for the diagram $\dv$ in $Q^j_k(j\geq 1)$. 
In this normalization, $Q_{2k} (Q_{2k+1})$ is even~(odd) under mirror reflection.
Our normalization is different from some of the previous studies.
For example, the fourth charge of~\cite{GRABOWSKI1995299} is $Q_4+U^2H$ in the spin variable notation~\cite{doublingnote}.
$(s,d)$-connected diagrams in $Q_k^j$ have the same coefficients if the lists are identical up to the permutation explained above.
$C^{j,m}_{n,d}\paren{\lambdabold_\dv}$ satisfies some other non-trivial identities given in~\cite{supp}.
\begin{theorem}\label{recursion}
$C^{j,m}_{n,d}(\lambdabold)$ is calculated from the following recursion equation.
    \begin{align}
        &C^{j,m}_{n,d}\paren{\lambdabold}
        =
        C^{j,m}_{n,d}\paren{\mathcal{T}\lambdabold}
        +
        \sum_{\Tilde{n}=0}^n
        \paren{n+1-\Tilde{n}}
        \nonumber\\
        &
        \quad
        \times
        \paren{
        C^{j-1,m-1}_{\Tilde{n},n+d-\Tilde{n}}\paren{\lambdabold_{\leftarrow 0}}
        -
        C^{j-1,m-1}_{\Tilde{n},n+d-\Tilde{n}}\paren{{}_{0\rightarrow}\mathcal{T}\lambdabold}
        }
        ,
    \end{align}
    where $\mathcal{T}\lambdabold=\bce{\lambda_1\!-\!1,\lambda_2,\ldots,\lambda_{l-1},\lambda_l\!+\!1}$, ${}_{0\rightarrow}\lambdabold=\bce{0,\lambda_1,\ldots}$,    $\lambdabold_{\leftarrow 0}=\bce{\ldots,\lambda_l,0}$, and $l\equiv j+1-2m$.
    For $\lambda_1=-1$ case, we define $C^{j,m}_{n,d}\paren{-1,\lambda_2,\mydots}\equiv C^{j,m}_{n,d-1}\paren{1,\lambda_2,\mydots}$ for $d>0$ and $C^{j,m}_{n,d=0}\paren{-1,\lambda_2,\mydots}\equiv
    C^{j,m}_{n-1,1}\paren{0,\lambda_2\!-\!1,\mydots}
    +C^{j-1,m}_{n,0}\paren{\lambda_2\!+\!1,\mydots}
    $.
    The initial condition is $C^{j=1,m=0}_{n,d}\paren{\lambdabold}=1$.
    $C^{j,m}_{n,d}\paren{\lambdabold}=0$ if $\lambda_i<0 (1 < i < l)$ or $n<0$ or $m<0$ or $m \geq \floor{j/2}$.
\end{theorem}

We obtained the general expressions of $C^{j,m}_{n,d}\paren{\lambdabold}$ for some cases.
For $n=0$ and $m=0$ case, we have
\begin{align}
    \label{coeffn=0}
    C^{j,m}_{n=0,d}\paren{\lambdabold}
    &=
    \binom{j-1+d}{m}-\binom{j-1+d}{m-1}
    ,
    \\
    \label{coeffm=0}
    C^{j,m=0}_{n,d}\paren{\lambdabold}
    &=
    \sum_{x_2=0}^{\lambda_2}\cdots \sum_{x_{j}=0}^{\lambda_{j}}
    \theta\paren{n-\sum_{i=2}^{j}x_i}
    ,
\end{align}
 where $j>1$ and $\theta(x)=1$ for $x\geq 0$ and $\theta(x)=0$ for $x< 0$. 
$C^{j,m}_{n=0,d}\paren{\lambdabold}$ is independent of $\lambdabold$ and $C^{j,m=0}_{n,d}\paren{\lambdabold}$ is independent of $\lambda_{1},\lambda_{j+1},d$.
We note that~\eqref{coeffn=0} is the generalized Catalan number~\cite{GRABOWSKI1995299,GrabowskiCatalantreepattern,doi:10.1142/S0217732394002057}.
The expressions are more complicated for the $n,m>0$ case.
For $j=3, m=1$ case, we have
\begin{align}
    &C^{j=3,m=1}_{n,d}\paren{\lambda_1,\lambda_2}
    =
    \sum_{\eta=\lambda_1,\lambda_2}\sum_{x=1}^{\eta}\binom{n+3-x}{3}
    \nonumber\\
    &
    \hspace{0em}
    +
    2\binom{n+4}{4}
    +
    (d-1)
    \bce{
    2
    \binom{n+3}{3}
    -
    \binom{n+2}{2}
    }.
    \label{coeffj3m1}
\end{align}
We also obtained the explicit expression of $Q^2_k,Q^3_k$ for all $k$ from~\eqref{coeffm=0} and~\eqref{coeffj3m1}.

Through the Jordan-Wigner transformation~\cite{GRABOWSKI1995299}, the 1D Hubbard model is mapped to coupled XX spin chains. 
A unit becomes the local conserved density in the XX chain: $\psi^{\pm,n(>0)}_{\mu}(i) \propto \sigma_{i,\mu}^{x}\sigma_{i+1,\mu}^{z}\cdots \sigma_{i+n-1,\mu}^{z}\sigma_{i+n,\mu}^{\bar{x}} + s \times (x \leftrightarrow y)$ and $\psi^{-,0}_{\mu}(i) = \sigma_{i,\mu}^{z}$ where $\sigma_{i,\mu}^{\alpha}$ is the Pauli matrix of flavor $\mu$, and $s = \mp (-1)^n$ and $\bar{x} = x(y)$ for $s = 1(-1)$~\cite{doublingnote}.

There are no other local conserved quantities independent of $Q_k$~\cite{fukai_complete_arxiv2023}.
This is shown as follows: if $F_k$ is a $k$-support charges, we can prove $F_k$ is written as $F_k = c_k Q_k + F_{k-1}$ where $F_{k-1}$ is a less than $k-1$-support charge and $c_k$ is some coefficient. 
Repeating this argument to $F_{k-1}$ and so on, we can see $F_k$ is a linear combination of $\bce{Q_l}_{l \leq k}$. 
The details of this proof are given in~\cite{fukai_complete_arxiv2023}.

From this completeness of our charges, we can see our $Q_k$ is written as a linear combination of local charges obtained from the transfer matrix, and we can confirm the mutual commutativity of our charges, $\bck{Q_k, Q_l}=0$, and their $\mathrm{SO}(4)$ symmetry~\cite{wadati1998}.
Our $Q_k$ coincides with the transfer matrix charges~\cite{Zhou1990} at least up to $Q_4$.

\textit{Summary and outlook.---}
We presented the exact expression for the local charges of the 1D Hubbard model $Q_k$. 
In Theorem~\ref{thm:qkj}, we proved $Q_k$ is constructed of connected diagrams, which represent the products of units, conserved densities of the XX chain in the spin variable notation, satisfying the conditions~(i) and~(ii).
The diagrams constructing $Q_k$ are accompanied by non-trivial coefficients~\cite{supp} for $k\geq 6$.
These coefficients can be calculated by the recursion equation in Theorem~\ref{recursion}.
Some of them are the generalized Catalan numbers~\eqref{coeffn=0}, which are also appearing in the local charges of the Heisenberg chain~\cite{anshelevich1980first,doi:10.1142/S0217732394002057, GRABOWSKI1995299, GrabowskiCatalantreepattern}.
Deriving the general explicit formula for the coefficients is the remaining task, which may be some further generalization of the Catalan number.
Our result is valid in both finite systems and the thermodynamic limit.

Our results have several applications: we can study the generalized Gibbs ensemble~\cite{pozsgay2013}, current mean value formula and the generalized hydrodynamics~\cite{pozsgay2017, sajat-currents, sajat-algebraic-currents, sajat-currents-review}, and factorization of correlation functions using local charges~\cite{fukai-TL-correlation-2023} in the 1D Hubbard model.
A model with fragmented Hilbert space can be derived by considering the strong coupling limit of the local charges in the XXZ chain~\cite{PhysRevE.104.044106}.
It is interesting to see what would happen in our case.
Recently, it has been shown that the quantum many-body scarring model can be constructed using odd-order charges $Q_{2k+1}$~\cite{PhysRevB.108.155102}.
We may also apply our results immediately in this direction.

To our knowledge, this is the first time revealing the structure of local conserved quantities in an integrable system without the boost operator, i.e.,~without a recursive way to construct them.

\textit{Acknowledgments.---}
The work was supported by FoPM, WINGS Program, JSR Fellowship, the University of Tokyo, and KAKENHI Grants No. JP21J20321 from the Japan Society for the Promotion of Science~(JSPS).

\let\oldaddcontentsline\addcontentsline
\renewcommand{\addcontentsline}[3]{}

\let\addcontentsline\oldaddcontentsline

\clearpage
\setcounter{equation}{0}
\makeatletter
\def\tagform@#1{\maketag@@@{(S#1)}}
\makeatother
\setcounter{figure}{0}
\renewcommand{\thefigure}{S\arabic{figure}}

\onecolumngrid

\vspace{0.5cm}
\begin{center} 
{\Large {\bf Supplemental Material  for ``All Local Conserved Quantities of the One-Dimensional Hubbard Model''}} 
\\
\vspace{0.3cm}
Kohei Fukai$ ^{\ast}$\\
\textit{The Institute for Solid State Physics, The University of Tokyo, Kashiwa, Chiba 277-8581, Japan}

\end{center}

\makeatletter
\def\l@subsection#1#2{}
\def\l@subsubsection#1#2{}
\makeatother

\makeatletter
\renewcommand{\@cite}[2]{[S#1]}
\renewcommand{\@biblabel}[1]{[S#1]}
\makeatother
\tableofcontents

\allowdisplaybreaks
\section*{S1. Commutation relations of diagram with Hamiltonian}
This section explains the commutation relation of a diagram and the Hamiltonian.
The commutators in this work have the additional factor of $1/2$: $\bck{A, B}\equiv \frac{1}{2} \paren{AB-BA}$.

The non-interacting part is written as 
$
H_0 
=
\sum_{\sigma = \uparrow, \downarrow}
\sum_{j=1}^{L}

_\sigma (j)
$.  
The commutators of units with the density of the non-interacting term
$
%
_\sigma (j)
$
are given by
\begin{align}
    \Bigl[
    %
    _{\sigma} (j-1)
    \times(-1)
    ,
\end{align}
and the other commutators are all zero.

The interaction term is written as 
$
H_\int
=
\sum_{j=1}^{L}
\interaction
(j)
$.
The commutators of units with the density of the interaction term
$
\interaction
(j)
$
are given by
\begin{align}
    \Bigl[
    %
    (j)
    ,
\end{align}
where $n\geq 1$.
The other commutators are all zero.

We introduce the notation to represent the commutator of a diagram with the Hamiltonian.
The two-row sequence representation of a diagram is denoted by 
$
%
$
indicates the commutator with 
$
%
_{\sigma}(j)
$
at the corresponding position~$j$ and flavor~$\sigma$.
For the two arrows in~\eqref{eq:graphical_Hint}, we often omit one of them in the following.
For the commutators with 
$%
_{\sigma}(j)$
which may increase the support, we introduce the following notations:
\begin{align}
    %
        _\downarrow (j-1)
    \Bigr]
    .
\end{align}

The examples of the explicit non-zero commutators with the density of $H_0$ are:
    \begin{align}
        %
        ,
        \nonumber
    \end{align}
where we only show the relevant parts in the diagram with the commutators, and the other unchanged parts are omitted.
The types of the units are unchanged by the commutator.

The following commutator generates two terms:
\begin{align}
    %
\end{align}
where in the second equality, we introduced the following notations:
\begin{align}
    %
    \times
    (-2)
    .
\end{align}

We introduce the unit with \textit{hole}.
\begin{align}
    %
    _{\sigma} (j)
    ,
\end{align}
where $1 \leq l \leq n-1$ and $2\leq n$ and the type of the unit with a hole in the first~(second) line is $+(-)$.
We call the part of 
$
%
$ as a hole.

The following commutator generates the unit with a hole:
\begin{align}
    %
    ,
\end{align}
where in the second line, we introduced the following notations:
\begin{align}
    %
    .
\end{align}
The type of the unit with a hole generated by the commutator is $-t_L t_R$ where $t_L$ and $t_R$ are the types of the units on the left and right which merge to form the unit with a hole by the commutator, respectively.
The other hole-making cases are
\begin{align}
    %
    }
    ,
\end{align}
where the type of the unit with a hole is determined in the same way as above and $t_L(t_R)=-1$ for the first~(second) equation for these cases.

The examples of the non-zero commutators with the density of $H_\int$ are
\begin{align}
    %
    \nonumber,
\end{align}
where the type of the units on the upper row is changed by the commutator.
The type of the units on the lower row in the third line is unchanged by the commutator, and the unit on the RHS has one hole.
The equations above hold when we exchange all the upper and lower rows.

We note that the following commutators are zero:
\begin{align}
    %
    =
    0.
\end{align}

We show examples of the commutator of diagrams with the Hamiltonian below.
We note that 
$
%
    ,
\end{align}
where we note that the second term from the end of~\eqref{eq:commuteeg} is a non-connected diagram.

We also note that the commutator of $\dv$ with $H_0$ generates connected diagrams whose lists are made by changing one element in $\lambdabold_\dv$ by one.

\newpage
\section{S2. Identities of $C^{j,m}_{n,d}\paren{\lambdabold}$}

In this section, we show the identities that $C^{j,m}_{n,d}\paren{\lambdabold}$ satisfies.
The proofs of the identities are given in section S4.

We redefine the notation for the list introduced in the main manuscript.
We shift the first index of the list from $1$ to $0$: $\bce{\lambda_1,\ldots,\lambda_{l}} \rightarrow \bce{\lambda_0,\ldots,\lambda_{l-1}}$.
In the following, for the list $\lambdabold=\bce{\lambda_0,\ldots,\lambda_{w+1}}$, we separate $\lambda_0(=\lambda_L)$ and $\lambda_{w+1}(=\lambda_R)$ from the other $\lambda_j (1\leq j \leq w)$ by semicolon, such as $\lambdabold=\{\lambda_L=\lambda_0;\lambda_1,\lambda_2,\ldots,\lambda_w;\lambda_{w+1}=\lambda_R\}$.
We introduce some notations for list $\lambdabold$.
$\lambdabold_{a\leftrightarrow b}$ denotes the configuration where $\lambda_a$ and $\lambda_b$ in $\lambdabold$ are swapped: $\bce{\ldots, \lambda_a, \ldots, \lambda_b, \ldots}_{a\leftrightarrow b} = \bce{\ldots, \lambda_b, \ldots, \lambda_a, \ldots}$.
$\lambdabold_{\hat{a}}$ denotes the configuration where $\lambda_a$ is removed from $\lambdabold$: $\bce{\ldots, \lambda_{a-1}, \lambda_{a}, \lambda_{a+1}, \ldots}_{\hat{a}} = \bce{\ldots, \lambda_{a-1},\lambda_{a+1}, \ldots}$.
$\lambdabold_{a\rightarrow A}$ denotes the configuration where $\lambda_a$ in $\lambdabold$ is replaced by $A$: $\bce{\ldots, \lambda_{a}, \ldots}_{a \rightarrow A} = \bce{\ldots, A , \ldots}$
.
We define $\lambdabold_{a:\pm \delta}\equiv \lambdabold_{a\rightarrow \lambda_a\pm \delta}$.
We note that $\mathcal{T}\lambdabold = \lambdabold_{L:-1, R:+1}$.
We also introduce $\vec{\lambda} \equiv \bce{\lambda_1,\lambda_2,\ldots,\lambda_w}$, and $\lambdabold$ is written as $\lambdabold = \{\lambda_L; \vec{\lambda} ; \lambda_R\}$.

$C^{j,m}_{n,d}(\lambdabold)$ satisfies the following identities:
\begin{align}
    C^{j,m}_{n,d}\paren{\lambdabold}
    &=
    C^{j,m}_{n,d}\paren{\lambdabold_{L\leftrightarrow R}}
    ,
    \label{eq:LRswap}
    \\
    &=
    C^{j,m}_{n,d}\paren{\lambdabold_{i_1\leftrightarrow i_2}}
    \quad(w\geq 2)
    ,
    \label{eq:swap}
    \\
    &=
    C^{j,m}_{n,d}\paren{\lambdabold_{a\rightarrow \min(\lambda_a,n)}}
    ,
    \label{eq:minprop}
    \\
    &=
    C^{j,m}_{n-1,d+2}\paren{\lambdabold_{{L(R)}:-1,i:-1}}
    +
    C^{j-1,m}_{n,d+1}(\lambdabold_{L(R)\rightarrow \lambda_{L(R)}+\lambda_{i}, \hat{i}})
     \quad(w\geq 1)
    ,
    \label{eq:betweenjLR}
    \\
    &=
    C^{j,m}_{n-1,d+2}\paren{\lambdabold_{i_1:-1,i_2:-1}}
    +
    C^{j-1,m}_{n,d+1}(\lambdabold_{i_1\rightarrow \lambda_{i_1}+\lambda_{i_2},\hat{i_2}})
    \quad(w\geq 2)
    ,
    \label{eq:betweenj}
    \\
    &=
    C^{j,m}_{n,d}\paren{\lambdabold_{a:-1,b:+1}}
    +
    C^{j,m}_{n-1,d+2}\paren{\lambdabold_{a:-1,b:-1}}-C^{j,m}_{n-1,d+2}\paren{\lambdabold_{a:-2}}
    \quad(\lambda_a>0)
    ,
    \label{eq:oxo}
\end{align}
where $\lambdabold = \{ \lambda_0,\ldots,\lambda_{w+1}\}$ ($w\equiv j-1-2m$), and $\lambda_a \geq 0$, and $1\leq i,i_1,i_2\leq w$ and $0\leq a,b\leq w+1$  ($a, b=L(R)$ means $a, b=0(w+1)$).
In the case of $\lambda_i =0$ in~\eqref{eq:betweenjLR}, we have
\begin{align}
    C^{j,m}_{n,d}\paren{\lambdabold_{i\rightarrow 0}}
    =
    C^{j-1,m}_{n,d+1}\paren{\lambdabold_{\hat{i}}}
    .
    \label{forholecase}
\end{align}
Using~\eqref{eq:oxo} repeatedly, we have ($p>0$):
\begin{align}
    \label{eq:oxo2}
    C^{j,m}_{n,d}\paren{\lambdabold}
    -
    C^{j,m}_{n,d}\paren{\lambdabold_{a:-p,b:p}}
    =
    C^{j,m}_{n-1,d+2}\paren{\lambdabold_{a:-1,b:-1}}-C^{j,m}_{n-1,d+2}\paren{\lambdabold_{a:-p-1, b:p-1}}
    \quad(\lambda_a\geq p)
    .
\end{align}

We review the recursion equation of Theorem 2.
\begin{align}
    \label{eq:basiceq3}
    C^{j,m}_{n,d}\paren{\lambdabold}
    &=
    C^{j,m}_{n,d}(\lambda_L-1;\vec{\lambda}; \lambda_R+1)
    \nonumber\\
    &\quad
    +
    \sum_{\Tilde{n}=0}^{n}
    (n+1-\Tilde{n})
    \paren{
    C^{j-1,m-1}_{\Tilde{n},n+d-\Tilde{n}}(\lambda_L;\vec{\lambda}, \lambda_R;0)
    -
    C^{j-1,m-1}_{\Tilde{n},n+d-\Tilde{n}}(0;\lambda_L-1,\vec{\lambda}; \lambda_R+1)
    }
    .
\end{align}
For the  $\lambda_L=-1$ case, we define 
\begin{align}
    \label{eq:Lm1def}
    C^{j,m}_{n,d}\paren{-1;\lambda_1,\ldots}
    \coloneqq
    \begin{cases}
        C^{j,m}_{n,d-1}\paren{1;\lambda_1,\ldots} &(d>0)
        \\
        C^{j,m}_{n-1,1}\paren{0;\lambda_1-1,\ldots}+C^{j-1,m}_{n,0}\paren{\lambda_1+1;\ldots} &(d=0)
    \end{cases}
    .
\end{align}
With this definition in the case of $\lambda_{L}=-1$, we can see~\eqref{eq:LRswap} and~\eqref{eq:swap} hold also in the case of $\lambda_{L(R)}=-1$.

The recursion equation can be rewritten as 
\begin{align}
    \Delta C^{j,m}_{n, d}\paren{\lambdabold}
    &=
    2\Delta C^{j,m}_{n-1, d+1}\paren{\lambdabold}
    -
    \Delta C^{j,m}_{n-2, d+2}\paren{\lambdabold}
    +
    C^{j-1,m-1}_{n, d}\paren{\lambdabold_{\leftarrow 0}}
    -
    C^{j-1,m-1}_{n, d}\paren{{}_{0\rightarrow}(\mathcal{T}\lambdabold)}
            ,
    \label{eq:basiceq}
\end{align}
where $\Delta C^{j,m}_{n, d}\paren{\lambdabold}=C^{j,m}_{n, d}\paren{\lambdabold}-C^{j,m}_{n, d}\paren{\mathcal{T}\lambdabold}$.

\begin{figure}[tb]
    \centering
    \includegraphics[width=0.7\linewidth]{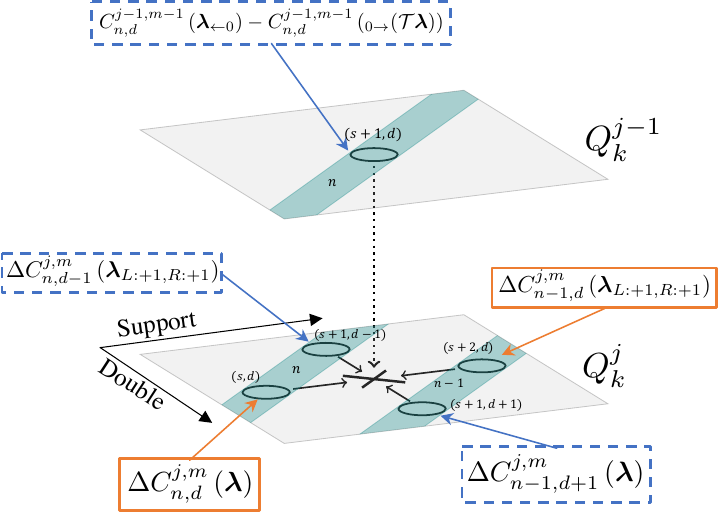}
    \caption{
    The general structure of the cancellation of diagrams in $\bck{Q_k, H}=0$.
    The upper plane represents $Q_k^{j-1}$ and lower plane represents $Q_k^{j}$.
    We show only the components of $Q_k$ that are relevant to the cancellation of diagrams in the crosses at $(s+1,d)$ in the plane of $Q_k^{j}$.
    The coefficients encircled with the orange solid (blue dotted) line are the coefficients of the diagrams in the circles indicated by each arrow tip and contribute to the recursion relation in the form of~\eqref{eq:basiceq2} with the factor of $+1(-1)$.
    }   \label{recursion_3d}
\end{figure}

The recursion equation reflects the local cancellation in $\bck{Q_k, H}=0$ (Fig.~\ref{recursion_3d}).
To see this, we further rewrite~\eqref{eq:basiceq} as
\begin{align}
    0
    =&
    \Delta C^{j,m}_{n, d}\paren{\lambdabold}
    -
    \Delta C^{j,m}_{n-1, d+1}\paren{\lambdabold}
    -
    \Delta C^{j,m}_{n, d-1}\paren{\lambdabold_{L:+1, R: +1}}
    +
    \Delta C^{j,m}_{n-1, d}\paren{\lambdabold_{L:+1, R: +1}}
    \nonumber\\
    &\hspace*{5em}
    -
    \paren{
    C^{j-1,m-1}_{n, d}\paren{\lambdabold_{\leftarrow 0}}
    -
    C^{j-1,m-1}_{n, d}\paren{{}_{0\rightarrow}
    (\mathcal{T}\lambdabold)}
    }
    ,
    \label{eq:basiceq2}
\end{align}
where we used the relation obtained  from~\eqref{eq:oxo}:
$\Delta C^{j,m}_{n-1, d+1}\paren{\lambdabold}
=
\Delta C^{j,m}_{n, d-1}\paren{\lambdabold_{L:+1, R: +1}} 
$
and $\Delta C^{j,m}_{n-2, d+2}\paren{\lambdabold} = \Delta C^{j,m}_{n-1, d}\paren{\lambdabold_{L:+1, R: +1}}$
.
The first, second, third,  and fourth terms in the RHS of~\eqref{eq:basiceq2} are the coefficients of the $(s,d), (s+1,d+1), (s+1,d-1), (s+2,d)$-diagrams in $Q_k^j$ respectively ($s=k-j-2n-d$), which are represented by the circles in the $Q_k^j$ plane in Fig.~\ref{recursion_3d}. 
The last term in the RHS of~\eqref{eq:basiceq2} is the coefficients of the $(s+1,d)$ diagrams in $Q_k^{j-1}$, which are represented by the circle in the $Q_k^{j-1}$ plane in Fig.~\ref{recursion_3d}. 
These commutators generate the diagrams in the cross at $(s+1,d)$ on the $Q_k^j$ plane in Fig.~\ref{recursion_3d}, which are to be canceled for the conservation law.

We can easily show the explicit expressions for $C^{j,m=0}_{n,d}(\lambdabold)$ and $C^{j,m}_{n=0,d}(\lambdabold)$ and $C^{j=3,m=1}_{n,d}(\lambdabold)$ in the main manuscript satisfy the recursion equations and the identities above.

We prove~\eqref{eq:basiceq}.
From the recursion equation, we have 
\begin{align}
    &
    \Delta C^{j,m}_{n, d}\paren{\lambdabold}
    -2\Delta C^{j,m}_{n-1, d+1}\paren{\lambdabold}
    +\Delta C^{j,m}_{n-2, d+2}\paren{\lambdabold}
    \nonumber\\
    =&
    \sum_{\Tilde{n}=0}^{n}
    (n+1-\Tilde{n})
    \left\{
    C^{j-1,m-1}_{\Tilde{n},n+d-\Tilde{n}}(\lambda_L;\vec{\lambda}, \lambda_R;0)
    -
    C^{j-1,m-1}_{\Tilde{n},n+d-\Tilde{n}}(0; \lambda_L-1;\vec{\lambda}; \lambda_R+1)
    \right\}
    \nonumber\\
    &
    -2
    \sum_{\Tilde{n}=0}^{n-1}
    (n-\Tilde{n})
    \left\{
    C^{j-1,m-1}_{\Tilde{n},n+d-\Tilde{n}}(\lambda_L;\vec{\lambda}, \lambda_R;0)
    -
    C^{j-1,m-1}_{\Tilde{n},n+d-\Tilde{n}}(0; \lambda_L-1;\vec{\lambda}; \lambda_R+1)
    \right\}
    \nonumber\\
    &+
    \sum_{\Tilde{n}=0}^{n-2}
    (n-1-\Tilde{n})
    \left\{
    C^{j-1,m-1}_{\Tilde{n},n+d-\Tilde{n}}(\lambda_L;\vec{\lambda}, \lambda_R;0)
    -
    C^{j-1,m-1}_{\Tilde{n},n+d-\Tilde{n}}(0; \lambda_L-1;\vec{\lambda}; \lambda_R+1)
    \right\}
    \nonumber\\
    =&
    C^{j-1,m-1}_{n,d}(\lambda_L;\vec{\lambda}, \lambda_R;0)
    -
    C^{j-1,m-1}_{n,d}(0;\lambda_L-1,\vec{\lambda}; \lambda_R+1)
    +2\bce{
        C^{j-1,m-1}_{n-1,d+1}(\lambda_L;\vec{\lambda}, \lambda_R;0)
        -
        C^{j-1,m-1}_{n-1,d+1}(0;\lambda_L-1,\vec{\lambda}; \lambda_R+1) } 
    \nonumber\\
    &
    -2\bce{
        C^{j-1,m-1}_{n-1,d+1}(\lambda_L;\vec{\lambda}, \lambda_R;0)
        -
        C^{j-1,m-1}_{n-1,d+1}(0;\lambda_L-1,\vec{\lambda}; \lambda_R+1)
    }
    \nonumber\\
    =&
    C^{j-1,m-1}_{n,d}(\lambda_L;\vec{\lambda}, \lambda_R;0)
    -
    C^{j-1,m-1}_{n,d}(0;\lambda_L-1,\vec{\lambda}; \lambda_R+1)
    ,
\end{align}
where in the first equality, we used~\eqref{eq:oxo}.
Thus, we have proved~\eqref{eq:basiceq}.

\newpage
\section{S3. Proof of conservation law of $Q_k$}
In this section, we prove Theorem 1 and Theorem 2 in the main manuscript by showing that $Q_k$ constructed from the result of  Theorem 1 and Theorem 2 is actually conserved, i.e.~we prove $\bck{Q_k,H}=0$.
We represent the $\bck{Q_k,H}$ as
\begin{align}
    \bck{Q_k,H}
    &=
    \sum_{j=0}^{j_f+1}U^{j}
    \bce{
        \bck{Q_k^j,H_0}+\bck{Q_k^{j-1},H_\mathrm{int}}
    }
    \nonumber\\
    &=
    \sum_{j=1}^{j_f+1}U^{j}
    \sum_{n=0}^{\floor{\frac{k-1-j}{2}}}
    \sum_{m=0}^{\ceil{\frac{j-1}{2}}}
    \sum_{d=0}^{\floor{\frac{k-1-j}{2}}-n}
    \sum_{g=0}^{d}
    \sum_{\dvx\in \mathcal{F}^{k,j,m}_{n,d,g}}
    \widetilde{D}^{k,j,m}_{n,d,g}(\dvx)\dvx
    ,
\end{align}
where $1\leq j\leq j_f\equiv 2\floor{k/2}-1$ and $Q_k^{-1}=Q_k^{j_f+1}=0$
and $\mathcal{F}^{k,j,m}_{n,d,g}$ is the set of $(k-j-2n-d+1,d)$-diagrams $\dvx$ whose unit number is $j+1-2m$ and gap number is $g$.
The $U^0$ terms cancel because of $\bck{Q_k^0, H_0}=0$.
Note that $\mathcal{F}^{k,j,m}_{n,d,g}$ include not only connected diagrams  but also non-connected diagrams.
What we have to prove is $\widetilde{D}^{k,j,m}_{n,d,g}(\dvx)=0$ for all $\dvx\in\mathcal{F}^{k,j,m}_{n,d,g}$.

\subsubsection{Cancellation of connected diagram}
We prove $\widetilde{D}^{k,j,m}_{n,d,g}(\dvx)=0$ in the case in which $\dvx\in\mathcal{F}^{k,j,m}_{n,d,g}$ is a connected diagram and $l_\dvx > 1$, i.e. $j > 2m$.
The list of $\dvx$ is denoted by $\lambdabold = \bce{\lambda_L \equiv \lambda_0;\lambda_1,\ldots,\lambda_w;\lambda_{w+1} \equiv \lambda_R}$, and $w \equiv j-1-2m$.
$\widetilde{D}^{k,j,m}_{n,d,g}(\dvx)$ is written as
\begin{align}
    \widetilde{D}^{k,j,m}_{n,d,g}(\dvx)
    &=
    A_0^{\sigma_0^{R}}(\dvx)
    +
    \sum_{i=1}^{w}
    A_{i}^{\sigma_i^{L},\sigma_i^{R}}(\dvx)
    +A_{w+1}^{\sigma_{w+1}^{L}}(\dvx)
    ,
\end{align}
where $A_{i}^{\sigma_i^{L},\sigma_i^{R}}(\dvx)$ is the contribution to the cancellation of $\dvx$ in $\bck{Q_k,H}$ from the diagram in $Q_k^j$ whose list is $\lambdabold_{i:\pm 1}$.
$A_{0}^{\sigma_0^{R}}(\dvx)$ is the contribution to the cancellation of $\dvx$ from the diagram in $Q_k^j$ whose list is $\lambdabold_{L:\pm 1}$ and the diagram in $Q_k^{j-1}$ whose list is ${}_{0\rightarrow}(\lambdabold_{L:-1})$ and the diagram in $Q_k^{j-1}$ whose list is $(\lambdabold_{\hat{L}})_{L:+1}$ in the case of $\lambda_L = 0$.
$A_{w+1}^{\sigma_0^{L}}(\dvx)$ is the contribution to the cancellation of $\dvx$ from the diagram in $Q_k^j$ whose list is $\lambdabold_{R:\pm 1}$ and the diagram in $Q_k^{j-1}$ whose list is $(\lambdabold_{R:-1})_{\leftarrow 0}$ and the diagram in $Q_k^{j-1}$ whose list is $(\lambdabold_{\hat{R}})_{R:+1}$ in the case of $\lambda_R = 0$.
There is no other contribution to the cancellation of $\dvx$ other than the above contribution.
$\sigma_i^{L}=+(-)$ if there is (is no) overlap or gap between the $i$-th coast and $i-1$-th coast.
$\sigma_i^{R}=+(-)$ if there is (is no) overlap or gap between the $i$-th coast and $(i+1)$-th coast.
Therefore, $\sigma_i^{R}=\sigma_{i+1}^{L}$ holds.
In the following, we write $A_0^{\sigma_0^{R}}(\dvx), A_{i}^{\sigma_i^{L},\sigma_i^{R}}(\dvx), A_{w+1}^{\sigma_{w+1}^{L}}(\dvx)$ simply as $A_0^{\sigma_0^{R}}, A_{i}^{\sigma_i^{L},\sigma_i^{R}}, A_{w+1}^{\sigma_{w+1}^{L}}$.

We give the examples of the case of $\sigma_i^{R}=\sigma_{i+1}^{L}=-$, i.e., the case in which there are no overlap or gap between $i$-th and $(i+1)$-th coasts:
\begin{align}
    \begin{tikzpicture}[baseline=-0.5*\rd]
            \dottedlineboth{-1}{0}
            \osu{0}{3}
            \lengtarrowdownoneside{0}{2.85}{$\hspace{-0.5em}\lambda_i+1$}
            \osu{3}{6}
            \zd{3}
            \lengtarrowdownoneside{6}{3.15}{$\hspace{1.5em}\lambda_{i+1}+1$}
            \dottedlineboth{6}{7}
    \end{tikzpicture}
    ,
    \quad
    \begin{tikzpicture}[baseline=-0.5*\rd]
            \dottedlineboth{-1}{0}
            \osu{0}{3}
            \lengtarrowdownoneside{0}{3}{$\hspace{-0.5em}\lambda_i+1$}
            \osd{3}{6}
            \lengtarrowuponeside{6}{3}{$\hspace{1.5em}\lambda_{i+1}+1$}
            \dottedlineboth{6}{7}
    \end{tikzpicture}
    .
    \nonumber
\end{align}

We give the example of the case of $\sigma_i^{R}=\sigma_{i+1}^{L}=+$, i.e., the case in which there is some overlap or gap between $i$-th and $(i+1)$-th coasts:
\begin{align}
    \begin{tikzpicture}[baseline=-0.5*\rd]
            \dottedlineboth{-1}{0}
            \osu{0}{3}
            \lengtarrowdownoneside{0}{3}{$\hspace{-0.5em}\lambda_i+1$}
            \osu{3}{8}
            \oszd{3}{5}
            \lengtarrowdownoneside{8}{5}{$\hspace{1.5em}\lambda_{i+1}+1$}
            \dottedlineboth{8}{9}
    \end{tikzpicture}
    ,
    \quad
    \begin{tikzpicture}[baseline=-0.5*\rd]
            \dottedlineboth{-1}{0}
            \osu{0}{3}
            \lengtarrowdownoneside{0}{3}{$\hspace{-1em}\lambda_i+1$}
            \gap{3}{5}
            \osd{5}{8}
            \lengtarrowuponeside{8}{5}{$\hspace{1.5em}\lambda_{i+1}+1$}
            \dottedlineboth{8}{9}
    \end{tikzpicture}
    .
    \nonumber
\end{align}

$A_i^{\sigma_i^{L}\sigma_i^{R}}$ satisfies the following relations:
\begin{align}
    &A_i^{++}=A_i^{--}=0,
    \label{A++}
    \\
    &A_i^{+-}=-A_i^{-+}
    =
    (-1)^{n+m+g}
    \left\{
    C^{j,m}_{n-1, d+1}\paren{\lambdabold_{i:-1}}
    +
    C^{j,m}_{n, d-1}\paren{\lambdabold_{i:+1}}
    \right\},
    \label{A+-}
    \\
    &
    A_0^{+}
    =
    (-1)^{n+m+g}
    \left\{
    C^{j,m}_{n-1, d+1}\paren{\lambdabold_{L:-1}}
    -
    C^{j,m}_{n, d}\paren{\lambdabold_{L:-1}}
    \right.
    \nonumber\\
    &
    \left.
    \hspace{13em}
    +
    C^{j,m}_{n-1, d}\paren{\lambdabold_{L:+1}}
    -
    C^{j,m}_{n, d-1}\paren{\lambdabold_{L:+1}}
    +
    C^{j-1,m-1}_{n, d}\paren{{}_{0\rightarrow}(\lambdabold_{L:-1})}
    \right\},
    \label{AL+}
    \\
    &
    A_{w+1}^{+}
    =
    (-1)^{n+m+g+1}
    \left\{
    C^{j,m}_{n-1, d+1}\paren{\lambdabold_{R:-1}}
    -
    C^{j,m}_{n, d}\paren{\lambdabold_{R:-1}}
    \right.
    \nonumber\\
    &
    \left.
    \hspace{13em}
    +
    C^{j,m}_{n-1, d}\paren{\lambdabold_{R:+1}}
    -
    C^{j,m}_{n, d-1}\paren{\lambdabold_{R:+1}}
    +
    C^{j-1,m-1}_{n, d}\paren{(\lambdabold_{R:-1})_{\leftarrow 0}}
    \right\},
    \label{AR+}
    \\
    &
    A_0^{-}
    =
    (-1)^{n+m+g}
    \left\{
    2
    C^{j,m}_{n-1, d+1}\paren{\lambdabold_{L:-1}}
    -
    C^{j,m}_{n, d}\paren{\lambdabold_{L:-1}}
    +
    C^{j,m}_{n-1, d}\paren{\lambdabold_{L:+1}}
    +
    C^{j-1,m-1}_{n, d}\paren{{}_{0\rightarrow}(\lambdabold_{L:-1})}
    \right\},
    \label{AL-}
    \\
    &
    A_{w+1}^{-}
    =
    (-1)^{n+m+g+1}
    \left\{
    2
    C^{j,m}_{n-1, d+1}\paren{\lambdabold_{R:-1}}
    -
    C^{j,m}_{n, d}\paren{\lambdabold_{R:-1}}
    +
    C^{j,m}_{n-1, d}\paren{\lambdabold_{R:+1}}
    +
    C^{j-1,m-1}_{n, d}\paren{(\lambdabold_{R:-1})_{\leftarrow 0}}
    \right\}
    \label{AR-}
    .
\end{align}
The proof of these equations is given in section S5.

Using the identity~\eqref{eq:oxo}, we have
\begin{align}
    A_{i_1}^{+-}+A_{i_2}^{-+}
    &=
    (-1)^{n+m+g}
    \left\{
    C^{j,m}_{n-1, d+1}\paren{\lambdabold_{i_1:-1}}
    +
    C^{j,m}_{n, d-1}\paren{\lambdabold_{i_1:+1}}
    -
    \paren{
    C^{j,m}_{n-1, d+1}\paren{\lambdabold_{i_2:-1}}
    +
    C^{j,m}_{n, d-1}\paren{\lambdabold_{i_2:+1}}
    }
    \right\}
    \nonumber\\
    &=
    (-1)^{n+m+g}
    \left\{
    C^{j,m}_{n-1, d+1}\paren{\lambdabold_{i_1:-1}}
    -
    C^{j,m}_{n-1, d+1}\paren{\lambdabold_{i_2:-1}}
    -
    \paren{
    C^{j,m}_{n, d-1}\paren{\lambdabold_{i_2:+1}}
    -
    C^{j,m}_{n, d-1}\paren{\lambdabold_{i_1:+1}}
    }
    \right\}
    \nonumber\\
    &=0
    ,
\end{align}
where we used~\eqref{eq:oxo} in the last equality.
In the same way, we also have $ A_{i_1}^{-+}+A_{i_2}^{+-}=0$.

Using the identity~\eqref{eq:oxo}, we have
\begin{align}
    A_0^{-}
    +
    A_{i}^{-+}
    &=
    (-1)^{n+m+g}
    \left\{
    2
    C^{j,m}_{n-1, d+1}\paren{\lambdabold_{L:-1}}
     -
    C^{j,m}_{n, d}\paren{\lambdabold_{L:-1}}
    +
    C^{j,m}_{n-1, d}\paren{\lambdabold_{L:+1}}
    +
    C^{j-1,m-1}_{n, d}\paren{{}_{0\rightarrow}(\lambdabold_{L:-1})}
    \right.
    \nonumber\\
    &
    \left.
    \hspace{15em}
    -\paren{
    C^{j,m}_{n-1, d+1}\paren{\lambdabold_{i:-1}}
    +
    C^{j,m}_{n, d-1}\paren{\lambdabold_{i:+1}}
    }
    \right\}
    \nonumber\\
    &=
    (-1)^{n+m+g}
    \left\{
    C^{j,m}_{n-1, d+1}\paren{\lambdabold_{L:-1}}
     -
    C^{j,m}_{n, d}\paren{\lambdabold_{L:-1}}
    +
    C^{j,m}_{n-1, d}\paren{\lambdabold_{L:+1}}
    +
    C^{j-1,m-1}_{n, d}\paren{{}_{0\rightarrow}(\lambdabold_{L:-1})}
    \right.
    \nonumber\\
    &
    \left.
    \hspace{15em}
    -\paren{
    C^{j,m}_{n-1, d+1}\paren{\lambdabold_{i:-1}}
    +
    C^{j,m}_{n, d-1}\paren{\lambdabold_{i:+1}}
    -
    C^{j,m}_{n-1, d+1}\paren{\lambdabold_{L:-1}}
    }
    \right\}
    \nonumber\\
    &=
    (-1)^{n+m+g}
    \left\{
    C^{j,m}_{n-1, d+1}\paren{\lambdabold_{L:-1}}
     -
    C^{j,m}_{n, d}\paren{\lambdabold_{L:-1}}
    +
    C^{j,m}_{n-1, d}\paren{\lambdabold_{L:+1}}
    +
    C^{j-1,m-1}_{n, d}\paren{{}_{0\rightarrow}(\lambdabold_{L:-1})}
    \right.
    \nonumber\\
    &
    \left.
    \hspace{14em}
    -C^{j,m}_{n, d-1}\paren{\lambdabold_{L:+1}}
    \right\}
    \nonumber\\
    &=
    A_{0}^+
    ,
\end{align}
where we used~\eqref{eq:oxo} in the third equality.

In the same way, we have 
\begin{align}
    A_i^{+-}+A_{w+1}^{-}=A_{w+1}^{+}.
\end{align}

Using the identity~\eqref{eq:oxo}, we have
\begin{align}
	 A_{0}^{+}+A_{i}^{+-}
   	 &=
    (-1)^{n+m+g}
    \left\{
    C^{j,m}_{n-1, d+1}\paren{\lambdabold_{L:-1}}
    -
    C^{j,m}_{n, d}\paren{\lambdabold_{L:-1}}
    +
    C^{j,m}_{n-1, d}\paren{\lambdabold_{L:+1}}
    -
    C^{j,m}_{n, d-1}\paren{\lambdabold_{L:+1}}
    +
    C^{j-1,m-1}_{n, d}\paren{{}_{0\rightarrow}(\lambdabold_{L:-1})}
    \right.
    \nonumber\\
    &
    \left.
    \hspace{13em}
    +
     C^{j,m}_{n-1, d+1}\paren{\lambdabold_{i:-1}}
    +
    C^{j,m}_{n, d-1}\paren{\lambdabold_{i:+1}}
    \right\}
    \nonumber\\
     &=
    (-1)^{n+m+g}
    \left\{
    C^{j,m}_{n-1, d+1}\paren{\lambdabold_{L:-1}}
    -
    C^{j,m}_{n, d}\paren{\lambdabold_{L:-1}}
    +
    C^{j,m}_{n-1, d}\paren{\lambdabold_{L:+1}}
      +
    C^{j-1,m-1}_{n, d}\paren{{}_{0\rightarrow}(\lambdabold_{L:-1})}
    \right.
    \nonumber\\
    &
    \left.
    \hspace{13em}
    +
     C^{j,m}_{n-1, d+1}\paren{\lambdabold_{i:-1}}
    +
    \paren{
    C^{j,m}_{n, d-1}\paren{\lambdabold_{i:+1}}
     -
    C^{j,m}_{n, d-1}\paren{\lambdabold_{L:+1}}
	}
    \right\}
     \nonumber\\
     &=
    (-1)^{n+m+g}
    \left\{
    C^{j,m}_{n-1, d+1}\paren{\lambdabold_{L:-1}}
    -
    C^{j,m}_{n, d}\paren{\lambdabold_{L:-1}}
    +
    C^{j,m}_{n-1, d}\paren{\lambdabold_{L:+1}}
      +
    C^{j-1,m-1}_{n, d}\paren{{}_{0\rightarrow}(\lambdabold_{L:-1})}
    \right.
    \nonumber\\
    &
    \left.
    \hspace{13em}
    +
     C^{j,m}_{n-1, d+1}\paren{\lambdabold_{i:-1}}
    +
    \paren{
    C^{j,m}_{n-1, d+1}\paren{\lambdabold_{L:-1}}
     -
    C^{j,m}_{n-1, d+1}\paren{\lambdabold_{i:-1}}
	}
    \right\}
     \nonumber\\
     &=
    (-1)^{n+m+g}
    \left\{
    2C^{j,m}_{n-1, d+1}\paren{\lambdabold_{L:-1}}
    -
    C^{j,m}_{n, d}\paren{\lambdabold_{L:-1}}
    +
    C^{j,m}_{n-1, d}\paren{\lambdabold_{L:+1}}
      +
    C^{j-1,m-1}_{n, d}\paren{{}_{0\rightarrow}(\lambdabold_{L:-1})}
    \right\}
    \nonumber\\
     &=
     A_0^-
     .
\end{align}
where we used~\eqref{eq:oxo} in  the third equality.

In the same way, we have
\begin{align}
 	A_{i}^{-+}+A_{w+1}^+=A_{w+1}^-
	.
\end{align}

Using the identity~\eqref{eq:oxo}, we have
\begin{align}
    &A_0^{-}+A_{w+1}^-
    \nonumber\\
    &=
    (-1)^{n+m+g}
    \left\{
    2
    C^{j,m}_{n-1, d+1}\paren{\lambdabold_{L:-1}}
    -
    C^{j,m}_{n, d}\paren{\lambdabold_{L:-1}}
    +
    C^{j,m}_{n-1, d}\paren{\lambdabold_{L:+1}}
    +
    C^{j-1,m-1}_{n, d}\paren{{}_{0\rightarrow}(\lambdabold_{L:-1})}
    \right.
    \nonumber\\
    &
    \hspace{4em}
    -
    \left.
    \left(
    2
    C^{j,m}_{n-1, d+1}\paren{\lambdabold_{R:-1}}
    -
    C^{j,m}_{n, d}\paren{\lambdabold_{R:-1}}
    +
    C^{j,m}_{n-1, d}\paren{\lambdabold_{R:+1}}
    +
    C^{j-1,m-1}_{n, d}\paren{(\lambdabold_{R:-1})_{\leftarrow 0}}
    \right)\right\}
    \nonumber\\
    &=
    (-1)^{n+m+g}
    \left\{
    C^{j,m}_{n-1, d+1}\paren{\lambdabold_{L:-1}}
    -
    C^{j,m}_{n, d-1}\paren{\lambdabold_{L:+1}}
    -
    C^{j,m}_{n, d}\paren{\lambdabold_{L:-1}}
    +
    C^{j,m}_{n-1, d}\paren{\lambdabold_{L:+1}}
    +
    C^{j-1,m-1}_{n, d}\paren{{}_{0\rightarrow}(\lambdabold_{L:-1})}
    \right.
    \nonumber\\
    &
    \hspace{4em}
    -
    \left.
    \left(
    C^{j,m}_{n-1, d+1}\paren{\lambdabold_{R:-1}}
    -
    C^{j,m}_{n, d-1}\paren{\lambdabold_{R:+1}}
    -
    C^{j,m}_{n, d}\paren{\lambdabold_{R:-1}}
    +
    C^{j,m}_{n-1, d}\paren{\lambdabold_{R:+1}}
    +
    C^{j-1,m-1}_{n, d}\paren{(\lambdabold_{R:-1})_{\leftarrow 0}}
    \right)\right\}
    \nonumber\\
    &=
    A_0^++A_{w+1}^+
    ,
\end{align}
where we used the identity 
$
 C^{j,m}_{n-1, d+1}\paren{\lambdabold_{L:-1}}
 -
 C^{j,m}_{n-1, d+1}\paren{\lambdabold_{R:-1}}
 =
  C^{j,m}_{n, d-1}\paren{\lambdabold_{R:+1}}
 -
 C^{j,m}_{n, d-1}\paren{\lambdabold_{L:+1}}
$
from~\eqref{eq:oxo} in the second equality.

From the above arguments, we can see
\begin{align}
 A_0^{\sigma}
 +
 A_{i}^{\sigma,-\sigma}
 &=
 A_0^{-\sigma},
 \nonumber\\
 A_0^{\sigma}+A_{w+1}^{\sigma}
 &=
  A_0^{+}+A_{w+1}^{+}.
\end{align}

Therefore,  in any case of a connected diagram $\dvx$, we have
\begin{align}
    \widetilde{D}^{k,j,m}_{n,d,g}(\dvx)&
    =
    A_0^{\sigma_0^{R}}
    +
    \sum_{i=1}^{w}
    A_{i}^{\sigma_i^{L},\sigma_i^{R}}
    +A_{w+1}^{\sigma_{w+1}^{L}}
     \nonumber\\
    &=
    A_0^{\sigma}
    +
    A_{p_1}^{\sigma,-\sigma}
    +
    A_{p_2}^{-\sigma,\sigma}
    +\cdots+
    A_{p_{M}}^{(-)^{M-1}\sigma,(-)^{M}\sigma}
    +
    A_{w+1}^{(-)^{M}\sigma}
    \nonumber\\
    &=
    A_0^{-\sigma}
    +
    A_{p_2}^{-\sigma,\sigma}
    +\cdots+
    A_{p_{M}}^{(-)^{M-1}\sigma,(-)^{M}\sigma}
    +
    A_{w+1}^{(-)^{M}\sigma}
     \nonumber\\
    &=
    A_0^{\sigma}
    +
    A_{p_3}^{\sigma,-\sigma}
    +\cdots+
    A_{p_{M}}^{(-)^{M-1}\sigma,(-)^{M}\sigma}
    +
    A_{w+1}^{(-)^{M}\sigma}
    \nonumber\\
    &\hspace{.5em}\vdots
    \nonumber\\
    &=
    A_0^{(-)^{M}\sigma}
    +A_{w+1}^{(-)^{M}\sigma}
     \nonumber\\
      &=
    A_0^{+}
    +A_{w+1}^{+}
     \nonumber\\
    &=
    (-1)^{n+m+g}
    \big\{
    C^{j,m}_{n-1, d+1}\paren{\lambdabold_{L:-1}}
    -
    C^{j,m}_{n, d}\paren{\lambdabold_{L:-1}}
    \nonumber\\
    &
    \hspace*{8em}
    +
    C^{j,m}_{n-1, d}\paren{\lambdabold_{L:+1}}
    -
    C^{j,m}_{n, d-1}\paren{\lambdabold_{L:+1}}
    +
    C^{j-1,m-1}_{n, d}\paren{{}_{0\rightarrow}(\lambdabold_{L:-1})}
    \nonumber\\
    &
    \hspace*{4em}
    -
    (
    C^{j,m}_{n-1, d+1}\paren{\lambdabold_{R:-1}}
    -
    C^{j,m}_{n, d}\paren{\lambdabold_{R:-1}}
    \nonumber\\
    &
    \hspace*{8em}
    +
    C^{j,m}_{n-1, d}\paren{\lambdabold_{R:+1}}
    -
    C^{j,m}_{n, d-1}\paren{\lambdabold_{R:+1}}
    +
    C^{j-1,m-1}_{n, d}\paren{(\lambdabold_{R:-1})_{\leftarrow 0}}
    )
    \big\}
    \nonumber\\
    &=
    (-1)^{n+m+g}
    \Big[
    \bce{ C^{j,m}_{n, d}\paren{\lambdabold_{R:-1}} - C^{j,m}_{n, d}\paren{\lambdabold_{L:-1}} }
    -
    \bce{ C^{j,m}_{n-1, d+1}\paren{\lambdabold_{R:-1}} - C^{j,m}_{n-1, d+1}\paren{\lambdabold_{L:-1}} }
    \nonumber\\
    &
    \hspace*{4em}
    - \bce{ C^{j,m}_{n, d-1}\paren{\lambdabold_{L:+1}} - C^{j,m}_{n, d-1}\paren{\lambdabold_{R:+1}} }
    +
    \bce{ C^{j,m}_{n-1, d}\paren{\lambdabold_{L:+1}} - C^{j,m}_{n-1, d}\paren{\lambdabold_{R:+1}} }
    \nonumber\\
    &
    \hspace*{8em}
    -\bce{C^{j-1,m-1}_{n, d}\paren{(\lambdabold_{R:-1})_{\leftarrow 0}} - C^{j-1,m-1}_{n, d}\paren{{}_{0\rightarrow}(\lambdabold_{L:-1})}
     }
    \Big]
    \nonumber\\
    &=
    (-1)^{n+m+g}
    \Big[
    \bce{ C^{j,m}_{n, d}\paren{\lambdabold_{R:-1}} - C^{j,m}_{n, d}\paren{\lambdabold_{L:-1}} }
    - 
    2 \bce{ C^{j,m}_{n-1, d+1}\paren{\lambdabold_{R:-1}} - C^{j,m}_{n-1, d+1}\paren{\lambdabold_{L:-1}} }
    \nonumber\\
    &
    \hspace*{4em}
    +
    \bce{ C^{j,m}_{n-2, d+2}\paren{\lambdabold_{R:-1}} - C^{j,m}_{n-2, d+2}\paren{\lambdabold_{L:-1}} }
    \nonumber\\
    &
    \hspace*{7em}
    -\bce{C^{j-1,m-1}_{n, d}\paren{(\lambdabold_{R:-1})_{\leftarrow 0}} - C^{j-1,m-1}_{n, d}\paren{{}_{0\rightarrow}(\lambdabold_{L:-1})}
     }
    \Big]
    \nonumber\\
    &=
    (-1)^{n+m+g}
    \left[
       \Delta C^{j,m}_{n, d}(\widetilde{\lambdabold})
       -2\Delta C^{j,m}_{n-1, d+1}(\widetilde{\lambdabold})
       +
       \Delta C^{j,m}_{n-2, d+2}(\widetilde{\lambdabold})
       \right.
       \nonumber\\
       &
       \hspace{13em}
       \left.
        -\bce{C^{j-1,m-1}_{n, d} (\widetilde{\lambdabold}_{\leftarrow 0}) - C^{j-1,m-1}_{n, d} ( {}_{0\rightarrow} (\mathcal{T}\widetilde{\lambdabold}) )
        }
    \right]
    \nonumber\\
    &=
    0
    ,
\end{align}
where $\widetilde{\lambdabold} \equiv \lambdabold_{R:-1}$, $\sigma=\sigma_0^{R}$, $p_i$ is the $i$ th $p$ from the smallest that satisfies $\sigma_p^{L}=-\sigma_{p}^{R}$, M is an integer that satisfies $M \le w$ in the third-to-last equality, we used the following identities from~\eqref{eq:oxo}
\begin{align}
    C^{j,m}_{n, d-1}\paren{\lambdabold_{L:+1}} - C^{j,m}_{n, d-1}\paren{\lambdabold_{R:+1}}
    &=
    C^{j,m}_{n-1, d+1}\paren{\lambdabold_{R:-1}} - C^{j,m}_{n-1, d+1}\paren{\lambdabold_{L:-1}}
    ,
    \\
    C^{j,m}_{n-1, d}\paren{\lambdabold_{L:+1}} - C^{j,m}_{n-1, d}\paren{\lambdabold_{R:+1}}
    &=
    C^{j,m}_{n-2, d+2}\paren{\lambdabold_{R:-1}} - C^{j,m}_{n-2, d+2}\paren{\lambdabold_{L:-1}},
\end{align}
and in the last equality, we used~\eqref{eq:basiceq}.
We note that $\mathcal{T}\widetilde{\lambdabold} = \lambdabold_{L:-1}$ and $(-)^M \sigma= \sigma^L_{w+1}$ .

We next prove $\widetilde{D}^{k,j,m}_{n,d,g}(\dvx)=0$ in the case in which $\dvx\in\mathcal{F}^{k,j,m}_{n,0,0}$ is a connected diagram and $l_\dvx=1$, i.e., $j=2m$.
In this case, the double number and the gap number of $\dvx$ are zero, such as
\begin{align}
    \dvx
    =
    \begin{tikzpicture}[baseline=-0.5*\rd]
        \dottedlineboth{1}{2}
        \osu{0}{1}
        \osu{2}{3}
        \Isd{0}{1}
        \Isd{2}{3}
        \ubrace{0}{3}{$l$}
    \end{tikzpicture}
    .
\end{align}
The contributions to the cancellation of $\dvx$ in $\bck{Q_k,H}$ comes only from  $\bck{Q^{j-1}_k,H_\int}$ because $Q^j_k$ $(j\geq 1)$ does not include the diagram with a unit number of $1$, given our normalization that fixes the freedom to add lower-order charges $Q_{k'<k}$.
The contributions to the cancellation of $\dvx$ are
\begin{align}
    (-1)^{n+m}
    \bce{
        -
        C^{j,m-1}_{n,0}\paren{0;l}
        \begin{tikzpicture}[baseline=-0.5*\rd]
            \dottedlineboth{1}{2}
            \osu{0}{1}
            \oszu{2}{3}
            \Isd{0}{1}
            \Isd{2}{3}
            \zd{0}
            \upverticalarrowtikz{0}
        \end{tikzpicture}
        -
        C^{j,m-1}_{n,0}\paren{l;0}
        \begin{tikzpicture}[baseline=-0.5*\rd]
            \dottedlineboth{1}{2}
            \osu{0}{1}
            \oszu{2}{3}
            \Isd{0}{1}
            \Isd{2}{3}
            \zd{3}
            \upverticalarrowtikz{3}
        \end{tikzpicture}
    }
    &=
    (-1)^{n+m}
    \paren{C^{j,m-1}_{n,0}\paren{0;l}-C^{j,m-1}_{n,0}\paren{l;0}}
    \begin{tikzpicture}[baseline=-0.5*\rd]
        \dottedlineboth{1}{2}
        \osu{0}{1}
        \osu{2}{3}
        \Isd{0}{1}
        \Isd{2}{3}
        \ubrace{0}{3}{$l$}
    \end{tikzpicture}
    \nonumber\\
    &=0,
\end{align}
where we used $C^{j,m-1}_{n,0}\paren{0;l}=C^{j,m-1}_{n,0}\paren{l;0}$ from~\eqref{eq:LRswap} and we have proved $\widetilde{D}^{k,j,m}_{n,d,g}(\dvx)=0$.

\subsubsection{Cancellation of non-connected diagram}
We prove $\widetilde{D}^{k,j,m}_{n,d,g}(\dvx)=0$ in the case in which $\dvx\in\mathcal{F}^{k,j,m}_{n,d,g}$ is a non-connected diagram.

We consider the case in which there is a hole in the region  between the $i$-th coast and $(i+1)$-th coasts, such as
\begin{align}
    \dvx
    =
    \begin{tikzpicture}[baseline=-0.5*\rd]
        \dottedlineboth{-1}{0}
        \osu{0}{4}
        \osd{0}{4}
        \zu{2}
        \dottedlineboth{4}{5}
    \end{tikzpicture}
    \,.
\end{align}
In this case, the contributions to the cancellation of $\dvx$ are 
\begin{align}
    &
    (-1)^{n+m+g}
    \left\{
    C^{j,m}_{n, d-1}\paren{\lambdabold_{i,(0),i+1}}
    \begin{tikzpicture}[baseline=-0.5*\rd]
        \encloseup{1}{2}
        \abovearrowtikz{1}{2}
        \dottedlineboth{-1}{0}
        \osu{0}{1}
        \Isu{1}{2}
        \osu{2}{4}
        \osd{0}{4}
        \dottedlineboth{4}{5}
    \end{tikzpicture}
    +
    C^{j,m}_{n, d-1}\paren{\lambdabold_{i,(0),i+1}}
    \begin{tikzpicture}[baseline=-0.5*\rd]
        \encloseup{2}{3}
        \abovearrowtikz{3}{2}
        \dottedlineboth{-1}{0}
        \osu{0}{2}
        \Isu{2}{3}
        \osu{3}{4}
        \osd{0}{4}
        \dottedlineboth{4}{5}
    \end{tikzpicture}
    \right\}
    \nonumber\\
    =
    &
    (-1)^{n+m+g+1}
    \left\{
    C^{j,m}_{n, d-1}\paren{\lambdabold_{i,(0),i+1}}
    -
    C^{j,m}_{n, d-1}\paren{\lambdabold_{i,(0),i+1}}
    \right\}
    \dvx
    \nonumber\\
    =
    &
    0
    ,
\end{align}
where 
$
\lambdabold_{i,(0),i+1}
\equiv
\{
\ldots,\lambda_i,0,\lambda_{i+1},\ldots
\}
$
and
$
\begin{tikzpicture}[baseline=-0.5*\rd]
    \dottedlineboth{-1}{0}
    \osu{0}{1}
    \Isu{1}{2}
    \osu{2}{4}
    \osd{0}{4}
    \dottedlineboth{4}{5}
\end{tikzpicture}
    \
    ,
    \quad
\begin{tikzpicture}[baseline=-0.5*\rd]
    \dottedlineboth{-1}{0}
    \osu{0}{2}
    \Isu{2}{3}
    \osu{3}{4}
    \osd{0}{4}
    \dottedlineboth{4}{5}
\end{tikzpicture}
    \in 
    \connectedset^{k,j,m}_{n,d-1,g}
$.
Thus we can see $\widetilde{D}^{k,j,m}_{n,d,g}(\dvx)=0$.

In the same way, we can prove $\widetilde{D}^{k,j,m}_{n,d,g}(\dvx)=0$ in the case of 
\begin{align}
    \dvx
    =
    \begin{tikzpicture}[baseline=-0.5*\rd]
        \dottedlineboth{-1}{0}
        \osu{1}{3}
        \Isu{0}{1}
        \osd{0}{3}
        \zu{2}
        \dottedlineboth{3}{4}
    \end{tikzpicture}
    \
    ,
    \quad
    \begin{tikzpicture}[baseline=-0.5*\rd,xscale=-1]
        \dottedlineboth{-1}{0}
        \osu{1}{3}
        \Isu{0}{1}
        \osd{0}{3}
        \zu{2}
        \dottedlineboth{3}{4}
    \end{tikzpicture}
    .
\end{align}

We  next consider the case in which there is a hole on the right of the $i$-th coast, such as
\begin{align}
    \dvx
    =
    \begin{tikzpicture}[baseline=-0.5*\rd]
        \dottedlineboth{-1}{0}
        \osu{0}{3}
        \osd{1}{3}
        \zu{1}
        \dottedlineboth{3}{4}
        \lengtarrowdownoneside{0}{1}{$\hspace{-1.5em}\lambda_i+1$}
    \end{tikzpicture}
    \ .
\end{align}
In this case, the contributions to the cancellation of $\dvx$ are
\begin{align}
    &
    (-1)^{n+m+g}
    \left\{
    C^{j,m}_{n, d-1}\paren{\lambdabold_{i,(0),i+1}}
    \begin{tikzpicture}[baseline=-0.5*\rd]
        \encloseup{1}{2}
        \abovearrowtikz{2}{1}
        \dottedlineboth{-1}{0}
        \osu{0}{1}
        \Isu{1}{2}
        \osu{2}{3}
        \osd{1}{3}
        \dottedlineboth{3}{4}
        \lengtarrowdownoneside{0}{1}{$\hspace{-1.5em}\lambda_i+1$}
    \end{tikzpicture}
    +
    C^{j-1,m}_{n, d}\paren{\lambdabold}
    \begin{tikzpicture}[baseline=-0.5*\rd]
        \downverticalarrowtikz{1}
        \dottedlineboth{-1}{0}
        \osu{0}{3}
        \osd{1}{3}
        \dottedlineboth{3}{4}
        \lengtarrowdownoneside{0}{1}{$\hspace{-2em}\lambda_i+1$}
    \end{tikzpicture}
    \right\}
    \nonumber\\
    =
    &
    (-1)^{n+m+g}
    \left\{
    C^{j,m}_{n, d-1}\paren{\lambdabold_{i,(0),i+1}}
    -
    C^{j-1,m}_{n, d}\paren{\lambdabold}
    \right\}
    \dvx
    \nonumber\\
    =
    &
    0
    ,
\end{align}
where 
$
\lambdabold_{i,(0),i+1}
\equiv
\{
\ldots,\lambda_i,0,\lambda_{i+1},\ldots
\}
$
and
$
\begin{tikzpicture}[baseline=-0.5*\rd]
    \dottedlineboth{-1}{0}
    \osu{0}{1}
    \Isu{1}{2}
    \osu{2}{3}
    \osd{1}{3}
    \dottedlineboth{3}{4}
    \lengtarrowdownoneside{0}{1}{$\hspace{-1.5em}\lambda_i+1$}
\end{tikzpicture}
\in 
\connectedset^{k,j,m}_{n,d-1,g}
,\quad
\begin{tikzpicture}[baseline=-0.5*\rd]
    \dottedlineboth{-1}{0}
    \osu{0}{3}
    \osd{1}{3}
    \dottedlineboth{3}{4}
    \lengtarrowdownoneside{0}{1}{$\hspace{-2em}\lambda_i+1$}
\end{tikzpicture}
\in 
\connectedset^{k,j-1,m}_{n,d,g}
$ and we used~\eqref{forholecase} in the last equality.
Thus we can see $\widetilde{D}^{k,j,m}_{n,d,g}(\dvx)=0$.

In the same way, we can prove $\widetilde{D}^{k,j,m}_{n,d,g}(\dvx)=0$ in the case of 
\begin{align}
    \dvx
    =
    \begin{tikzpicture}[baseline=-0.5*\rd,xscale=-1]
        \dottedlineboth{-1}{0}
        \osu{0}{3}
        \osd{1}{3}
        \zu{1}
        \dottedlineboth{3}{4}
        \lengtarrowdownoneside{0}{1}{$\hspace{1.5em}\lambda_i+1$}
    \end{tikzpicture}
    .
\end{align}

We next consider the cancellation of a non-connected diagram with a gap, such as
\begin{align}
    \dvx
    =
    \begin{tikzpicture}[baseline=-0.5*\rd]
        \dottedlineboth{-1}{0}
        \osd{0}{1}
        \dottedlinedown{1}{2}
        \osd{2}{3}
        \spacelengthup{0}{3}{$\lambda_i+1$}
        \gap{3}{4}
        \dottedlineboth{4}{5}
        \gap{5}{6}
        \osu{6}{8}
        \osd{6}{8}
        \dottedlineboth{8}{9}
        \encloseboth{3.1}{5.9}{black}{densely dashed}
    \end{tikzpicture}
    ,
\end{align}
where we enclose the gap with the dashed line.
In this case, the contributions to the cancellation of $\dvx$  are
\begin{align}
    &
    (-1)^{n+m+g}
    \left\{
    -
    C^{j,m}_{n, d-1}\paren{\lambdabold_{i,(0),i+1}}
    \begin{tikzpicture}[baseline=-0.5*\rd]
        \encloseup{5}{6}
        \dottedlineboth{-1}{0}
        \osd{0}{1}
        \dottedlinedown{1}{2}
        \osd{2}{3}
        \spacelengthup{0}{3}{$\lambda_i+1$}
        \gap{3}{4}
        \dottedlineboth{4}{5}
        \Isd{5}{6}
        \osu{5}{8}
        \osd{6}{8}
        \dottedlineboth{8}{9}
        \encloseboth{3.1}{4.9}{black}{densely dashed}
    \end{tikzpicture}
    +
    C^{j,m}_{n, d-1}\paren{\lambdabold_{i,(0),i+1}}
    \begin{tikzpicture}[baseline=-0.5*\rd]
        \enclosedown{6}{7}
        \dottedlineboth{-1}{0}
        \osd{0}{1}
        \dottedlinedown{1}{2}
        \osd{2}{3}
        \spacelengthup{0}{3}{$\lambda_i+1$}
        \gap{3}{4}
        \dottedlineboth{4}{5}
        \gap{5}{6}
        \osu{6}{8}
        \Isd{6}{7}
        \osd{7}{8}
        \dottedlineboth{8}{9}
        \encloseboth{3.1}{5.9}{black}{densely dashed}
    \end{tikzpicture}
    \right\}
    \nonumber\\
    =
    &
    (-1)^{n+m+g}
    \left\{
    C^{j,m}_{n, d-1}\paren{\lambdabold_{i,(0),i+1}}
    -
    C^{j,m}_{n, d-1}\paren{\lambdabold_{i,(0),i+1}}
    \right\}
    \dvx
    \nonumber\\
    =
    &
    0,
\end{align}
where 
$
\lambdabold_{i,(0),i+1}
\equiv
\{
\ldots,\lambda_i,1,\lambda_{i+1},\ldots
\}
$
and
$
\begin{tikzpicture}[baseline=-0.5*\rd]
        \dottedlineboth{-1}{0}
        \osd{0}{1}
        \dottedlinedown{1}{2}
        \osd{2}{3}
        \spacelengthup{0}{3}{$\lambda_i+1$}
        \gap{3}{4}
        \dottedlineboth{4}{5}
        \Isd{5}{6}
        \osu{5}{8}
        \osd{6}{8}
        \dottedlineboth{8}{9}
        \encloseboth{3.1}{4.9}{black}{densely dashed}
    \end{tikzpicture}
    \in 
    \connectedset^{k,j,m}_{n,d-1,g-1}
    ,
    \quad
\begin{tikzpicture}[baseline=-0.5*\rd]
        \dottedlineboth{-1}{0}
        \osd{0}{1}
        \dottedlinedown{1}{2}
        \osd{2}{3}
        \spacelengthup{0}{3}{$\lambda_i+1$}
        \gap{3}{4}
        \dottedlineboth{4}{5}
        \gap{5}{6}
        \osu{6}{8}
        \Isd{6}{7}
        \osd{7}{8}
        \dottedlineboth{8}{9}
        \encloseboth{3.1}{5.9}{black}{densely dashed}
    \end{tikzpicture}
    \in 
    \connectedset^{k,j,m}_{n,d-1,g}
$.
Thus we can see $\widetilde{D}^{k,j,m}_{n,d,g}(\dvx)=0$.

In the same way, we can prove $\widetilde{D}^{k,j,m}_{n,d,g}(\dvx)=0$ in the case of 
\begin{align}
    \dvx
    =
    \begin{tikzpicture}[baseline=-0.5*\rd,xscale=-1]
        \dottedlineboth{-1}{0}
        \osd{0}{1}
        \dottedlinedown{1}{2}
        \osd{2}{3}
        \spacelengthup{0}{3}{$\lambda_i+1$}
        \gap{3}{4}
        \dottedlineboth{4}{5}
        \gap{5}{6}
        \osu{6}{7}
        \osd{6}{7}
        \dottedlineboth{7}{8}
        \encloseboth{6}{9}{black}{densely dashed}
    \end{tikzpicture}
    .
\end{align}

We next consider the cancellation of a non-connected diagram $\dvx$ with a gap such as
\begin{align}
    \dvx
    =
    \begin{tikzpicture}[baseline=-0.5*\rd]
        \dottedlineboth{-1}{0}
        \osd{0}{1}
        \dottedlinedown{1}{2}
        \osd{2}{3}
        \spacelengthup{0}{3}{$\lambda_i+1$}
        \gap{3}{4}
        \dottedlineboth{4}{5}
        \gap{5}{6}
        \osu{6}{7}
        \encloseboth{3.1}{5.7}{black}{densely dashed}
        \zd{6}
        \dottedlineboth{7}{9}
        \lengtarrowdownoneside{7}{6.1}{$\hspace{3em}\lambda_{i+1}+1$}
    \end{tikzpicture}
    \,.
\end{align}
In this case, the contributions to the cancellation of $\dvx$ are
\begin{align}
    &
    (-1)^{n+m+g}
    \left\{
    -
    C^{j,m}_{n, d-1}\paren{\lambdabold_{i,(0),i+1}}
    \begin{tikzpicture}[baseline=-0.5*\rd]
        \encloseup{5}{6}
        \dottedlineboth{-1}{0}
        \osd{0}{1}
        \dottedlinedown{1}{2}
        \osd{2}{3}
        \spacelengthup{0}{3}{$\lambda_i+1$}
        \gap{3}{4}
        \dottedlineboth{4}{5}
        \Isd{5}{6}
        \osu{5}{7}
        \encloseboth{3.1}{4.9}{black}{densely dashed}
        \zd{6}
        \dottedlineboth{7}{9}
        \lengtarrowdownoneside{7}{6}{$\hspace{3em}\lambda_{i+1}+1$}
    \end{tikzpicture}
    +
    C^{j-1,m}_{n, d}\paren{\lambdabold}
    \begin{tikzpicture}[baseline=-0.5*\rd]
        \upverticalarrowtikz{6}
        \dottedlineboth{-1}{0}
        \osd{0}{1}
        \dottedlinedown{1}{2}
        \osd{2}{3}
        \spacelengthup{0}{3}{$\lambda_i+1$}
        \gap{3}{4}
        \dottedlineboth{4}{5}
        \gap{5}{6}
        \osu{6}{7}
        \encloseboth{3.1}{5.7}{black}{densely dashed}
        \dottedlineboth{7}{9}
        \lengtarrowdownoneside{7}{6}{$\hspace{3em}\lambda_{i+1}+1$}
    \end{tikzpicture}
    \right\}
    \nonumber\\
    =
    &
    (-1)^{n+m+g}
    \left\{
    C^{j,m}_{n, d-1}\paren{\lambdabold_{i,(0),i+1}}
    -
    C^{j-1,m}_{n, d}\paren{\lambdabold}
    \right\}
    \dvx
    \nonumber\\
    =
    &
    0
    ,
\end{align}
where 
$
\lambdabold_{i,(0),i+1}
\equiv
\{
\ldots,\lambda_i,1,\lambda_{i+1},\ldots
\}
$
and
$
\begin{tikzpicture}[baseline=-0.5*\rd]
    \dottedlineboth{-1}{0}
    \osd{0}{1}
    \dottedlinedown{1}{2}
    \osd{2}{3}
    \spacelengthup{0}{3}{$\lambda_i+1$}
    \gap{3}{4}
    \dottedlineboth{4}{5}
    \Isd{5}{6}
    \osu{5}{7}
    \encloseboth{3.1}{4.9}{black}{densely dashed}
    \zd{6}
    \dottedlineboth{7}{9}
    \lengtarrowdownoneside{7}{6}{$\hspace{3em}\lambda_{i+1}+1$}
\end{tikzpicture}
    \in 
    \connectedset^{k,j,m}_{n,d-1,g-1}
    ,
    \quad
\begin{tikzpicture}[baseline=-0.5*\rd]
        \dottedlineboth{-1}{0}
        \osd{0}{1}
        \dottedlinedown{1}{2}
        \osd{2}{3}
        \spacelengthup{0}{3}{$\lambda_i+1$}
        \gap{3}{4}
        \dottedlineboth{4}{5}
        \gap{5}{6}
        \osu{6}{7}
        \encloseboth{3.1}{5.7}{black}{densely dashed}
        \dottedlineboth{7}{9}
        \lengtarrowdownoneside{7}{6}{$\hspace{3em}\lambda_{i+1}+1$}
    \end{tikzpicture}
    \in 
    \connectedset^{k,j-1,m}_{n,d,g}
$ and we used~\eqref{forholecase} in the last equality.
Thus we can see $\widetilde{D}^{k,j,m}_{n,d,g}(\dvx)=0$.

The same argument holds in the case in which the upper and lower rows of $\dvx$ above are exchanged.

In other cases of the non-connected diagram $\dvx$, $\widetilde{D}^{k,j,m}_{n,d,g}(\dvx)=0$ trivially holds because such a $\dvx$ is not generated in $\bck{Q_k,H}$.

Therefore, we have proved $\bck{Q_k, H}=0$.

\newpage
\section{S4. Proof of identities of $C^{j,m}_{n,d}\paren{\lambdabold}$ 
}
In this section, we prove the identities of $C^{j,m}_{n,d}\paren{\lambdabold}$~\eqref{eq:LRswap}--\eqref{eq:oxo} by induction.
Let $P(j,n)$ be the statement~\eqref{eq:LRswap}--\eqref{eq:oxo} for $d\geq 0$ and $\floor{j/2}>m \geq 0$.

We show that if we suppose $P(j-1,n)$ and $P(j,n^\prime)$ for $n^\prime \leq n-1$, then $P(j,n)$ also holds.
The base cases are $P(j=1,n)$ for any $n\geq 0$, and $P(j,n=0)$ for any $j\geq 1$.
We can see $P(j=1,n)$ holds trivially because of $C^{j=1,m=0}_{n,d}(\lambdabold)=1$.
We can also easily prove $P(j,n=0)$ from the general expression of $C_{n=0, d}^{j, m}(\lambdabold) = \binom{j-1+d}{m}-\binom{j-1+d}{m-1}$.
We define $w\equiv j-2m-1$ in the following.

\subsection{Induction for~\eqref{eq:LRswap}}
Calculating $\sum_{p=0}^{\lambda_L} \Delta C^{j,m}_{n,d}(\mathcal{T}^{p}\lambdabold)$ with~\eqref{eq:basiceq}, we have  
\begin{align}
    \label{eq:basicLR}
    &
    C^{j,m}_{n,d}(\lambda_L; \vec{\lambda} ; \lambda_R)
    -
    C^{j,m}_{n,d}(-1; \vec{\lambda} ; \lambda_{L+R}+1)
    \nonumber\\
    =&
    2\paren{
        C^{j, m}_{n-1,d+1}(\lambda_L; \vec{\lambda} ; \lambda_R)
        -
        C^{j, m}_{n-1, d}(1; \vec{\lambda} ; \lambda_{L+R}+1)
    }
    -
    \paren{
        C^{j, m}_{n-1,d}(\lambda_L+1; \vec{\lambda} ; \lambda_R+1)
        -
        C^{j, m}_{n-1, d}(0; \vec{\lambda} ; \lambda_{L+R}+2)
    }
    \nonumber\\
    &\hspace*{3em}
    +
    \sum_{p=0}^{\lambda_L}
    \paren{
        C^{j-1, m-1}_{n, d}(p; \lambda_{L+R}-p, \vec{\lambda}; 0)
        -
        C^{j-1, m-1}_{n, d}(\lambda_{L+R}-p+1; p-1, \vec{\lambda} ; 0)
    }
    ,
\end{align}
where $\lambda_{L+R} \equiv \lambda_L + \lambda_R$ and when deriving the term with summation over $p$ in the last line, we used~\eqref{eq:LRswap} and~\eqref{eq:swap} of $P(j-1,n)$ and variable transformation $p\rightarrow \lambda_L-p$.
We also used $C^{j, m}_{n-1, d+1}(0; \vec{\lambda} ; \lambda_{L+R}+1) =  C^{j, m}_{n-1, d}(1; \vec{\lambda} ; \lambda_{L+R}+1)$.

Using~\eqref{eq:basicLR} interchanging the role of $\lambda_L$ and $\lambda_R$, we have
\begin{align}
    \label{eq:LRswapproof}
    &
    C^{j,m}_{n,d}(\lambda_L; \vec{\lambda} ; \lambda_R)
    -
    C^{j,m}_{n,d}(\lambda_R; \vec{\lambda} ; \lambda_L)
    \nonumber\\
    =&
    s
    \sum_{p=\lambda_{-}+1}^{\lambda_{+}}
    \paren{
        C^{j-1, m}_{n, d}(p; \lambda_{L+R}-p, \vec{\lambda}; 0)
        -
        C^{j-1, m}_{n, d}(\lambda_{L+R}-p+1; p-1, \vec{\lambda} ; 0)
    }
    \nonumber\\
    =&
    0
    ,
\end{align}
where $\lambda_+ = \max(\lambda_L, \lambda_R)$ and $\lambda_- = \min(\lambda_L, \lambda_R)$ and $s = \pm 1$ for $\lambda_L = \lambda_\pm$, and we used~\eqref{eq:LRswap} of $P(j,n-1)$ and we can see the last equality holds with the variable transformation $p \rightarrow \lambda_{L+R}-p+1$ in the second term in the summation on the second line.

Then, we have proved~\eqref{eq:LRswap} of $P(j,n)$.

We note that from~\eqref{eq:LRswapproof} and~\eqref{eq:basiceq2}, we have
\begin{align}
    \label{eq:basicLR2}
    &
    C^{j,m}_{n,d}(\lambda_L; \vec{\lambda} ; \lambda_R)
    -
    C^{j,m}_{n,d}(-1; \vec{\lambda} ; \lambda_{L+R}+1)
    \nonumber\\
    =&
    2\paren{
        C^{j, m}_{n-1,d+1}(\lambda_L; \vec{\lambda} ; \lambda_R)
        -
        C^{j, m}_{n-1, d}(1; \vec{\lambda} ; \lambda_{L+R}+1)
    }
    -
    \paren{
        C^{j, m}_{n-1,d}(\lambda_L+1; \vec{\lambda} ; \lambda_R+1)
        -
        C^{j, m}_{n-1, d}(0; \vec{\lambda} ; \lambda_{L+R}+2)
    }
    \nonumber\\
    &
    \hspace*{5em}
    +
    \sum_{p=0}^{\lambda_R}
    \paren{
        C^{j-1, m-1}_{n, d}(p; \lambda_{L+R}-p, \vec{\lambda}; 0)
        -
        C^{j-1, m-1}_{n, d}(\lambda_{L+R}-p+1; p-1, \vec{\lambda} ; 0)
    }
    .
\end{align}
 
 \subsection{ Induction for~\eqref{eq:swap}}
 Throughout this subsection, we define 
 $
 F^{j,m}_{n,d}\paren{\lambdabold} 
 =
 C^{j,m}_{n,d}\paren{\lambdabold}
 -
 C^{j,m}_{n,d}\paren{\lambdabold_{i_1\leftrightarrow i_2}}
 $
where $j\geq 2m+3~(w\geq 2)$ and $1\leq i_1<i_2\leq w$ is assumed.
We prove $F^{j,m}_{n,d}\paren{\lambdabold} =0$ in the following.
From the recursion equation~\eqref{eq:basiceq} and~\eqref{eq:swap} of $P(j-1,n)$ and $P(j,n-1)$, we have
\begin{align}
    \label{eq:swapproof1}
    &F^{j,m}_{n,d}\paren{\lambdabold} - F^{j,m}_{n,d}\paren{\mathcal{T}\lambdabold}
    \nonumber\\
    =&
    \Delta C^{j,m}_{n,d}\paren{\lambdabold}
    -
    \Delta C^{j,m}_{n,d}\paren{\lambdabold_{i_1\leftrightarrow i_2}}
    \nonumber\\
    =&
    2\bck{
        \Delta C^{j,m}_{n-1,d+1}\paren{\lambdabold} - \Delta C^{j,m}_{n-1,d+1}\paren{\lambdabold_{i_1\leftrightarrow i_2}}
    }
    -
    \bck{
        \Delta C^{j,m}_{n-2,d+2}\paren{\lambdabold} - \Delta C^{j,m}_{n-2,d+2}\paren{\lambdabold_{i_1\leftrightarrow i_2}}
    }
    \nonumber\\
    &\hspace*{3em}
    +
    \bck{
        C^{j-1,m-1}_{n,d}\paren{\lambdabold_{\leftarrow 0}} - C^{j-1,m-1}_{n,d}\paren{(\lambdabold_{i_1\leftrightarrow i_2})_{\leftarrow 0}}
    }
    -
    \bck{
        C^{j-1,m-1}_{n,d}\paren{{}_{0\rightarrow}(\mathcal{T}\lambdabold)} - C^{j-1,m-1}_{n,d}\paren{{}_{0\rightarrow}(\mathcal{T}(\lambdabold_{i_1\leftrightarrow i_2}))}
    }
    \nonumber\\
    &=
    0
    .
\end{align}
Using~\eqref{eq:swapproof1} repeatedly, we have 
\begin{align}
    F^{j,m}_{n,d}(\lambda_L; \vec{\lambda}; \lambda_R)
    &
    =
    F^{j,m}_{n,d}(-1; \vec{\lambda}; \lambda_{L+R}+1)
    =
    F^{j,m}_{n,d-1}(1; \vec{\lambda}; \lambda_{L+R}+1)
    \nonumber\\
    &
    =
    F^{j,m}_{n,d-1}(-1; \vec{\lambda}; \lambda_{L+R}+3)
    =
    F^{j,m}_{n,d-1}(1; \vec{\lambda}; \lambda_{L+R}+5)
    \nonumber\\
    &
    \quad\vdots
    \nonumber\\
    &
    =
    F^{j,m}_{n,0}(-1; \vec{\lambda}; \lambda_{L+R}+1+2d)
    ,
\end{align}
and then what we have to prove is $F^{j,m}_{n,0}(-1; \lambda_1,\ldots,\lambda_w; A) = 0$ for $A>0$.

For the $1<i_1$ case, we have $F^{j,m}_{n,0}(-1; \lambda_1,\ldots,\lambda_w; A) =  0$ straightforwardly from the definition of~\eqref{eq:Lm1def} and~\eqref{eq:swap} of $P(j-1,n)$ and $P(j,n-1)$.
For the $i_1 = 1$ case, we have 
\begin{align}
    &
    F^{j,m}_{n,0}(-1; \lambda_1,\ldots,\lambda_{i_2},\ldots)
    \nonumber\\
    =&
    C^{j,m}_{n,0}(-1; \lambda_1,\ldots,\lambda_{i_2},\ldots)
    -
    C^{j,m}_{n,0}(-1; \lambda_{i_2},\ldots,\lambda_{1},\ldots)
    \nonumber\\
    =&
    C^{j,m}_{n-1,1}(0; \lambda_1-1,\ldots,\lambda_{i_2},\ldots)
    -
    C^{j,m}_{n-1,1}(0; \lambda_{i_2}-1,\ldots,\lambda_{1},\ldots)
    +
    C^{j-1,m}_{n,0}(\lambda_1+1;\ldots,\lambda_{i_2},\ldots)
    -
    C^{j-1,m}_{n,0}(\lambda_{i_2}+1;\ldots,\lambda_{1},\ldots)
    \nonumber\\
    =&
    C^{j,m}_{n-2,3}(-1; \lambda_1-1,\ldots,\lambda_{i_2}-1,\ldots)
    -
    C^{j,m}_{n-2,3}(-1; \lambda_{i_2}-1,\ldots,\lambda_{1}-1,\ldots)
    +
    C^{j-1,m}_{n-1,2}(\lambda_{i_2}; \lambda_1-1,\ldots,\cancel{\lambda_{i_2}},\ldots)
    \nonumber\\
    &\hspace*{3em}
    -
    C^{j-1,m}_{n-1,2}(\lambda_{1}; \lambda_{i_2}-1,\ldots,\cancel{\lambda_{1}},\ldots)
    +
    C^{j-1,m}_{n,0}(\lambda_1+1;\ldots,\lambda_{i_2},\ldots)
    -
    C^{j-1,m}_{n,0}(\lambda_{i_2}+1;\ldots,\lambda_{1},\ldots)
    \nonumber\\
    =&
    C^{j-1,m}_{n,0}(\lambda_1+1;\ldots,\lambda_{i_2},\ldots)
    -
    C^{j-1,m}_{n,0}(\lambda_{i_2}+1;\ldots,\lambda_{1},\ldots)
    -
    \paren{
        C^{j-1,m}_{n-1,2}(\lambda_1;\ldots,\lambda_{i_2}-1,\ldots)
        -
        C^{j-1,m}_{n-1,2}(\lambda_{i_2};\ldots,\lambda_{1}-1,\ldots)
    }
    \nonumber\\
    =&
    0,
\end{align}
where we omitted the irrelevant elements in the list, $\cancel{\lambda_i}$ denotes the absence of the element at the corresponding position: $\{\ldots, \lambda_{i-1},\cancel{\lambda_{i}}, \lambda_{i+1},\ldots \} = \{\ldots, \lambda_{i-1}, \lambda_{i+1},\ldots \}$, in the third equality, we used~\eqref{eq:betweenjLR} of $P(j,n-1)$, in the fourth equality, we used~\eqref{eq:swap} of $P(j,n-2)$, and in the last equality, we used~\eqref{eq:oxo} of $P(j-1,n)$.

Then, we have proved~\eqref{eq:swap} of $P(j,n)$.

\subsection{ Induction for~\eqref{eq:minprop}}
We next consider~\eqref{eq:minprop}. 
It is enough to consider the $\lambda_a>n$ case.

We first consider the $a=L$ ($\lambda_L>n$) and $w>0$ case.
Using~\eqref{eq:basicLR2}, we have
\begin{align}
    &
    C^{j,m}_{n,d}(\lambda_L; \vec{\lambda} ; \lambda_R)
    -
    C^{j,m}_{n,d}(n; \vec{\lambda} ; \lambda_R)
    \nonumber\\
    =&
    C^{j,m}_{n,d}(-1; \vec{\lambda} ; \lambda_{L}+\lambda_{R}+1)
    -
    C^{j,m}_{n,d}(-1; \vec{\lambda} ; n+\lambda_{R}+1)
    \nonumber\\
    =&
    C^{j, m}_{n,d-1}(1; \vec{\lambda} ; \lambda_{L}+\lambda_{R}+1)
    -
    C^{j, m}_{n,d-1}(1; \vec{\lambda} ; n+\lambda_{R}+1)
    \nonumber\\
    =&
    \paren{
        C^{j, m}_{n,d-1}(\lambda_{L}+\lambda_{R}+1; \vec{\lambda} ;1)
        -
        C^{j, m}_{n,d-1}(n; \vec{\lambda} ;1)
    }
    -
    \paren{
        C^{j, m}_{n,d-1}(n+\lambda_{R}+1; \vec{\lambda} ; 1)
        -
        C^{j, m}_{n,d-1}(n; \vec{\lambda} ;1)
    }
    \nonumber\\
    =&
    \paren{
        C^{j, m}_{n,d-1}(-1; \vec{\lambda} ; \lambda_{L}+\lambda_{R}+3)
        -
        C^{j, m}_{n,d-1}(-1; \vec{\lambda} ; n+\lambda_{R}+3)
    }
    -
    \paren{
        C^{j, m}_{n,d-1}(-1; \vec{\lambda} ; n+\lambda_{R}+3)
        -
        C^{j, m}_{n,d-1}(-1; \vec{\lambda} ; n+\lambda_{R}+3)
    }
    \nonumber\\
    =&
    C^{j, m}_{n,d-1}(-1; \vec{\lambda} ; \lambda_{L}+\lambda_{R}+3)
    -
    C^{j, m}_{n,d-1}(-1; \vec{\lambda} ; n+\lambda_{R}+3)
    \nonumber\\
    &\quad\vdots 
    \nonumber\\
    =&
    C^{j, m}_{n,0}(-1; \vec{\lambda} ; \lambda_{L}+\lambda_{R}+1+2d)
    -
    C^{j, m}_{n,0}(-1; \vec{\lambda} ; n+\lambda_{R}+1+2d)
    \nonumber\\
    =&
    C^{j, m}_{n-1,1}(0; \lambda_1-1, \lambda_2, \ldots ; \lambda_{L}+\lambda_{R}+1+2d)
    -
    C^{j, m}_{n-1,1}(0; \lambda_1-1, \lambda_2, \ldots  ; n+\lambda_{R}+1+2d)
    \nonumber\\
    &\hspace*{3em}
    +
    C^{j-1,m}_{n,0}(\lambda_1+1; \lambda_2, \ldots ; \lambda_{L}+\lambda_{R}+1+2d)
    -
    C^{j-1,m}_{n,0}(\lambda_1+1; \lambda_2, \ldots; n+\lambda_{R}+1+2d)
    \nonumber\\
    =&
    0
    ,
\end{align}
where in the first equality, we used~\eqref{eq:minprop} of $P(j-1, n)$ and $P(j,n-1)$, and then we can see the contributions from the RHS of~\eqref{eq:basicLR2} cancels each other, and in the third equality, we used~\eqref{eq:LRswap} of $P(j,n)$, which we have already proved,  and in the fourth equality, we used the same argument as the first and second equality for the two terms, and in the last equality, we used~\eqref{eq:minprop} of $P(j-1, n)$ and $P(j,n-1)$.

In the same way, we can prove $C^{j,m}_{n,d}(\lambdabold)-C^{j,m}_{n,d}(\lambdabold_{L\rightarrow n}) = 0$ for the $w=0$.
From~\eqref{eq:LRswap} of $P(j,n)$ which we have already proved, 
we can see $C^{j,m}_{n,d}(\lambdabold)-C^{j,m}_{n,d}(\lambdabold_{R\rightarrow n}) = 0$ also holds for the $\lambda_R>n$ case.

We next consider the case of $a = i$ ($1\leq i \leq w$), $\lambda_i >n$.
From~\eqref{eq:LRswap} of $P(j,n)$ proved above, we can set $i=1$ without losing generality.
Using~\eqref{eq:basicLR}, we have
\begin{align}
    &C^{j,m}_{n,d}\paren{\lambdabold}
    -
    C^{j,m}_{n,d}\paren{\lambdabold_{\lambda_1\rightarrow n}}
    \nonumber\\
    =&
    C^{j,m}_{n,d}\paren{-1;\lambda_1,\ldots;\lambda_{L+R}+1}
    -
    C^{j,m}_{n,d}\paren{-1;n,\ldots;\lambda_{L+R}+1}
    \nonumber\\
    =&
    C^{j, m}_{n, d-1}\paren{1;\lambda_1,\ldots;\lambda_{L+R}+1}
    -
    C^{j, m}_{n, d-1}\paren{1;n,\ldots;\lambda_{L+R}+1}
    \nonumber\\
    =&
    C^{j, m}_{n, d-1}\paren{-1;\lambda_1,\ldots;\lambda_{L+R}+3}
    -
    C^{j, m}_{n, d-1}\paren{-1;n,\ldots;\lambda_{L+R}+3}
    \nonumber\\
    &\quad 
    \vdots 
    \nonumber\\
    =&
    C^{j, m}_{n, 0}\paren{-1;\lambda_1,\ldots;\lambda_{L+R}+1+2d}
    -
    C^{j, m}_{n, 0}\paren{-1;n,\ldots;\lambda_{L+R}+1+2d}
    \nonumber\\
    =&
    C^{j, m}_{n-1, 1}\paren{0;\lambda_1-1,\ldots;\lambda_{L+R}+1+2d}
    -
    C^{j, m}_{n-1, 1}\paren{0;n-1,\ldots;\lambda_{L+R}+1+2d}
    \nonumber\\
    &\hspace*{3em}
    +
    C^{j-1, m}_{n, 0}\paren{\lambda_1+1,\ldots;\lambda_{L+R}+1+2d}
    -
    C^{j-1, m}_{n, 0}\paren{n+1,\ldots;\lambda_{L+R}+1+2d}
    \nonumber\\
    =&
    0
    ,
\end{align}
where in the first equality, we used~\eqref{eq:minprop} of $P(j-1, n)$ and $P(j,n-1)$, and then we can see the contributions from the RHS of~\eqref{eq:basicLR} cancels each other, and in the last equality, we used~\eqref{eq:minprop} of $P(j-1, n)$ and $P(j,n-1)$.

Thus, we have proved~\eqref{eq:minprop} of $P(j,n)$.

 \subsection{ Induction for~\eqref{eq:betweenjLR} and~\eqref{eq:betweenj}}
Throughout this subsection, we define
\begin{align}
   F^{j,m}_{n,d}\paren{\lambdabold}_{(a,b)}
\equiv 
 C^{j,m}_{n,d}\paren{\lambdabold}
   -
   C^{j,m}_{n-1,d+2}\paren{\lambdabold_{a:-1,b:-1}}
   -
   C^{j-1,m}_{n,d+1}(\lambdabold_{a\rightarrow \lambda_{a}+\lambda_{b},\hat{b}})
   ,
\end{align} 
where $0\leq a<b\leq w+1$ and  $w > 0$ is assumed.
First, we prove $F^{j,m}_{n,d}\paren{\lambdabold}_{L(R), i} = 0$ ($1\leq i \leq w$), and complete the proof of~\eqref{eq:betweenjLR} of $P(j,n)$.
Because we have proved~\eqref{eq:LRswap} and~\eqref{eq:swap} of $P(j,n)$, we can set $a=L, b=1$ without losing generality.
Using~\eqref{eq:basicLR2}, we have
\begin{align}
    \label{eq:betweenjLRproof}
    &
    F^{j, m}_{n,d}(\lambda_L;\vec{\lambda};\lambda_R)_{(L, 1)}
    \nonumber\\
    =&
    C^{j, m}_{n,d}(-1;\vec{\lambda};\lambda_{L+R}+1)
    -
    C^{j, m}_{n-1,d+2}(-1;\vec{\lambda}_{1:-1};\lambda_{L+R})
    -
    C^{j-1,m}_{n,d+1}(-1;\vec{\lambda}_{\hat{1}};\lambda_{L+R}+\lambda_1+1)
    \nonumber\\
    &\hspace*{2em}
    +
    2  \paren{ F^{j,m}_{n-1,d+1}(\lambdabold)_{(L, 1)} - F^{j,m}_{n-1,d}(\mathcal{T}\etabold)_{(L, 1)}  }
    -
    \paren{
        F^{j,m}_{n-1,d}(\lambdabold_{L:+1, R:+1})_{(L, 1)}
        -
        F^{j,m}_{n-1,d}(\etabold)_{(L, 1)} 
    }
    \nonumber\\
    &\hspace*{4em}
    +\sum_{p=0}^{\lambda_R}
    \paren{
        F^{j-1,m}_{n,d}(\xibold^{(p)})_{(1, 2)}
        -
        F^{j-1,m}_{n,d}(\xibold^{(\lambda_{L+R}-p+1)})_{(L, 2)}
    }
    \nonumber\\
    =&
    C^{j, m}_{n,d-1}(1;\vec{\lambda};\lambda_{L+R}+1)
    -
    C^{j, m}_{n-1,d+1}(1;\vec{\lambda}_{1:-1};\lambda_{L+R})
    -
    C^{j-1,m}_{n,d}(1;\vec{\lambda}_{\hat{1}};\lambda_{L+R}+\lambda_1+1)
    \nonumber\\
    =&
    F^{j, m}_{n, d-1} (\lambda_{L+R}+1; \vec{\lambda}; 1)_{(L, 1)}
    \nonumber\\
    =&
    F^{j, m}_{n, d-2} (\lambda_{L+R}+3; \vec{\lambda}; 1)_{(L, 1)}
    \nonumber\\
    &\vdots
    \nonumber\\
    =&
    F^{j, m}_{n, 0} (\lambda_{L+R}+2d-1; \vec{\lambda}; 1)_{(L, 1)}
    \nonumber\\
    =&
    C^{j, m}_{n,0}(-1;\vec{\lambda};A+1)
    -
    C^{j, m}_{n-1,2}(-1;\vec{\lambda}_{1:-1};A)
    -
    C^{j-1,m}_{n,1}(-1;\vec{\lambda}_{\hat{1}};A+\lambda_1+1)
    \nonumber\\
    =&
    C^{j, m}_{n-1,1}(0;\vec{\lambda}_{1:-1};A+1)
    -
    C^{j, m}_{n-1,1}(1;\vec{\lambda}_{1:-1};A)
    +
    \paren{
        C^{j-1,m}_{n,0}(\lambda_1+1;\vec{\lambda}_{\hat{1}};A+1)
        -
        C^{j-1,m}_{n,0}(1;\vec{\lambda}_{\hat{1}};A+\lambda_1+1)
    }
    \nonumber\\
    =&
    C^{j, m}_{n-1,1}(0;\vec{\lambda}_{1:-1};A+1)
    -
    C^{j, m}_{n-1,1}(1;\vec{\lambda}_{1:-1};A)
    +
    \paren{
        C^{j-1,m}_{n-1,2}(\lambda_1;\vec{\lambda}_{\hat{1}};A)
        -
        C^{j-1,m}_{n-1,2}(0;\vec{\lambda}_{\hat{1}};A+\lambda_1)
    }
    \nonumber\\
    =&
    C^{j, m}_{n-1,1}(0;\vec{\lambda}_{1:-1};A+1)
    +
    \paren{
    -
    C^{j, m}_{n-1,1}(1;\vec{\lambda}_{1:-1};A)
    +
    C^{j-1,m}_{n-1,2}(\lambda_1;\vec{\lambda}_{\hat{1}};A)
    }
    -
    C^{j-1,m}_{n-1,2}(0;\vec{\lambda}_{\hat{1}};A+\lambda_1)
    \nonumber\\
    =&
    C^{j, m}_{n-1,1}(0;\vec{\lambda}_{1:-1};A+1)
    -
    C^{j, m}_{n-2,3}(0;\vec{\lambda}_{1:-2};A)
    -
    C^{j-1,m}_{n-1,2}(0;\vec{\lambda}_{\hat{1}};A+\lambda_1)
    \nonumber\\
    =&
    0
    ,
\end{align}
where $\etabold = \{\lambda_{L+R}+2;\vec{\lambda};0\}$ and $\xibold^{(p)} = \{p; \lambda_{L+R}-p; \vec{\lambda}; 0\}$, and $A= \lambda_{L+R}+2d$, and in the second equality, we used~\eqref{eq:betweenjLR} and~\eqref{eq:betweenj} of $P(j, n-1)$ and $P(j-1, n)$, and in the eighth equality, we used~\eqref{eq:oxo} of $P(j-1, n)$, and in the eleven-th equality, we used~\eqref{eq:betweenjLR} of $P(j, n-1)$, and in the last equality, we used~\eqref{eq:betweenjLR} of $P(j, n-1)$.
Thus, we have proved~\eqref{eq:betweenjLR} of $P(j,n)$.

Next, we will prove~\eqref{eq:betweenj} of $P(j,n)$ below.
The proof is similar to that of~\eqref{eq:betweenjLR} of $P(j,n)$ above.
We assume  $w \geq 2$.
In the following, we will prove $F^{j,m}_{n,d}\paren{\lambdabold}_{(i_1, i_2)} = 0$ ($1\leq i_1< i_2 \leq w$).
Because we have proved~\eqref{eq:swap} of $P(j,n)$, we can set $i_1=1, i_2=2$ without losing generality.


In the similar manner with the proof of~\eqref{eq:betweenjLR} of $P(j,n)$, we have
\begin{align}
    &
    F^{j, m}_{n,d}\paren{\lambda_L;\vec{\lambda};\lambda_R}_{(1, 2)}
    \nonumber\\
    =&
    C^{j, m}_{n,0}(-1;\vec{\lambda};A)
    -
    C^{j, m}_{n-1,2}(-1;\vec{\lambda}_{1:-1, 2:-1};A)
    -
    C^{j-1,m}_{n,1}(-1;\vec{\lambda}_{1:\rightarrow \lambda_{1,2}, \hat{2}};A)
    \nonumber\\
    =&
    \paren{
    C^{j, m}_{n-1,1}(0;\vec{\lambda}_{1:-1};A) 
    +
    C^{j-1,m}_{n,0}(\lambda_{1}+1;\vec{\lambda}_{\hat{1}};A)
    }
    -
    C^{j, m}_{n-1,1}(1;\vec{\lambda}_{1:-1, 2:-1};A)
    -
    C^{j-1,m}_{n,0}(1;\vec{\lambda}_{1:\rightarrow \lambda_{1,2}, \hat{2}};A)
    \nonumber\\
    =&
    \paren{
    C^{j-1,m}_{n,0}(\lambda_{1}+1;\lambda_{2},\ldots;A)
    -
    C^{j-1,m}_{n,0}(1;\lambda_{1}+\lambda_{2},\ldots ;A)
    }
    +
    C^{j, m}_{n-1,1}(0;\vec{\lambda}_{1:-1};A) 
    -
    C^{j, m}_{n-1,1}(1;\vec{\lambda}_{1:-1, 2:-1};A)
    \nonumber\\
    =&
    \paren{
    C^{j-1,m}_{n-1, 2}(\lambda_{1};\lambda_{2}-1,\ldots;A)
    -
    C^{j-1,m}_{n-1, 2}(0;\lambda_{1}+\lambda_{2}-1,\ldots ;A)
    }
    \nonumber\\
    &\hspace*{9em}
    +
    C^{j, m}_{n-1,1}(0;\lambda_{1}-1, \lambda_{2}, \ldots ;A) 
    -
    C^{j, m}_{n-1,1}(1;\lambda_{1}-1, \lambda_{2}-1, \ldots ;A)
    \nonumber\\
    =&
    \paren{
    C^{j, m}_{n-1,1}(0;\lambda_{1}-1, \lambda_{2}, \ldots ;A) 
    -
    C^{j-1,m}_{n-1, 2}(0;\lambda_{1}+\lambda_{2}-1,\ldots ;A)
    }
    \nonumber\\
    &\hspace*{9em}
    -
    \paren{
        C^{j, m}_{n-1,1}(1;\lambda_{1}-1, \lambda_{2}-1, \ldots ;A)
        -
    C^{j-1,m}_{n-1, 2}(\lambda_{1};\lambda_{2}-1,\ldots;A)
    }
    \nonumber\\
    =&
    C^{j, m}_{n-2,3}(0;\lambda_{1}-2, \lambda_{2}-1, \ldots ;A) 
    -
    C^{j, m}_{n-2,3}(0;\lambda_{1}-2, \lambda_{2}-1, \ldots ;A) 
    \nonumber\\
    =&
    0
    ,
\end{align}
where we can prove the first equality in the same way as the first to sixth equality in~\eqref{eq:betweenjLRproof}, and $A= \lambda_{L+R}+2d+1$, and $\lambda_{1,2} = \lambda_{1} + \lambda_{2}$, and the dotted line $\ldots$ represents $\lambda_3,\ldots,\lambda_w$, and in the fourth equality, we used~\eqref{eq:oxo2} of $P(j-1, n)$, and in the sixth equality, we used~\eqref{eq:betweenj} and~\eqref{eq:betweenjLR} of $P(j, n-1)$.

Thus, we have proved~\eqref{eq:betweenjLR} of $P(j,n)$.

\subsection{ Induction for~\eqref{eq:oxo}}
Throughout this subsection, we define 
\begin{align}
   F^{j,m}_{n,d}(\lambdabold)_{(a,b)}
\equiv 
   C^{j,m}_{n,d}(\lambdabold)
   -
   C^{j,m}_{n,d}(\lambdabold_{a:-1,b:+1})
   -
   \bce{
        C^{j,m}_{n-1,d+2}(\lambdabold_{a:-1, b:-1})
        -
        C^{j,m}_{n-1,d+2}(\lambdabold_{a:-2})
   }
   ,
\end{align}
where $0\leq a, b\leq w+1$. 
In the following, we will prove $F^{j,m}_{n,d}\paren{\lambdabold}_{(a,b)} = 0$.

We first consider the case of $a=L~(\text{or }R), b=i (1\leq i\leq w)$ and $\lambda_L>0$.
Because we have proved~\eqref{eq:LRswap} and~\eqref{eq:swap}  of $P(j,n)$, we can set $a=L, b=1$ without losing generality.

Using~\eqref{eq:basicLR2}, we have
\begin{align}
    \label{eq:oxoL1proof}
    &F^{j,m}_{n,d}(\lambda_{L};\vec{\lambda};\lambda_{R})_{(L,1)}
    \nonumber\\
    =&
    C^{j, m}_{n,d}(-1;\vec{\lambda};\lambda_{L+R}+1)
    -
    C^{j, m}_{n,d}(-1;\vec{\lambda}_{1:+1};\lambda_{L+R})
    -
    \bce{ 
    C^{j, m}_{n-1,d+2}(-1;\vec{\lambda}_{1:-1};\lambda_{L+R})
    -
    C^{j, m}_{n-1,d+2}(-1;\vec{\lambda};\lambda_{L+R}-1)
    }
    \nonumber\\
    &\hspace*{2em}
    +
    2  \paren{ F^{j,m}_{n-1,d+1}(\lambdabold)_{(L,1)} - F^{j,m}_{n-1,d}(\mathcal{T}\etabold)_{(L,1)}  }
    -
    \paren{
        F^{j,m}_{n-1,d}(\lambdabold_{L:+1, R:+1})_{(L,1)}
        -
        F^{j,m}_{n-1,d}(\etabold)_{(L,1)}
    }
    \nonumber\\
    &\hspace*{4em}
    +\sum_{p=0}^{\lambda_R}
    \paren{
        F^{j-1,m}_{n,d}(\xibold^{(p)})_{(1,2)}
        -
        F^{j-1,m}_{n,d}(\xibold^{(\lambda_{L+R}-p+1)})_{(L,2)}
    }
    \nonumber\\
    =&
    C^{j, m}_{n,d-1}(1;\vec{\lambda};\lambda_{L+R}+1)
    -
    C^{j, m}_{n,d-1}(1;\vec{\lambda}_{1:+1};\lambda_{L+R})
    -
    \bce{ 
    C^{j, m}_{n-1,d+1}(1;\vec{\lambda}_{1:-1};\lambda_{L+R})
    -
    C^{j, m}_{n-1,d+1}(1;\vec{\lambda};\lambda_{L+R}-1)
    }
    \nonumber\\
    =&
    F^{j,m}_{n,d-1}(\lambda_{L+R}+1;\vec{\lambda};1)_{(L,1)}
    \nonumber\\
    =&
    F^{j,m}_{n,d-2}(\lambda_{L+R}+3;\vec{\lambda};1)_{(L,1)}
    \nonumber\\
    \quad\vdots 
    \nonumber\\
    =&
    F^{j,m}_{n,0}(\lambda_{L+R}+2d-1;\vec{\lambda};1)_{(L,1)}
    \nonumber\\
    =&
    C^{j, m}_{n,0}(-1;\vec{\lambda};A+1)
    -
    C^{j, m}_{n,0}(-1;\vec{\lambda}_{1:+1}; A)
    -
    \bce{ 
    C^{j, m}_{n-1,2}(-1;\vec{\lambda}_{1:-1};A)
    -
    C^{j, m}_{n-1,2}(-1;\vec{\lambda};A-1)
    }
    \nonumber\\
    =&
    C^{j, m}_{n-1,1}(0;\lambda_1-1,\ldots;A+1)
    -
    C^{j, m}_{n-1,1}(0;\lambda_1, \ldots; A)
    +
    \paren{
    C^{j-1, m}_{n,0}(\lambda_1+1; \ldots;A+1)
    -
    C^{j-1, m}_{n,0}(\lambda_1+2; \ldots;A)
    }
    \nonumber\\
    &\hspace*{4em}
    -
    \bce{ 
    C^{j, m}_{n-1,1}(1;\lambda_1-1,\ldots;A)
    -
    C^{j, m}_{n-1,1}(1;\lambda_1,\ldots;A-1)
    }
    \nonumber\\
    =&
    C^{j, m}_{n-1,1}(0;\lambda_1-1,\ldots;A+1)
    -
    C^{j, m}_{n-1,1}(0;\lambda_1, \ldots; A)
    +
    \paren{
    C^{j-1, m}_{n-1,2}(\lambda_1; \ldots;A)
    -
    C^{j-1, m}_{n-1,2}(\lambda_1+1; \ldots;A-1)
    }
    \nonumber\\
    &\hspace*{4em}
    -
    \bce{ 
    C^{j, m}_{n-1,1}(1;\lambda_1-1,\ldots;A)
    -
    C^{j, m}_{n-1,1}(1;\lambda_1,\ldots;A-1)
    }
    \nonumber\\
    =&
    \paren{
    C^{j, m}_{n-1,1}(1;\lambda_1,\ldots;A-1)
    -
    C^{j-1, m}_{n-1,2}(\lambda_1+1; \ldots;A-1)
    }
    -
    \paren{
    C^{j, m}_{n-1,1}(0;\lambda_1, \ldots; A)
    -
    C^{j-1, m}_{n-1,2}(\lambda_1; \ldots;A)
    }
    \nonumber\\
    &\hspace*{4em}
    -
    \paren{ 
    C^{j, m}_{n-1,1}(1;\lambda_1-1,\ldots;A)
    -
    C^{j, m}_{n-1,1}(0;\lambda_1-1,\ldots;A+1)
    }
    \nonumber\\
    =&
    C^{j, m}_{n-2,3}(0;\lambda_1-1,\ldots;A-1)
    -
    C^{j, m}_{n-2,3}(-1;\lambda_1-1, \ldots; A)
    -
    \paren{ 
    C^{j, m}_{n-1,1}(1;\lambda_1-1,\ldots;A)
    -
    C^{j, m}_{n-1,1}(0;\lambda_1-1,\ldots;A+1)
    }
    \nonumber\\
    =&
    \paren{ 
    C^{j, m}_{n-1,1}(1;\lambda_1-1,\ldots;A)
    -
    C^{j, m}_{n-1,1}(0;\lambda_1-1, \ldots; A+1)
    }
    -
    \paren{ 
    C^{j, m}_{n-1,1}(1;\lambda_1-1,\ldots;A)
    -
    C^{j, m}_{n-1,1}(0;\lambda_1-1,\ldots;A+1)
    }
    \nonumber\\
    =&0,
\end{align}
where $\etabold = \{\lambda_{L+R}+2;\vec{\lambda};0\}$, $\xibold^{(p)} = \{p; \lambda_{L+R}-p, \vec{\lambda}; 0\}$,  $A= \lambda_{L+R}+2d$, in the second equality, we used~\eqref{eq:oxo} of $P(j, n-1)$ and $P(j-1, n)$, in the third equality, we used~\eqref{eq:LRswap} of $P(j,n)$ we have proved above, the dotted line $\ldots$ above denotes $\lambda_2,\ldots,\lambda_w$, in the eighth equality, we used~\eqref{eq:oxo} of $P(j-1,n)$, in the tenth equality, we used~\eqref{eq:betweenjLR} of $P(j, n-1)$, and eleventh equality, we used~\eqref{eq:oxo} of $P(j,n-1)$.

We next consider the case of $b=L (\text{ or }R), a=i (1\leq i\leq w)$ and $\lambda_i>0$.
Because we have proved~\eqref{eq:LRswap} and~\eqref{eq:swap}  of $P(j,n)$, we can set $b=L, i=1$ without losing generality.
Using~\eqref{eq:basicLR2}, we have
\begin{align}
    \label{eq:oxo1Lproof}
    &F^{j,m}_{n,d}(\lambda_{L};\vec{\lambda};\lambda_{R})_{(1,L)}
    \nonumber\\
    =&
    C^{j, m}_{n,0}(-1;\vec{\lambda};A)
    -
    C^{j, m}_{n,0}(-1;\vec{\lambda}_{1:-1}; A+1)
    -
    \bce{ 
    C^{j, m}_{n-1,2}(-1;\vec{\lambda}_{1:-1};A-1)
    -
    C^{j, m}_{n-1,2}(-1;\vec{\lambda}_{i:-2};A)
    }
    \nonumber\\
    =&
    C^{j, m}_{n-1,1}(0;\lambda_1-1,\ldots;A)
    -
    C^{j, m}_{n-1,1}(0;\lambda_1-2, \ldots; A+1)
    +
    \paren{
    C^{j-1, m}_{n,0}(\lambda_1+1; \ldots;A)
    -
    C^{j-1, m}_{n,0}(\lambda_1; \ldots;A+1)
    }
    \nonumber\\
    &\hspace*{4em}
    -
    \bce{ 
    C^{j, m}_{n-1,1}(1;\lambda_1-1,\ldots;A-1)
    -
    C^{j, m}_{n-1,1}(1;\lambda_1-2,\ldots;A)
    }
    \nonumber\\
    =&
    C^{j, m}_{n-1,1}(0;\lambda_1-1,\ldots;A)
    -
    C^{j, m}_{n-1,1}(0;\lambda_1-2, \ldots; A+1)
    +
    \paren{
    C^{j-1, m}_{n-1,2}(\lambda_1; \ldots;A-1)
    -
    C^{j-1, m}_{n-1,2}(\lambda_1-1; \ldots;A)
    }
    \nonumber\\
    &\hspace*{4em}
    -
    \bce{ 
    C^{j, m}_{n-1,1}(1;\lambda_1-1,\ldots;A-1)
    -
    C^{j, m}_{n-1,1}(1;\lambda_1-2,\ldots;A)
    }
    \nonumber\\
    =&
    C^{j, m}_{n-1,1}(0;\lambda_1-1,\ldots;A)
    -
    C^{j, m}_{n-1,1}(0;\lambda_1-2, \ldots; A+1)
    +
    \paren{
    C^{j-1, m}_{n-1,2}(\lambda_1; \ldots;A-1)
    -
    C^{j-1, m}_{n-1,2}(\lambda_1-1; \ldots;A)
    }
    \nonumber\\
    &\hspace*{4em}
    -
    \bce{ 
    C^{j, m}_{n-1,1}(1;\lambda_1-1,\ldots;A-1)
    -
    C^{j, m}_{n-1,1}(1;\lambda_1-2,\ldots;A)
    }
    \nonumber\\
    =&
    C^{j, m}_{n-1,1}(1;\lambda_1-2,\ldots;A)
    -
    C^{j, m}_{n-1,1}(0;\lambda_1-2, \ldots; A+1)
    +
    \paren{
    C^{j, m}_{n-1,1}(0;\lambda_1-1,\ldots;A)
    -
    C^{j-1, m}_{n-1,2}(\lambda_1-1; \ldots;A)
    }
    \nonumber\\
    &\hspace*{4em}
    -
    \bce{ 
    C^{j, m}_{n-1,1}(1;\lambda_1-1,\ldots;A-1)
    -
    C^{j-1, m}_{n-1,2}(\lambda_1; \ldots;A-1)
    }
    \nonumber\\
    =&
    C^{j, m}_{n-1,1}(1;\lambda_1-2,\ldots;A)
    -
    C^{j, m}_{n-1,1}(0;\lambda_1-2, \ldots; A+1)
    -
    \paren{
    C^{j, m}_{n-2,3}(0;\lambda_1-2,\ldots;A-1)
    -
    C^{j, m}_{n-2,3}(-1;\lambda_1-2,\ldots;A)
    }
    \nonumber\\
    =&
    C^{j, m}_{n-1,1}(1;\lambda_1-2,\ldots;A)
    -
    C^{j, m}_{n-1,1}(0;\lambda_1-2, \ldots; A+1)
    -
    \paren{
    C^{j, m}_{n-1,1}(1;\lambda_1-2,\ldots;A)
    -
    C^{j, m}_{n-1,1}(0;\lambda_1-2,\ldots;A+1)
    }
    \nonumber\\
    =&
    0
    ,
\end{align}
where $A=\lambda_{L+R}+2d+1$, we can prove the first equality in the same way as the first to sixth equality in~\eqref{eq:oxoL1proof}, the dotted line $\ldots$ represents $\{\lambda_2,\ldots,\lambda_w\}$, in the third equality, we used~\eqref{eq:oxo} of $P(j-1, n)$, in the sixth equality, we used~\eqref{eq:betweenjLR} of $P(j, n-1)$ and in the seventh equality, we used~\eqref{eq:oxo} of $P(j, n-1)$.

We next consider the case of $(a,b) = (L, R), (R, L)$, and $w=0 \ (j=2m+1)$.
Because we have proved~\eqref{eq:LRswap} of $P(j,n)$, we can set $a=L, b=R$ without losing generality.
Using~\eqref{eq:basicLR}, we have
\begin{align}
    \label{eq:oxoLRproof}
    &F^{j,m}_{n,d}(\lambda_{L};\lambda_{R})_{(L, R)}
    \nonumber\\
    =&
    2  F^{j,m}_{n-1,d+1}(\lambda_{L};\lambda_{R})_{(L,R)} 
    -
    F^{j,m}_{n-1,d}(\lambda_{L}+1;\lambda_{R}+1)_{(L,R)} 
    +
    \paren{
        C^{j-1,m-1}_{n,d}(\lambda_{L};\lambda_{R};0)
        -
        C^{j-1,m-1}_{n,d}(\lambda_{R}+1;\lambda_{L}-1;0)
    }
    \nonumber\\
    &\hspace*{7em}
    -
    \paren{
        C^{j-1,m-1}_{n-1,d+2}(\lambda_{L}-1;\lambda_{R}-1;0)
        -
        C^{j-1,m-1}_{n-1,d+2}(\lambda_{R};\lambda_{L}-2;0)
    }
    \nonumber\\
    =&
    0
    ,
\end{align}
where in the first equality, we apply~\eqref{eq:basicLR} to every term in $F^{j,m}_{n,d}(\lambda_{L};\lambda_{R})_{(L, R)}$ and we can see there are trivial cancellations with straightforward calculation, and in the second equality, we used~\eqref{eq:oxo} of $P(j, n-1)$, and used~\eqref{eq:oxo2} of $P(j-1, n)$.

The other cases of~\eqref{eq:oxo} of $P(j,n)$ can be easily proved from the above results.
In the case of $w\geq 2$ and $a=i_1, b=i_2 (1\leq i_1, i_2\leq w)$ and $\lambda_a >0$, we have
\begin{align}
    F^{j,m}_{n,d}(\lambdabold)_{(i_1, i_2)}
    =
    F^{j,m}_{n,d}(\lambdabold)_{(i_1, L)}
    +
    F^{j,m}_{n,d}(\lambdabold_{L:+1, i_1:-1})_{(L, i_2)}
    =
    0
    ,
\end{align}
where we used the above result~\eqref{eq:oxoL1proof} and~\eqref{eq:oxo1Lproof}.
In the case of $w > 0$ and $a=L, b=R$, we have
\begin{align}
    F^{j,m}_{n,d}(\lambdabold)_{(L, R)}
    =
    F^{j,m}_{n,d}(\lambdabold)_{(L, i)}
    +
    F^{j,m}_{n,d}(\lambdabold_{i:+1, L:-1})_{(i, R)}
    =
    0
    ,
\end{align}
where we used the above result~\eqref{eq:oxoL1proof} and~\eqref{eq:oxo1Lproof}.

Thus, we have proved~\eqref{eq:oxo} of $P(j,n)$.

Therefore, we have proved~\eqref{eq:LRswap}--\eqref{eq:oxo} by induction.

\subsection{Proof of positivity of $C^{j,m}_{n,d}(\lambdabold)$}
\label{app:positivity}
In this subsection, we prove the positivity of the coefficient: $C^{j,m}_{n,d}(\lambdabold)>0$.
we prove this positivity by induction. 
we show $C^{j,m}_{n,d}(\lambdabold)>0$ if we suppose $C^{j-1,m}_{n,d}(\lambdabold)>0$ and $C^{j,m}_{n-1,d}(\lambdabold)>0$ hold.
The base cases trivially hold from $C^{1,m}_{n,d} = 1$ and from~\eqref{coeffn=0}.
From~\eqref{eq:LRswap} and~\eqref{eq:minprop}, we can set $0\le \lambda_L \le \lambda_R \le n$ without losing generality.

First, we prove $\Delta C^{j,m}_{n,d}(\lambdabold)>0$.
From~\eqref{eq:basiceq3}, we have
\begin{align}
    \Delta C_{n,d}^{j,m}(\lambdabold) 
    &=
    \sum_{\widetilde{n}=0}^{n}
    (n+1-\widetilde{n})
    \paren{
    C^{j-1,m-1}_{\widetilde{n}, n+d-\widetilde{n}}\paren{\lambdabold_{\leftarrow 0}}
    -
    C^{j-1,m-1}_{\widetilde{n}, n+d-\widetilde{n}}\paren{{}_{0\rightarrow}(\mathcal{T}\lambdabold)}
    }
    \nonumber\\
    &=
    \sum_{\widetilde{n}=0}^{n}
    (n+1-\widetilde{n})
    \paren{
    C^{j-1,m-1}_{\widetilde{n}, n+d-\widetilde{n}}\paren{0;\vec{\lambda},\lambda_R;\lambda_L}
    -
    C^{j-1,m-1}_{\widetilde{n}, n+d-\widetilde{n}}\paren{0;\vec{\lambda},\lambda_L-1;\lambda_R+1}
    }
    \nonumber\\
    &=
    \sum_{\widetilde{n}=0}^{n}
    (n+1-\widetilde{n})
    \paren{
    C^{j-1,m-1}_{\widetilde{n}-\lambda_L, n+d-\widetilde{n}+2 \lambda_L}\paren{0;\vec{\lambda},\lambda_R-\lambda_L;0}
    \linebreaklr
    \hspace*{12em}
    -
    C^{j-1,m-1}_{\widetilde{n}-\lambda_L, n+d-\widetilde{n}+2 \lambda_L}\paren{0;\vec{\lambda},-1;\lambda_R-\lambda_L+1}
    }
    \nonumber\\
    &=
    \sum_{\widetilde{n}=0}^{n}
    (n+1-\widetilde{n})
    C^{j-1,m-1}_{\widetilde{n}-\lambda_L, n+d-\widetilde{n}+2 \lambda_L}\paren{0;\vec{\lambda},\lambda_R-\lambda_L;0}
    \nonumber\\
    &\geq
    C^{j-1,m-1}_{n-\lambda_L, d+2 \lambda_L}\paren{0;\vec{\lambda},\lambda_R-\lambda_L;0}
    \nonumber\\
    &>0
    ,
\end{align}
where in the third equality, we used~\eqref{eq:oxo} repeatedly, and the last and second-to-last inequalities hold due to the assumption of induction.  
Then we proved $\Delta C^{j,m}_{n,d}(\lambdabold)>0$.

From the above argument, we have
\begin{align}
    C^{j,m}_{n,d}(\lambdabold)
    &=
    \Delta C^{j,m}_{n,d}(\lambdabold) + C^{j,m}_{n,d}(\lambda_L-1 ;\vec{\lambda}; \lambda_R+1)
    \nonumber\\
    &>
    C^{j,m}_{n,d}(\lambda_L-1 ;\vec{\lambda}; \lambda_R+1)
    \nonumber\\
    & \hspace*{0.5em}\vdots
    \nonumber\\
    &>
    C^{j,m}_{n,d}(-1 ;\vec{\lambda}; \lambda_{L+R}+1)
    \nonumber\\
    &=
    C^{j,m}_{n,d-1}(1 ;\vec{\lambda}; \lambda_{L+R}+1)
    \nonumber\\
    &>
    C^{j,m}_{n,d-2}(1 ;\vec{\lambda}; \lambda_{L+R}+3)
    \nonumber\\
    & \hspace*{0.5em}\vdots
    \nonumber\\
    &>
    C^{j,m}_{n,0}(1 ;\vec{\lambda}; \lambda_{L+R}+2d-1)
    \nonumber\\
    &>
    C^{j,m}_{n,0}(-1 ;\vec{\lambda}; \lambda_{L+R}+2d+1)
    \nonumber\\
    &=
    C^{j,m}_{n-1,1}(0 ;\lambda_1-1,\ldots; \lambda_{L+R}+2d+1)
    +
    C^{j-1,m}_{n,0}(\lambda_1+1,\ldots; \lambda_{L+R}+2d+1)
    \nonumber\\
    &> 0,
\end{align} 
where we used the definition~\eqref{eq:Lm1def} repeatedly, and the last inequality holds due to the assumption of induction.  
Thus, we have proved the positivity of the coefficient $C^{j,m}_{n,d}(\lambdabold)>0$.

\newpage

\section{S5. Proof of identities of $A^{\sigma\mu}_i$}
In this section, we prove the identity for $A_i^{\sigma,\nu}(\dvx)$~\eqref{A++}--\eqref{AR-}.
Note that the diagrams generated by the commutator of $H$ and connected diagrams satisfy rule~(ii) for the connected diagram.
Thus, the types of such diagrams are determined uniquely.
In the following, we do not explicitly write the type of unit unless mentioned.
We assume $\dvx \in \mathcal{F}_{n,d,g}^{k,j,m}$ in the following.
We use the notation $A_i^{\sigma,\nu}(\dvx)=A_i^{\sigma,\nu}$ for simplicity in the following.
We call a coast in a diagram looking in the north~(south) if the symbols~~\Ifig~in the coast is on the upper~(lower) row.
For example, a coast looking in the north is like:
$
\begin{tikzpicture}[baseline=-0.5*\rd]
    \osd{0}{2}
    \Isu{0}{2}
    \dottedlinemiddle{-1}{0}
    \dottedlinemiddle{2}{3}
\end{tikzpicture}
$.
We omit~\Ifig~in the coast and indicate the length of the coast by an arrow, for example:
$
\begin{tikzpicture}[baseline=-0.5*\rd]
    \dottedlineboth{-1}{0}
    \osboth{0}{1}
    \osu{1}{4}
    \Isd{1}{4}
    \osboth{4}{5}
    \dottedlineboth{5}{6}
\end{tikzpicture}
=
\begin{tikzpicture}[baseline=-0.5*\rd]
    \dottedlineboth{-1}{0}
    \osboth{0}{1}
    \osu{1}{4}
    \osboth{4}{5}
    \dottedlineboth{5}{6}
    \spacelengthdown{1}{4}{$3$}
\end{tikzpicture}
$\ .
 
\subsection{Proof of~\eqref{A++}}
In this subsection, we prove~\eqref{A++}, $A_i^{++}=A_i^{--}=0$.
We firstly prove $A_i^{++}=0$.
Let $\dvx$ be a diagram where the $i$-th coast and $(i\pm 1)$-th coasts are not adjacent.

We consider the case in which the left and right sides of the $i$-th coast are overlaps, such as
\begin{align}
    \dvx
    =
    \begin{tikzpicture}[baseline=-0.5*\rd]
        \dottedlineup{-1}{0}
        \dottedlinedown{-1}{0}
        \osd{0}{2}
        \osd{7}{9}
        \osu{0}{4}
        \dottedlineup{4}{5}
        \osu{5}{9}
        \dottedlineup{9}{10}
        \dottedlinedown{9}{10}
        \spacelengthdown{2}{7}{$\lambda_i+1$}
    \end{tikzpicture}
    ,
\end{align}
where 
$
\begin{tikzpicture}[baseline=0.5*\rd]
    \osu{0}{3}
    \dottedlineup{3}{4}
    \osu{4}{7}
\end{tikzpicture}
$ represents the sequence of \ofig.
In this case, we can see $\sigma^L_i=\sigma^R_i=+$, and we have
\begin{align}
    \label{eq:app_proof_first}
A_i^{++}\dvx=
    &
    (-1)^{n+m+g}
    \left\{
        -
        C^{j,m}_{n-1, d+1}\paren{\lambdabold_{i:-1}}
        \begin{tikzpicture}[baseline=-0.5*\rd]
        \dottedlineup{-1}{0}
        \dottedlinedown{-1}{0}
        \osd{0}{3}
        \osd{7}{9}
        \osu{0}{4}
        \dottedlineup{4}{5}
        \osu{5}{9}
        \dottedlineup{9}{11}
        \dottedlinedown{9}{11}
        \spacelengthdown{3}{7}{$\lambda_i$}
        \enclosedown{2}{3}
    \end{tikzpicture}
        -
        C^{j,m}_{n-1, d+1}\paren{\lambdabold_{i:-1}}
        \begin{tikzpicture}[baseline=-0.5*\rd]
        \dottedlineup{-1}{0}
        \dottedlinedown{-1}{0}
        \osd{0}{2}
        \osd{6}{9}
        \osu{0}{4}
        \dottedlineup{4}{5}
        \osu{5}{9}
        \dottedlineup{9}{11}
        \dottedlinedown{9}{11}
        \spacelengthdown{2}{6}{$\lambda_i$}
        \enclosedown{6}{7}
    \end{tikzpicture}
    \right.
    \nonumber\\
    &
    \left.
    +
    C^{j,m}_{n, d-1}\paren{\lambdabold_{i:+1}}
    \begin{tikzpicture}[baseline=-0.5*\rd]
        \enclosedown{1}{2}
        \dottedlineup{-1}{0}
        \dottedlinedown{-1}{0}
        \osd{0}{1}
        \osd{7}{9}
        \osu{0}{4}
        \dottedlineup{4}{5}
        \osu{5}{9}
        \dottedlineup{9}{11}
        \dottedlinedown{9}{11}
        \spacelengthdown{1}{7}{$\lambda_i+2$}
    \end{tikzpicture}
    +
    C^{j,m}_{n, d-1}\paren{\lambdabold_{i:+1}}
    \begin{tikzpicture}[baseline=-0.5*\rd]
        \enclosedown{7}{8}
        \dottedlineup{-1}{0}
        \dottedlinedown{-1}{0}
        \osd{0}{2}
        \osd{8}{9}
        \osu{0}{4}
        \dottedlineup{4}{5}
        \osu{5}{9}
        \dottedlineup{9}{11}
        \dottedlinedown{9}{11}
        \spacelengthdown{2}{8}{$\lambda_i+2$}
    \end{tikzpicture}
    \right\}
    \nonumber\\
    =&
    (-1)^{n+m+g}
    \left\{
        -
        C^{j,m}_{n-1, d+1}\paren{\lambdabold_{i:-1}}
        +
        C^{j,m}_{n-1, d+1}\paren{\lambdabold_{i:-1}}
    +
    C^{j,m}_{n, d-1}\paren{\lambdabold_{i:+1}}
    -
    C^{j,m}_{n, d-1}\paren{\lambdabold_{i:+1}}
    \right\}
    \dvx
    \nonumber\\
    =&
    0.
\end{align}
We can see from where the contributions to $A_i^{\sigma, \mu}$ come by observing the changes in the support, double, gap-number, and unit-number between the $\dvx$ and the diagram in $Q_k^j$ that contribute to $A_i^{\sigma, \mu}$.
For example, for the above case~\eqref{eq:app_proof_first}, we can see
$
\begin{tikzpicture}[baseline=-0.5*\rd]
        \dottedlineup{-1}{0}
        \dottedlinedown{-1}{0}
        \osd{0}{3}
        \osd{7}{9}
        \osu{0}{4}
        \dottedlineup{4}{5}
        \osu{5}{9}
        \dottedlineup{9}{11}
        \dottedlinedown{9}{11}
        \spacelengthdown{3}{7}{$\lambda_i$}
    \end{tikzpicture}
       ,\quad
    \begin{tikzpicture}[baseline=-0.5*\rd]
        \dottedlineup{-1}{0}
        \dottedlinedown{-1}{0}
        \osd{0}{2}
        \osd{6}{9}
        \osu{0}{4}
        \dottedlineup{4}{5}
        \osu{5}{9}
        \dottedlineup{9}{11}
        \dottedlinedown{9}{11}
        \spacelengthdown{2}{6}{$\lambda_i$}
    \end{tikzpicture}
    \in 
    \connectedset^{k,j,m}_{n-1,d+1,g}
    ,\quad
    \begin{tikzpicture}[baseline=-0.5*\rd]
        \dottedlineup{-1}{0}
        \dottedlinedown{-1}{0}
        \osd{0}{1}
        \osd{7}{9}
        \osu{0}{4}
        \dottedlineup{4}{5}
        \osu{5}{9}
        \dottedlineup{9}{11}
        \dottedlinedown{9}{11}
        \spacelengthdown{1}{7}{$\lambda_i+2$}
    \end{tikzpicture}
    ,\quad
    \begin{tikzpicture}[baseline=-0.5*\rd]
        \dottedlineup{-1}{0}
        \dottedlinedown{-1}{0}
        \osd{0}{2}
        \osd{8}{9}
        \osu{0}{4}
        \dottedlineup{4}{5}
        \osu{5}{9}
        \dottedlineup{9}{11}
        \dottedlinedown{9}{11}
        \spacelengthdown{2}{8}{$\lambda_i+2$}
    \end{tikzpicture}
    \in 
    \connectedset^{k,j,m}_{n,d-1,g}
$, and the sign factor of each term in the RHS on the first line of~\eqref{eq:app_proof_first} is determined by where the contribution comes from.
If the diagram $\dv \in \connectedset^{k,j,m}_{n,d,g}$ contribute to $A^{\sigma, \nu}_{i}$, $\dv$ is accompanied by the sign $(-1)^{n+m+g}$, which can be seen from Theorem~1.
In the following, we omit the explanation for where the contribution comes from unless mentioned.

We next consider the case in which the left and right sides of the $i$-th coast are both gaps, such as
\begin{align}
    \dvx
    =
    \begin{tikzpicture}[baseline=-0.5*\rd]
        \dottedlineboth{-1}{0}
        \gap{0}{2}
        \osu{2}{4}
        \dottedlineup{4}{5}
        \osu{5}{7}
        \gap{7}{9}
        \dottedlineboth{9}{10}
        \spacelengthdown{2}{7}{$\lambda_i+1$}
    \end{tikzpicture}
    .
\end{align}
In this case, we can see $\sigma^L_i=\sigma^R_i=+$, and we obtain the value of  $A_i^{\sigma^L_i,\sigma^R_i}=A_i^{++}$ from the following calculation,
\begin{align}
A_i^{++}\dvx=
    &
    (-1)^{n+m+g}
    \left\{
    -
    C^{j,m}_{n-1, d+1}\paren{\lambdabold_{i:-1}}
    \begin{tikzpicture}[baseline=-0.5*\rd]
        \encloseup{2}{3}
        \dottedlineboth{-1}{0}
        \gap{0}{3}
        \osu{3}{4}
        \dottedlineup{4}{5}
        \osu{5}{7}
        \gap{7}{9}
        \dottedlineboth{9}{10}
        \spacelengthdown{3}{7}{$\lambda_i$}
    \end{tikzpicture}
    -
    C^{j,m}_{n-1, d+1}\paren{\lambdabold_{i:-1}}
    \begin{tikzpicture}[baseline=-0.5*\rd]
        \encloseup{6}{7}
        \dottedlineboth{-1}{0}
        \gap{0}{2}
        \osu{2}{4}
        \dottedlineup{4}{5}
        \osu{5}{6}
        \gap{6}{9}
        \dottedlineboth{9}{10}
        \spacelengthdown{2}{6}{$\lambda_i$}
    \end{tikzpicture}
    \right.
    \nonumber\\
    &
    \left.
    -
    C^{j,m}_{n, d-1}\paren{\lambdabold_{i:+1}}
    \begin{tikzpicture}[baseline=-0.5*\rd]
        \encloseup{1}{2}
        \dottedlineboth{-1}{0}
        \gap{0}{1}
        \osu{1}{4}
        \dottedlineup{4}{5}
        \osu{5}{7}
        \gap{7}{9}
        \dottedlineboth{9}{10}
        \spacelengthdown{1}{7}{$\lambda_i+2$}
    \end{tikzpicture}
    -
    C^{j,m}_{n, d-1}\paren{\lambdabold_{i:+1}}
    \begin{tikzpicture}[baseline=-0.5*\rd]
        \encloseup{7}{8}
        \dottedlineboth{-1}{0}
        \gap{0}{2}
        \osu{2}{4}
        \dottedlineup{4}{5}
        \osu{5}{8}
        \gap{8}{9}
        \dottedlineboth{9}{10}
        \spacelengthdown{2}{8}{$\lambda_i+2$}
    \end{tikzpicture}
    \right\}
    \nonumber\\
    =
    &
    (-1)^{n+m+g}
    \left\{
    C^{j,m}_{n-1, d+1}\paren{\lambdabold_{i:-1}}
    -
    C^{j,m}_{n-1, d+1}\paren{\lambdabold_{i:-1}}
    +
    C^{j,m}_{n, d-1}\paren{\lambdabold_{i:-1}}
    -
    C^{j,m}_{n, d-1}\paren{\lambdabold_{i:-1}}
    \right\}
    \dvx
    \nonumber\\
    =
    &
    0.
\end{align}

We next consider the case in which the left side of the $i$-th coast is a gap, and the right side is an overlap, such as
\begin{align}
    \dvx
    =
    \begin{tikzpicture}[baseline=-0.5*\rd]
        \dottedlineboth{-1}{0}
        \gap{0}{2}
        \osu{2}{4}
        \dottedlineup{4}{5}
        \osu{5}{7}
        \osboth{7}{9}
        \dottedlineboth{9}{10}
        \spacelengthdown{2}{7}{$\lambda_i+1$}
    \end{tikzpicture}
    .
\end{align}
In this case, we can see $\sigma^L_i=\sigma^R_i=+$, and we obtain the value of  $A_i^{\sigma^L_i,\sigma^R_i}=A_i^{++}$ from the following calculation,
\begin{align}
A_i^{++}\dvx=
    &
    (-1)^{n+m+g}
    \left\{
    C^{j,m}_{n-1, d+1}\paren{\lambdabold_{i:-1}}
    \begin{tikzpicture}[baseline=-0.5*\rd]
        \encloseup{2}{3}
        \dottedlineboth{-1}{0}
        \gap{0}{3}
        \osu{3}{4}
        \dottedlineup{4}{5}
        \osu{5}{7}
        \osboth{7}{9}
        \dottedlineboth{9}{10}
        \spacelengthdown{3}{7}{$\lambda_i$}
    \end{tikzpicture}
    +
    C^{j,m}_{n-1, d+1}\paren{\lambdabold_{i:-1}}
    \begin{tikzpicture}[baseline=-0.5*\rd]
        \enclosedown{6}{7}
        \dottedlineboth{-1}{0}
        \gap{0}{2}
        \osu{2}{4}
        \dottedlineup{4}{5}
        \osu{5}{6}
        \osboth{6}{9}
        \dottedlineboth{9}{10}
        \spacelengthdown{2}{6}{$\lambda_i$}
    \end{tikzpicture}
    \right.
    \nonumber\\
    &
    \left.
    -
    C^{j,m}_{n, d-1}\paren{\lambdabold_{i:+1}}
    \begin{tikzpicture}[baseline=-0.5*\rd]
        \encloseup{1}{2}
        \dottedlineboth{-1}{0}
        \gap{0}{1}
        \osu{1}{4}
        \dottedlineup{4}{5}
        \osu{5}{7}
        \osboth{7}{9}
        \dottedlineboth{9}{10}
        \spacelengthdown{1}{7}{$\lambda_i+2$}
    \end{tikzpicture}
    +
    C^{j,m}_{n, d-1}\paren{\lambdabold_{i:+1}}
    \begin{tikzpicture}[baseline=-0.5*\rd]
        \enclosedown{7}{8}
        \dottedlineboth{-1}{0}
        \gap{0}{2}
        \osu{2}{4}
        \dottedlineup{4}{5}
        \osu{5}{8}
        \osboth{8}{9}
        \dottedlineboth{9}{10}
        \spacelengthdown{2}{8}{$\lambda_i+2$}
    \end{tikzpicture}
    \right\}
    \nonumber\\
    =
    &
    (-1)^{n+m+g}
    \left\{
    -
    C^{j,m}_{n-1, d+1}\paren{\lambdabold_{i:-1}}
    +
    C^{j,m}_{n-1, d+1}\paren{\lambdabold_{i:-1}}
    +
    C^{j,m}_{n, d-1}\paren{\lambdabold_{i:-1}}
    -
    C^{j,m}_{n, d-1}\paren{\lambdabold_{i:-1}}
    \right\}
    \dvx
    \nonumber\\
    =
    &
    0.
\end{align}

In the same way, in the case in which the right side of the $i$-th coast is a gap, and the left side of that is an overlap,  such as
\begin{align}
    \dvx
    =
    \begin{tikzpicture}[baseline=-0.5*\rd]
        \dottedlineboth{-1}{0}
        \osboth{0}{2}
        \osu{2}{4}
        \dottedlineup{4}{5}
        \osu{5}{7}
        \gap{7}{9}
        \dottedlineboth{9}{10}
        \spacelengthdown{2}{7}{$\lambda_i+1$}
    \end{tikzpicture}
    \ ,
\end{align}
we can see $\sigma^L_i=\sigma^R_i=+$, and $A_i^{\sigma^L_i,\sigma^R_i}\dvx=A_i^{+,+}\dvx=0$.

Thus, we have proved  $A_i^{++}=0$ for all the cases.

We next prove $A_i^{--}=0$.
Let $\dvx$ be a diagram where the $i$-th coast and $(i\pm 1)$-th coasts are adjacent.

We first consider the case in which the $i$-th and $(i\pm 1)$-th coast are adjacent and looking in the same direction, such as 
\begin{align}
    \dvx
    =
    \begin{tikzpicture}[baseline=-0.5*\rd]
        \dottedlineup{-1}{0}
        \dottedlinedown{-1}{0}
        \Isd{0}{1}
        \Isd{8}{9}
        \zd{1}
        \zd{8}
        \osu{0}{4}
        \dottedlineup{4}{5}
        \osu{5}{9}
        \dottedlineup{9}{10}
        \dottedlinedown{9}{10}
        \spacelengthdown{1}{8}{$\lambda_i+1$}
    \end{tikzpicture}
    .
\end{align}
In this case, we can see $\sigma^L_i=\sigma^R_i=-$, and we have
\begin{align}
A_i^{--}\dvx=
    &
    (-1)^{n+m+g}
    \bce{
        -
        C^{j,m}_{n-1, d+1}\paren{\lambdabold_{i:-1}}
        \begin{tikzpicture}[baseline=-0.5*\rd]
        \enclosedown{1}{2}
        \belowarrowtikz{2}{1}
        \dottedlineup{-1}{0}
        \dottedlinedown{-1}{0}
        \Isd{0}{1}
        \Isd{8}{9}
        \oszd{1}{2}
        \zd{8}
        \osu{0}{4}
        \dottedlineup{4}{5}
        \osu{5}{9}
        \dottedlineup{9}{10}
        \dottedlinedown{9}{10}
        \spacelengthdown{2}{8}{$\lambda_i$}
    \end{tikzpicture}
        -
        C^{j,m}_{n-1, d+1}\paren{\lambdabold_{i:-1}}
        \begin{tikzpicture}[baseline=-0.5*\rd]
        \enclosedown{7}{8}
        \belowarrowtikz{7}{8}
        \dottedlineup{-1}{0}
        \dottedlinedown{-1}{0}
        \Isd{0}{1}
        \Isd{8}{9}
        \oszd{7}{8}
        \zd{1}
        \osu{0}{4}
        \dottedlineup{4}{5}
        \osu{5}{9}
        \dottedlineup{9}{10}
        \dottedlinedown{9}{10}
        \spacelengthdown{1}{7}{$\lambda_i$}
    \end{tikzpicture}
    }
    \nonumber\\
    =&
    (-1)^{n+m+g}
    2
    \bce{
        C^{j,m}_{n-1, d+1}\paren{\lambdabold_{i:-1}}
        -
        C^{j,m}_{n-1, d+1}\paren{\lambdabold_{i:-1}}
    }
    \nonumber\\
    =&
    0.
\end{align}

We consider the case in which the $i$-th and $(i\pm 1)$-th coast are adjacent, the $i$-th and $(i+1)$-th coasts are looking in the same direction, and the $i$-th and $(i-1)$-th coasts are not looking in the same direction, such as
\begin{align}
    \dvx
    =
    \begin{tikzpicture}[baseline=-0.5*\rd]
        \dottedlineboth{-1}{0}
        \lengtarrowuponeside{0}{2}{$\hspace{-1.5em}\lambda_{i\!-\!1}\!+\!1$}
        \osd{0}{2}
        \osu{2}{4}
        \dottedlineup{4}{5}
        \osu{5}{8}
        \zd{7}
        \Isd{7}{8}
        \dottedlineboth{8}{9}
        \spacelengthdown{2}{7}{$\lambda_i+1$}
    \end{tikzpicture}
    .
\end{align}
In this case, we can see $\sigma^L_i=\sigma^R_i=-$, and we have
\begin{align}
A_i^{--}\dvx=
    &
    (-1)^{n+m+g}
    \left\{
    -
    C^{j,m}_{n-1, d+1}\paren{\lambdabold_{i:-1}}
    \begin{tikzpicture}[baseline=-0.5*\rd]
        \enclosedown{2}{3}
        \dottedlineboth{-1}{0}
        \lengtarrowuponeside{0}{2}{}
        \osd{0}{3}
        \osu{2}{4}
        \dottedlineup{4}{5}
        \osu{5}{8}
        \zd{7}
        \Isd{7}{8}
        \dottedlineboth{8}{9}
        \spacelengthdown{3}{7}{$\lambda_i$}
    \end{tikzpicture}
    -
    C^{j,m}_{n-1, d+1}\paren{\lambdabold_{i:-1}}
    \begin{tikzpicture}[baseline=-0.5*\rd]
        \enclosedown{6}{7}
        \belowarrowtikz{6}{7}
        \dottedlineboth{-1}{0}
        \lengtarrowuponeside{0}{2}{}
        \osd{0}{2}
        \osu{2}{4}
        \dottedlineup{4}{5}
        \osu{5}{8}
        \oszd{6}{7}
        \Isd{7}{8}
        \dottedlineboth{8}{9}
        \spacelengthdown{2}{6}{$\lambda_i$}
    \end{tikzpicture}
    \right.
    \nonumber\\
    &
    \left.
    +
    C^{j,m}_{n-1, d+1}\paren{\lambdabold_{i:-1}}
    \begin{tikzpicture}[baseline=-0.5*\rd]
        \encloseup{2}{3}
        \dottedlineboth{-1}{0}
        \lengtarrowuponeside{0}{2}{}
        \osd{0}{2}
        \gap{2}{3}
        \osu{3}{4}
        \dottedlineup{4}{5}
        \osu{5}{8}
        \zd{7}
        \Isd{7}{8}
        \dottedlineboth{8}{9}
        \spacelengthdown{3}{7}{$\lambda_i$}
    \end{tikzpicture}
    \right\}
    \nonumber\\
    =
    &
    (-1)^{n+m+g}
    \left\{
    -
    C^{j,m}_{n-1, d+1}\paren{\lambdabold_{i:-1}}
    +2
    C^{j,m}_{n-1, d+1}\paren{\lambdabold_{i:-1}}
    -
    C^{j,m}_{n-1, d+1}\paren{\lambdabold_{i:-1}}
    \right\}
    \dvx
    \nonumber\\
    =
    &
    0.
\end{align}

In the same way, in the case in which the $i$-th and $(i\pm 1)$-th coasts are adjacent, the  $i$-th and $(i-1)$-th coast are looking in the same direction and the  $i$-th and $(i+1)$-th coast are not looking in the same direction, such as 
\begin{align}
    \dvx
    =
    \begin{tikzpicture}[baseline=-0.5*\rd, xscale=-1]
        \dottedlineboth{-1}{0}
        \lengtarrowuponeside{0}{2}{$\hspace{1.5em}\lambda_{i\!+\!1}\!+\!1$}
        \osd{0}{2}
        \osu{2}{4}
        \dottedlineup{4}{5}
        \osu{5}{8}
        \zd{7}
        \Isd{7}{8}
        \dottedlineboth{8}{9}
        \spacelengthdown{2}{7}{$\lambda_i+1$}
    \end{tikzpicture}
    \ ,
\end{align}
we can see $\sigma^L_i=\sigma^R_i=-$, and we have $A_i^{\sigma^L_i,\sigma^R_i}=A_i^{--}=0$.

We consider the case in which the $i$-th and $(i\pm 1)$-th coast are adjacent and they are not looking in the same direction, such as
\begin{align}
    \dvx
    =
    \begin{tikzpicture}[baseline=-0.5*\rd]
        \dottedlineboth{-1}{0}
        \lengtarrowuponeside{0}{2}{$\hspace{-1.5em}\lambda_{i\!-\!1}\!+\!1$}
        \lengtarrowuponeside{9}{7}{$\hspace{1.5em}\lambda_{i\!+\!1}\!+\!1$}
        \osd{0}{2}
        \osu{2}{4}
        \dottedlineup{4}{5}
        \osu{5}{7}
        \osd{7}{9}
        \dottedlineboth{9}{10}
        \spacelengthdown{2}{7}{$\lambda_i+1$}
    \end{tikzpicture}
    .
\end{align}
In this case, we can see $\sigma^L_i=\sigma^R_i=-$, and we have
\begin{align}
A_i^{--}\dvx=
    &
    (-1)^{n+m+g}
    \left\{
    -
    C^{j,m}_{n-1, d+1}\paren{\lambdabold_{i:-1}}
    \begin{tikzpicture}[baseline=-0.5*\rd]
        \enclosedown{2}{3}
        \dottedlineboth{-1}{0}
        \lengtarrowuponeside{0}{2}{}
        \lengtarrowuponeside{9}{7}{}
        \osd{0}{3}
        \osu{2}{4}
        \dottedlineup{4}{5}
        \osu{5}{7}
        \osd{7}{9}
        \dottedlineboth{9}{10}
        \spacelengthdown{3}{7}{$\lambda_i$}
    \end{tikzpicture}
    -
    C^{j,m}_{n-1, d+1}\paren{\lambdabold_{i:-1}}
    \begin{tikzpicture}[baseline=-0.5*\rd]
        \enclosedown{6}{7}
        \dottedlineboth{-1}{0}
        \lengtarrowuponeside{0}{2}{}
        \lengtarrowuponeside{9}{7}{}
        \osd{0}{2}
        \osu{2}{4}
        \dottedlineup{4}{5}
        \osu{5}{7}
        \osd{6}{9}
        \dottedlineboth{9}{10}
        \spacelengthdown{2}{6}{$\lambda_i$}
    \end{tikzpicture}
    \right.
    \nonumber\\
    &
    \left.
    +
     C^{j,m}_{n-1, d+1}\paren{\lambdabold_{i:-1}}
    \begin{tikzpicture}[baseline=-0.5*\rd]
        \encloseup{2}{3}
        \dottedlineboth{-1}{0}
        \lengtarrowuponeside{0}{2}{}
        \lengtarrowuponeside{9}{7}{}
        \osd{0}{2}
        \osu{3}{4}
        \dottedlineup{4}{5}
        \osu{5}{7}
        \gap{2}{3}
        \osd{7}{9}
        \dottedlineboth{9}{10}
        \spacelengthdown{3}{7}{$\lambda_i$}
    \end{tikzpicture}
    +
    C^{j,m}_{n-1, d+1}\paren{\lambdabold_{i:-1}}
    \begin{tikzpicture}[baseline=-0.5*\rd]
        \encloseup{6}{7}
        \dottedlineboth{-1}{0}
        \lengtarrowuponeside{0}{2}{}
        \lengtarrowuponeside{9}{7}{}
        \osd{0}{2}
        \osu{2}{4}
        \dottedlineup{4}{5}
        \osu{5}{6}
        \gap{6}{7}
        \osd{7}{9}
        \dottedlineboth{9}{10}
        \spacelengthdown{2}{6}{$\lambda_i$}
    \end{tikzpicture}
    \right\}
    \nonumber\\
    =
    &
    (-1)^{n+m+g}
    \left\{
    -
    C^{j,m}_{n-1, d+1}\paren{\lambdabold_{i:-1}}
    +
    C^{j,m}_{n-1, d+1}\paren{\lambdabold_{i:-1}}
    +
    C^{j,m}_{n-1, d+1}\paren{\lambdabold_{i:-1}}
    -
    C^{j,m}_{n-1, d+1}\paren{\lambdabold_{i:-1}}
    \right\}
    \dvx
    \nonumber\\
    =
    &
    0.
\end{align}

Thus, we have proved  $A_i^{--}=0$ for all the cases.

This concludes the proof of~\eqref{A++}.

\subsection{Proof of~\eqref{A+-}}
Within this subsection, we define
$R
\equiv
(-1)^{n+m+g}
\left\{
C^{j,m}_{n-1, d+1}\paren{\lambdabold_{i:-1}}
+
C^{j,m}_{n, d-1}\paren{\lambdabold_{i:+1}}
\right\}
$ and let $\dvx$ be a diagram where the $i$-th coast and $(i-1)$-th coasts are adjacent and the $i$-th coast and $(i+1)$-th coasts are not adjacent.
We prove~\eqref{A+-}, i.e. $A_i^{+-}=-A_i^{-+}=R$ in the following.

We first consider the case in which the $i$-th and $(i+1)$-th coasts are adjacent and looking in the same direction, and the left side of the $i$-th coast is an overlap, such as
\begin{align}
    \dvx
    =
    \begin{tikzpicture}[baseline=-0.5*\rd]
        \dottedlineup{-1}{0}
        \dottedlinedown{-1}{0}
        \osd{0}{2}
        \Isd{7}{8}
        \zd{7}
        \osu{0}{4}
        \dottedlineup{4}{5}
        \osu{5}{8}
        \dottedlineup{8}{9}
        \dottedlinedown{8}{9}
        \spacelengthdown{2}{7}{$\lambda_i+1$}
    \end{tikzpicture}
    .
\end{align}
In this case, we can see $\sigma^L_i=+ $ and $ \sigma^R_i=-$, and we have
\begin{align}
A_i^{+-}\dvx=
    &
    (-1)^{n+m+g}
    \left\{
    -
    C^{j,m}_{n-1, d+1}\paren{\lambdabold_{i:-1}}
    \begin{tikzpicture}[baseline=-0.5*\rd]
        \enclosedown{2}{3}
        \dottedlineup{-1}{0}
        \dottedlinedown{-1}{0}
        \osd{0}{3}
        \Isd{7}{8}
        \zd{7}
        \osu{0}{4}
        \dottedlineup{4}{5}
        \osu{5}{8}
        \dottedlineup{8}{9}
        \dottedlinedown{8}{9}
        \spacelengthdown{3}{7}{$\lambda_i$}
    \end{tikzpicture}
    -
    C^{j,m}_{n-1, d+1}\paren{\lambdabold_{i:-1}}
    \begin{tikzpicture}[baseline=-0.5*\rd]
    \enclosedown{6}{7}
        \dottedlineup{-1}{0}
        \dottedlinedown{-1}{0}
        \osd{0}{2}
        \Isd{7}{8}
        \oszd{6}{7}
        \belowarrowtikz{6}{7}
        \osu{0}{4}
        \dottedlineup{4}{5}
        \osu{5}{8}
        \dottedlineup{8}{9}
        \dottedlinedown{8}{9}
        \spacelengthdown{2}{6}{$\lambda_i$}
    \end{tikzpicture}
    \right.
    \nonumber\\
    &
    \hspace{23em}
    \left.
    +
    C^{j,m}_{n, d-1}\paren{\lambdabold_{i:+1}}
    \begin{tikzpicture}[baseline=-0.5*\rd]
        \enclosedown{1}{2}
        \dottedlineup{-1}{0}
        \dottedlinedown{-1}{0}
        \osd{0}{1}
        \Isd{7}{8}
        \zd{7}
        \osu{0}{4}
        \dottedlineup{4}{5}
        \osu{5}{8}
        \dottedlineup{8}{9}
        \dottedlinedown{8}{9}
        \spacelengthdown{1}{7}{$\lambda_i+2$}
    \end{tikzpicture}
    \right\}
    \nonumber\\
    =
    &
    (-1)^{n+m+g}
    \left\{
    -
    C^{j,m}_{n-1, d+1}\paren{\lambdabold_{i:-1}}
    +2
    C^{j,m}_{n-1, d+1}\paren{\lambdabold_{i:-1}}
    +
    C^{j,m}_{n, d-1}\paren{\lambdabold_{i:+1}}
    \right\}
    \dvx
    \nonumber\\
    =
    &
    R
    \dvx
    .
\end{align}

We next consider the case in which the $i$-th and $(i+1)$-th coasts are adjacent and looking in the same direction, and the left side of the $i$-th coast is a gap, such as
\begin{align}
    \dvx
    =
    \begin{tikzpicture}[baseline=-0.5*\rd]
        \dottedlineboth{-1}{0}
        \gap{0}{2}
        \osu{2}{4}
        \dottedlineup{4}{5}
        \osu{5}{7}
        \osu{7}{8}
        \zd{7}
        \Isd{7}{8}
        \dottedlineboth{8}{9}
        \spacelengthdown{2}{7}{$\lambda_i+1$}
    \end{tikzpicture}
    \ .
\end{align}
In this case, we can see $\sigma^L_i=+$ and $ \sigma^R_i=-$,  and we have
\begin{align}
A_i^{+-}\dvx=
    &
    (-1)^{n+m+g}
    \left\{
    C^{j,m}_{n-1, d+1}\paren{\lambdabold_{i:-1}}
    \begin{tikzpicture}[baseline=-0.5*\rd]
        \encloseup{2}{3}
        \dottedlineboth{-1}{0}
        \gap{0}{3}
        \osu{3}{4}
        \dottedlineup{4}{5}
        \osu{5}{7}
        \osu{7}{8}
        \zd{7}
        \Isd{7}{8}
        \dottedlineboth{8}{9}
        \spacelengthdown{3}{7}{$\lambda_i$}
    \end{tikzpicture}
    -
    C^{j,m}_{n-1, d+1}\paren{\lambdabold_{i:-1}}
    \begin{tikzpicture}[baseline=-0.5*\rd]
        \enclosedown{6}{7}
         \belowarrowtikz{6}{7}
        \dottedlineboth{-1}{0}
        \gap{0}{2}
        \osu{2}{4}
        \dottedlineup{4}{5}
        \osu{5}{7}
        \osu{7}{8}
        \oszd{6}{7}
        \Isd{7}{8}
        \dottedlineboth{8}{9}
        \spacelengthdown{2}{6}{$\lambda_i$}
    \end{tikzpicture}
    \right.
    \nonumber\\
    &
    \left.
    -
    C^{j,m}_{n, d-1}\paren{\lambdabold_{i:+1}}
    \begin{tikzpicture}[baseline=-0.5*\rd]
         \encloseup{1}{2}
        \dottedlineboth{-1}{0}
        \gap{0}{1}
        \osu{1}{4}
        \dottedlineup{4}{5}
        \osu{5}{7}
        \osu{7}{8}
        \zd{7}
        \Isd{7}{8}
        \dottedlineboth{8}{9}
        \spacelengthdown{1}{7}{$\lambda_i+2$}
    \end{tikzpicture}
    \right\}
    \nonumber\\
    =
    &
    (-1)^{n+m+g}
    \left\{
    -
    C^{j,m}_{n-1, d+1}\paren{\lambdabold_{i:-1}}
    +
    2
    C^{j,m}_{n-1, d+1}\paren{\lambdabold_{i:-1}}
    +
    C^{j,m}_{n, d-1}\paren{\lambdabold_{i:+1}}
    \right\}
    \dvx
    \nonumber\\
    =
    &
     R \dvx
    .
\end{align}

We next consider the case in which the $i$-th and $(i+1)$-th coasts are adjacent and looking in different direction, and the left side of the $i$-th coast is an overlap, such as
\begin{align}
    \dvx
    =
    \begin{tikzpicture}[baseline=-0.5*\rd]
        \dottedlineboth{-1}{0}
        \osboth{0}{2}
        \osu{2}{4}
        \dottedlineup{4}{5}
        \osu{5}{7}
        \osd{7}{9}
        \dottedlineboth{9}{10}
        \spacelengthdown{2}{7}{$\lambda_i+1$}
        \lengtarrowuponeside{9}{7}{$\ \ \ \   \lambda_{i+1}\!+\!1$}
    \end{tikzpicture}
    .
\end{align}
In this case, we can see $\sigma^L_i=+ $ and $ \sigma^R_i=-$,  and we have
\begin{align}
A_i^{+-}\dvx=
    &
    (-1)^{n+m+g}
    \left\{
    -
    C^{j,m}_{n-1, d+1}\paren{\lambdabold_{i:-1}}
    \begin{tikzpicture}[baseline=-0.5*\rd]
        \enclosedown{2}{3}
        \dottedlineboth{-1}{0}
        \osboth{0}{3}
        \osu{3}{4}
        \dottedlineup{4}{5}
        \osu{5}{7}
        \osd{7}{9}
        \dottedlineboth{9}{10}
        \spacelengthdown{3}{7}{$\lambda_i$}
        \lengtarrowuponeside{9}{7}{}
    \end{tikzpicture}
    -
    C^{j,m}_{n-1, d+1}\paren{\lambdabold_{i:-1}}
    \begin{tikzpicture}[baseline=-0.5*\rd]
        \enclosedown{6}{7}
        \dottedlineboth{-1}{0}
        \osboth{0}{2}
        \osu{2}{4}
        \dottedlineup{4}{5}
        \osu{5}{7}
        \osd{6}{9}
        \dottedlineboth{9}{10}
        \spacelengthdown{2}{6}{$\lambda_i$}
        \lengtarrowuponeside{9}{7}{}
    \end{tikzpicture}
    \right.
    \nonumber\\
    &
    \left.
    +
    C^{j,m}_{n-1, d+1}\paren{\lambdabold_{i:-1}}
    \begin{tikzpicture}[baseline=-0.5*\rd]
        \encloseup{6}{7}
        \dottedlineboth{-1}{0}
        \osboth{0}{2}
        \osu{2}{4}
        \dottedlineup{4}{5}
        \osu{5}{6}
        \gap{6}{7}
        \osd{7}{9}
        \dottedlineboth{9}{10}
        \spacelengthdown{2}{6}{$\lambda_i$}
        \lengtarrowuponeside{9}{7}{}
    \end{tikzpicture}
    +
    C^{j,m}_{n, d-1}\paren{\lambdabold_{i:+1}}
    \begin{tikzpicture}[baseline=-0.5*\rd]
        \enclosedown{1}{2}
        \dottedlineboth{-1}{0}
        \osboth{0}{1}
        \osu{1}{4}
        \dottedlineup{4}{5}
        \osu{5}{7}
        \osd{7}{9}
        \dottedlineboth{9}{10}
        \spacelengthdown{1}{7}{$\lambda_i+2$}
        \lengtarrowuponeside{9}{7}{}
    \end{tikzpicture}
    \right\}
    \nonumber\\
    =
    &
    (-1)^{n+m+g}
    \left\{
    -
    C^{j,m}_{n-1, d+1}\paren{\lambdabold_{i:-1}}
    +
    C^{j,m}_{n-1, d+1}\paren{\lambdabold_{i:-1}}
    +
    C^{j,m}_{n-1, d+1}\paren{\lambdabold_{i:-1}}
    +
    C^{j,m}_{n, d-1}\paren{\lambdabold_{i:+1}}
    \right\}
    \dvx
    \nonumber\\
    =
    &
    R
    \dvx
    .
\end{align}

We next consider the case in which the $i$-th and $(i+1)$-th coasts are adjacent and looking in different directions, and the left side of the $i$-th coast is a gap, such as
\begin{align}
    \dvx
    =
    \begin{tikzpicture}[baseline=-0.5*\rd]
        \dottedlineboth{-1}{0}
        \gap{0}{2}
        \osu{2}{4}
        \dottedlineup{4}{5}
        \osu{5}{7}
        \osd{7}{9}
        \dottedlineboth{9}{10}
        \spacelengthdown{2}{7}{$\lambda_i+1$}
        \lengtarrowuponeside{9}{7}{$\ \ \ \   \lambda_{i+1}\!+\!1$}
    \end{tikzpicture}
    .
\end{align}
In this case, $\sigma^L_i=+$ and $ \sigma^R_i=-$,  and we have
\begin{align}
A_i^{+-}\dvx=
    &
    (-1)^{n+m+g}
    \left\{
    C^{j,m}_{n-1, d+1}\paren{\lambdabold_{i:-1}}
    \begin{tikzpicture}[baseline=-0.5*\rd]
        \encloseup{2}{3}
        \dottedlineboth{-1}{0}
        \gap{0}{3}
        \osu{3}{4}
        \dottedlineup{4}{5}
        \osu{5}{7}
        \osd{7}{9}
        \dottedlineboth{9}{10}
        \spacelengthdown{3}{7}{$\lambda_i$}
        \lengtarrowuponeside{9}{7}{}
    \end{tikzpicture}
    +
    C^{j,m}_{n-1, d+1}\paren{\lambdabold_{i:-1}}
    \begin{tikzpicture}[baseline=-0.5*\rd]
        \encloseup{6}{7}
        \dottedlineboth{-1}{0}
        \gap{0}{2}
        \osu{2}{4}
        \dottedlineup{4}{5}
        \osu{5}{6}
        \gap{6}{7}
        \osd{7}{9}
        \dottedlineboth{9}{10}
        \spacelengthdown{2}{6}{$\lambda_i$}
        \lengtarrowuponeside{9}{7}{}
    \end{tikzpicture}
    \right.
    \nonumber\\
    &
    \left.
    -
    C^{j,m}_{n, d-1}\paren{\lambdabold_{i:+1}}
    \begin{tikzpicture}[baseline=-0.5*\rd]
        \encloseup{1}{2}
        \dottedlineboth{-1}{0}
        \gap{0}{1}
        \osu{1}{4}
        \dottedlineup{4}{5}
        \osu{5}{7}
        \osd{7}{9}
        \dottedlineboth{9}{10}
        \spacelengthdown{1}{7}{$\lambda_i+2$}
        \lengtarrowuponeside{9}{7}{}
    \end{tikzpicture}
    -
    C^{j,m}_{n-1, d+1}\paren{\lambdabold_{i:+1}}
    \begin{tikzpicture}[baseline=-0.5*\rd]
        \enclosedown{6}{7}
        \dottedlineboth{-1}{0}
        \gap{0}{2}
        \osu{2}{4}
        \dottedlineup{4}{5}
        \osu{5}{7}
        \osd{6}{9}
        \dottedlineboth{9}{10}
        \spacelengthdown{2}{6}{$\lambda_i$}
        \lengtarrowuponeside{9}{7}{}
    \end{tikzpicture}
    \right\}
    \nonumber\\
    =
    &
    (-1)^{n+m+g}
    \left\{
    -
    C^{j,m}_{n-1, d+1}\paren{\lambdabold_{i:-1}}
    +
    C^{j,m}_{n-1, d+1}\paren{\lambdabold_{i:-1}}
    +
    C^{j,m}_{n, d-1}\paren{\lambdabold_{i:+1}}
    -
    C^{j,m}_{n-1, d+1}\paren{\lambdabold_{i:-1}}
    \right\}
    \dvx
    \nonumber\\
    =
    &
    R
    \dvx
    .
\end{align}

Thus, we have proved 
$A_i^{+-}=R$ for all the cases.
In the same way, we can prove $A_i^{-+}=-R$ for all the cases.

This concludes the proof of~\eqref{A+-}.

\subsection{Proof of~\eqref{AL+} and~\eqref{AR+}}
Within this subsection, we define
$R
\equiv
(-1)^{n+m+g}
\{
C^{j,m}_{n-1, d+1}\paren{\lambdabold_{L:-1}}
-
C^{j,m}_{n, d}\paren{\lambdabold_{L:-1}}
+
C^{j,m}_{n-1, d}\paren{\lambdabold_{L:+1}}
-
C^{j,m}_{n, d-1}\paren{\lambdabold_{L:+1}}
+
C^{j-1,m-1}_{n, d}\paren{{}_{0\rightarrow}(\lambdabold_{L:-1})}
\}
$ and let $\dvx$ be a diagram where the leftmost coast(the $0$-th coast, corresponding to $\lambda_0$) and the $1$-th coast, corresponding to $\lambda_1$ are not adjacent.
In this subsection, we prove~\eqref{AL+}, i.e.  $A_0^{+} = R$.
\eqref{AR+} is proved in the same way.

We consider the case in which the right side of the leftmost coast is an overlap and $\lambda_L>0$, such as
\begin{align}
    \dvx
    =
    \begin{tikzpicture}[baseline=-0.5*\rd]
        \osu{0}{2}
        \dottedlineup{2}{3}
        \osu{3}{7}
        \osd{5}{7}
        \dottedlineboth{7}{8}
        \spacelengthdown{0}{5}{$\lambda_L$}
    \end{tikzpicture}
    .
\end{align}
In this case, we can see $\sigma^R_0=+$,  and we have
\begin{align}
A_0^{+}\dvx=
    &
    (-1)^{n+m+g}
    \left\{
    C^{j,m}_{n, d}\paren{\lambdabold_{L:-1}}
    \begin{tikzpicture}[baseline=-0.5*\rd]
        \encloseup{0}{1}
        \virtualgap{0}{1}
        \osu{1}{2}
        \dottedlineup{2}{3}
        \osu{3}{7}
        \osd{5}{7}
        \dottedlineboth{7}{8}
        \spacelengthdown{1}{5}{$\lambda_L-1$}
    \end{tikzpicture}
    -
    C^{j,m}_{n-1, d+1}\paren{\lambdabold_{L:-1}}
    \begin{tikzpicture}[baseline=-0.5*\rd]
        \enclosedown{4}{5}
        \osu{0}{2}
        \dottedlineup{2}{3}
        \osu{3}{7}
        \osd{4}{7}
        \dottedlineboth{7}{8}
        \spacelengthdown{0}{4}{$\lambda_L-1$}
    \end{tikzpicture}
    \right.
    \nonumber\\
    &
    \left.
    \hspace{7em}
    -
    C^{j,m}_{n-1, d}\paren{\lambdabold_{L:+1}}
    \begin{tikzpicture}[baseline=-0.5*\rd]
        \encloseup{-1}{0}
        \osu{-1}{2}
        \dottedlineup{2}{3}
        \osu{3}{7}
        \osd{5}{7}
        \dottedlineboth{7}{8}
        \spacelengthdown{-1}{5}{$\lambda_L+1$}
    \end{tikzpicture}
    +
    C^{j,m}_{n, d-1}\paren{\lambdabold_{L:+1}}
    \begin{tikzpicture}[baseline=-0.5*\rd]
        \enclosedown{5}{6}
        \osu{0}{2}
        \dottedlineup{2}{3}
        \osu{3}{7}
        \osd{6}{7}
        \dottedlineboth{7}{8}
        \spacelengthdown{0}{6}{$\lambda_L+1$}
    \end{tikzpicture}
    \right.
    \nonumber\\
    &
    \left.
    \hspace{14em}
    -
    C^{j-1,m-1}_{n, d}\paren{{}_{0\rightarrow}(\lambdabold_{L:-1})}
    \begin{tikzpicture}[baseline=-0.5*\rd]
        \osu{0}{2}
        \zu{0}
        \zd{0}
        \upverticalarrowtikz{0}
        \dottedlineup{2}{3}
        \osu{3}{7}
        \osd{5}{7}
        \dottedlineboth{7}{8}
        \spacelengthdown{0}{5}{$\lambda_L$}
    \end{tikzpicture}
    \right\}
    \nonumber\\
    =
    &
    R
    \dvx
    ,
\end{align}
where we note that 
$
    \begin{tikzpicture}[baseline=-0.5*\rd]
        \osu{0}{2}
        \zu{0}
        \zd{0}
        \dottedlineup{2}{3}
        \osu{3}{7}
        \osd{5}{7}
        \dottedlineboth{7}{8}
        \spacelengthdown{0}{5}{$\lambda_L$}
    \end{tikzpicture}
    \in 
    \connectedset^{k, j-1,m-1}_{n,d,g}
$.
The gap enclosed by the dotted line is \textit{virtual gap} on the leftmost or rightmost of the diagram.
We note that a virtual gap is not counted for the value of $g_\dv$ and $d_\dv$.

We consider the case in which $\lambda_L>0$ and the right side of the leftmost coast is a gap, such as, 
\begin{align}
    \dvx
    =
    \begin{tikzpicture}[baseline=-0.5*\rd]
        \osu{0}{2}
        \dottedlineup{2}{3}
        \osu{3}{5}
        \gap{5}{7}
        \dottedlineboth{7}8
        \spacelengthdown{0}{5}{$\lambda_L$}
    \end{tikzpicture}
    .
\end{align}
In this case, we can see $\sigma^R_0=+$, and we have
\begin{align}
A_0^{+}\dvx=
    &
    (-1)^{n+m+g}
    \left\{
    -
    C^{j,m}_{n-1, d}\paren{\lambdabold_{L:+1}}
    \begin{tikzpicture}[baseline=-0.5*\rd]
        \encloseup{-1}{0}
        \osu{-1}{2}
        \dottedlineup{2}{3}
        \osu{3}{5}
        \gap{5}{7}
        \dottedlineboth{7}8
        \spacelengthdown{-1}{5}{$\lambda_L+1$}
    \end{tikzpicture}
    +
    C^{j,m}_{n, d}\paren{\lambdabold_{L:-1}}
    \begin{tikzpicture}[baseline=-0.5*\rd]
        \encloseup{0}{1}
        \virtualgap{0}{1}
        \osu{1}{2}
        \dottedlineup{2}{3}
        \osu{3}{5}
        \gap{5}{7}
        \dottedlineboth{7}8
        \spacelengthdown{1}{5}{$\lambda_L-1$}
    \end{tikzpicture}
    \right.
    \nonumber\\
    &
    \left.
    +
    C^{j,m}_{n-1, d+1}\paren{\lambdabold_{L:-1}}
    \begin{tikzpicture}[baseline=-0.5*\rd]
        \encloseup{4}{5}
        \osu{0}{2}
        \dottedlineup{2}{3}
        \osu{3}{4}
        \gap{4}{7}
        \dottedlineboth{7}{8}
        \spacelengthdown{0}{4}{$\lambda_L-1$}
    \end{tikzpicture}
    -
    C^{j,m}_{n, d-1}\paren{\lambdabold_{L:+1}}
    \begin{tikzpicture}[baseline=-0.5*\rd]
        \encloseup{5}{6}
        \osu{0}{2}
        \dottedlineup{2}{3}
        \osu{3}{6}
        \gap{6}{7}
        \dottedlineboth{7}{8}
        \spacelengthdown{0}{6}{$\lambda_L+1$}
    \end{tikzpicture}
    \right.
    \nonumber\\
    &
    \hspace{17em}
    \left.
    -
    C^{j-1,m-1}_{n, d}\paren{{}_{0\rightarrow}(\lambdabold_{L:-1})}
    \begin{tikzpicture}[baseline=-0.5*\rd]
        \upverticalarrowtikz{0}
        \zboth{0}
        \osu{0}{2}
        \dottedlineup{2}{3}
        \osu{3}{5}
        \gap{5}{7}
        \dottedlineboth{7}8
        \spacelengthdown{0}{5}{$\lambda_L$}
    \end{tikzpicture}
    \right\}
    \nonumber\\
    =&
    R\dvx
    .
\end{align}

We consider the case in which $\lambda_L=0$ and the leftmost of the diagram is an overlap, such as
\begin{align}
    \dvx
    =
    \begin{tikzpicture}[baseline=-0.5*\rd]
        \osu{0}{2}
        \osd{0}{2}
        \dottedlineboth{2}{3}
    \end{tikzpicture}
    \ .
\end{align}
In this case, we can see $\sigma^R_0=+$ and we have
\begin{align}
A_0^{+}\dvx=
    &
    (-1)^{n+m+g}
    \left\{
    -
    C^{j,m}_{n-1, d}\paren{\lambdabold_{L:+1}}
    \begin{tikzpicture}[baseline=-0.5*\rd]
        \encloseup{-1}{0}
        \osu{-1}{2}
        \Isd{-1}{0}
        \osd{0}{2}
        \dottedlineboth{2}{3}
    \end{tikzpicture}
    -
    C^{j,m}_{n-1, d}\paren{\lambdabold_{L:+1}}
    \begin{tikzpicture}[baseline=-0.5*\rd]
        \enclosedown{-1}{0}
        \osd{-1}{2}
        \Isu{-1}{0}
        \osu{0}{2}
        \dottedlineboth{2}{3}
    \end{tikzpicture}
    \right.
    \nonumber\\
    &
    \left.
    \hspace{7em}
    +
    C^{j,m}_{n, d-1}\paren{\lambdabold_{L:+1}}
    \begin{tikzpicture}[baseline=-0.5*\rd]
        \enclosedown{0}{1}
        \osu{0}{2}
        \osd{1}{2}
        \Isd{0}{1}
        \dottedlineboth{2}{3}
    \end{tikzpicture}
    +
    C^{j,m}_{n, d-1}\paren{\lambdabold_{L:+1}}
    \begin{tikzpicture}[baseline=-0.5*\rd]
        \encloseup{0}{1}
        \osd{0}{2}
        \osu{1}{2}
        \Isu{0}{1}
        \dottedlineboth{2}{3}
    \end{tikzpicture}
    \right\}
    \nonumber\\
    =
    &
    (-1)^{n+m+g}
    2
    \left\{
    C^{j,m}_{n-1, d}\paren{\lambdabold_{L:+1}}
    -
    C^{j,m}_{n, d-1}\paren{\lambdabold_{L:+1}}
    \right\}
    \dvx
    \nonumber\\
    =
    &
    R
    \dvx
    ,
\end{align}
where we used 
$
C^{j,m}_{n-1, d}\paren{\lambdabold_{L:+1}}=C^{j,m}_{n-1, d+1}\paren{\lambdabold_{L:-1}}
$
and
$
    C^{j,m}_{n, d-1}\paren{\lambdabold_{L:+1}}
    =
     C^{j,m}_{n, d}\paren{\lambdabold_{L:-1}}
$
for $\lambda_L=0$
and 
$
C^{j-1,m-1}_{n, d}\paren{{}_{0\rightarrow}(\lambdabold_{L:-1})}
   =0
$ for $\lambda_L=0$.

We consider the case in which $\lambda_L=0$, the leftmost of the diagram is the zero-length unit on either the upper or lower row, and its right is a gap, such as
\begin{align}
    \dvx
    =
    \begin{tikzpicture}[baseline=-0.5*\rd]
        \zu{0}
        \gap{0}{2}
        \dottedlineboth{2}{3}
    \end{tikzpicture}
    \ .
\end{align}
In this case, we can see $\sigma^R_{0}=+$ and we have
\begin{align}
A_{0}^{+}\dvx
=
    &
    (-1)^{n+m+g}
    \left\{
    -
    C^{j,m}_{n-1, d}\paren{\lambdabold_{L:+1}}
    \begin{tikzpicture}[baseline=-0.5*\rd]
        \encloseup{-1}{0}
        \abovearrowtikz{-1}{0}
        \oszu{-1}{0}
        \Isd{-1}{0}
        \gap{0}{2}
        \dottedlineboth{2}{3}
    \end{tikzpicture}
    -
    C^{j,m}_{(n-1)-(g-1), d-1}\paren{\lambdabold_{L:+1}}
    \begin{tikzpicture}[baseline=-0.5*\rd]
        \encloseup{0}{1}
        \abovearrowtikz{1}{0}
        \oszu{0}{1}
        \Isd{0}{1}
        \gap{1}{2}
        \dottedlineboth{2}{3}
    \end{tikzpicture}
    \right\}
    \nonumber\\
    =&
    (-1)^{n+m+g}
    2
    \left\{
        C^{j,m}_{n-1, d}\paren{\lambdabold_{L:+1}}
        -
        C^{j,m}_{n, d-1}\paren{\lambdabold_{L:+1}}
    \right\}\dvx
    \nonumber\\
    =&
    R\dvx
    ,
\end{align}
where we used 
$
C^{j,m}_{n-1, d}\paren{\lambdabold_{L:+1}}
=
C^{j,m}_{n-1, d+1}\paren{\lambdabold_{L:-1}}
$
and 
$
C^{j,m}_{n, d-1}\paren{\lambdabold_{L:+1}}
=
C^{j,m}_{n, d}\paren{\lambdabold_{L:-1}}
$
and 
$
C^{j-1,m-1}_{n, d}\paren{{}_{0^\rightarrow}\lambdabold}
=0
$
for $\lambda_L=0$.

Thus, we have proved~\eqref{AL+} for all the cases.
In the same way, we can prove~\eqref{AR+} for all the cases.

\subsection{Proof of~\eqref{AL-} and~\eqref{AR-}}
Within this subsection, we define 
$R
\equiv
(-1)^{n+m+g}
\{
2
C^{j,m}_{n-1, d+1}\paren{\lambdabold_{L:-1}}
-
C^{j,m}_{n, d}\paren{\lambdabold_{L:-1}}
+
C^{j,m}_{n-1, d}\paren{\lambdabold_{L:+1}}
+
C^{j-1,m-1}_{n, d}\paren{{}_{0\rightarrow}(\lambdabold_{L:-1})}
\}
$ and let $\dvx$ be a diagram where the leftmost coast~(the $0$-th coast) and the $1$-th coast are adjacent.
In this subsection, we prove~\eqref{AL-}, i.e.  $A_0^{-} = R$.
\eqref{AR-} is proved in the same way.

We firstly consider the case in which the leftmost coast~(the $0$-th coast) and the $1$-th coast are looking in the same direction, and $\lambda_L>0$, such as
\begin{align}
    \dvx
    =
    \begin{tikzpicture}[baseline=-0.5*\rd]
        \osu{0}{2}
        \dottedlineup{2}{3}
        \osu{3}{6}
        \zd{5}
        \Isd{5}{6}
        \dottedlineboth{6}{7}
        \spacelengthdown{0}{5}{$\lambda_L$}
    \end{tikzpicture}
    .
\end{align}
In this case, we can see $\sigma^R_{0}=-$, and we have
\begin{align}
A_{0}^{+}\dvx=
    &
    (-1)^{n+m+g}
    \left\{
    C^{j,m}_{n, d}\paren{\lambdabold_{L:-1}}
    \begin{tikzpicture}[baseline=-0.5*\rd]
        \encloseup{0}{1}
        \virtualgap{0}{1}
        \osu{1}{2}
        \dottedlineup{2}{3}
        \osu{3}{6}
        \zd{5}
        \Isd{5}{6}
        \dottedlineboth{6}{7}
        \spacelengthdown{1}{5}{$\lambda_L-1$}
    \end{tikzpicture}
    -
    C^{j,m}_{n-1, d+1}\paren{\lambdabold_{L:-1}}
    \begin{tikzpicture}[baseline=-0.5*\rd]
        \enclosedown{4}{5}
        \osu{0}{2}
        \dottedlineup{2}{3}
        \osu{3}{6}
        \oszd{4}{5}
        \Isd{5}{6}
        \dottedlineboth{6}{7}
        \spacelengthdown{0}{4}{$\lambda_L-1$}
    \end{tikzpicture}
    \right.
    \nonumber\\
    &
    \left.
    \hspace{7em}
    -
    C^{j,m}_{n-1, d}\paren{\lambdabold_{L:+1}}
    \begin{tikzpicture}[baseline=-0.5*\rd]
        \encloseup{-1}{0}
        \osu{-1}{2}
        \dottedlineup{2}{3}
        \osu{3}{6}
        \zd{5}
        \Isd{5}{6}
        \dottedlineboth{6}{7}
        \spacelengthdown{-1}{5}{$\lambda_L+1$}
    \end{tikzpicture}
    -
    C^{j-1,m-1}_{n, d}\paren{{}_{0\rightarrow}(\lambdabold_{L:-1})}
    \begin{tikzpicture}[baseline=-0.5*\rd]
        \osu{0}{2}
        \zu{0}
        \zd{0}
        \upverticalarrowtikz{0}
        \dottedlineup{2}{3}
        \osu{3}{6}
        \zd{5}
        \Isd{5}{6}
        \dottedlineboth{6}{7}
        \spacelengthdown{0}{5}{$\lambda_L$}
    \end{tikzpicture}
    \right\}
    \nonumber\\
    =
    &
    R
    \dvx
    .
\end{align}

We next consider the case in which the leftmost coast~(the $0$-th coast) and the $1$-th coast are adjacent and looking in the different direction, and $\lambda_L>0$, such as
\begin{align}
    \dvx
    =
    \begin{tikzpicture}[baseline=-0.5*\rd]
        \osu{0}{2}
        \dottedlineup{2}{3}
        \oszu{3}{5}
        \osd{5}{7}
        \dottedlineboth{7}{8}
        \spacelengthdown{0}{5}{$\lambda_L$}
        \lengtarrowuponeside{7}{5}{$\hspace{1em}\lambda_1+1$}
    \end{tikzpicture}
    .
\end{align}
In this case, we can see $\sigma^R_{0}=-$,and we have
\begin{align}
A_{0}^{-}\dvx=
    &
    (-1)^{n+m+g}
    \left\{
    C^{j,m}_{n, d}\paren{\lambdabold_{L:-1}}
    \begin{tikzpicture}[baseline=-0.5*\rd]
        \encloseup{0}{1}
        \virtualgap{0}{1}
        \osu{1}{2}
        \dottedlineup{2}{3}
        \oszu{3}{5}
        \osd{5}{7}
        \dottedlineboth{7}{8}
        \spacelengthdown{1}{5}{$\lambda_L-1$}
        \lengtarrowuponeside{7}{5}{}
    \end{tikzpicture}
    -
    C^{j,m}_{n-1, d+1}\paren{\lambdabold_{L:-1}}
    \begin{tikzpicture}[baseline=-0.5*\rd]
        \enclosedown{4}{5}
        \osu{0}{2}
        \dottedlineup{2}{3}
        \oszu{3}{5}
        \osd{4}{7}
        \dottedlineboth{7}{8}
        \spacelengthdown{0}{4}{$\lambda_L-1$}
        \lengtarrowuponeside{7}{5}{}
    \end{tikzpicture}
    \right.
    \nonumber\\
    &
    \left.
    \hspace{7em}
    -
    C^{j,m}_{n-1, d}\paren{\lambdabold_{L:+1}}
    \begin{tikzpicture}[baseline=-0.5*\rd]
        \encloseup{-1}{0}
        \osu{-1}{2}
        \dottedlineup{2}{3}
        \oszu{3}{5}
        \osd{5}{7}
        \dottedlineboth{7}{8}
        \spacelengthdown{-1}{5}{$\lambda_L+1$}
        \lengtarrowuponeside{7}{5}{}
    \end{tikzpicture}
    +
    C^{j,m}_{n-1, d+1}\paren{\lambdabold_{L:+1}}
    \begin{tikzpicture}[baseline=-0.5*\rd]
        \encloseup{4}{5}
        \osu{0}{2}
        \dottedlineup{2}{3}
        \oszu{3}{4}
        \gap{4}{5}
        \osd{5}{7}
        \dottedlineboth{7}{8}
        \spacelengthdown{0}{4}{$\lambda_L-1$}
        \lengtarrowuponeside{7}{5}{}
    \end{tikzpicture}
    \right.
    \nonumber\\
    &
    \hspace{21em}
    \left.
    -
    C^{j-1,m-1}_{n, d}\paren{{}_{0\rightarrow}(\lambdabold_{L:-1})}
    \begin{tikzpicture}[baseline=-0.5*\rd]
        \upverticalarrowtikz{0}
        \zboth{0}
        \osu{0}{2}
        \dottedlineup{2}{3}
        \oszu{3}{5}
        \osd{5}{7}
        \dottedlineboth{7}{8}
        \spacelengthdown{0}{5}{$\lambda_L$}
        \lengtarrowuponeside{7}{5}{}
    \end{tikzpicture}
    \right\}
    \nonumber\\
    =
    &
    R
    \dvx
    ,
\end{align}
where we note that
$
    \begin{tikzpicture}[baseline=-0.5*\rd]
        \virtualgap{0}{1}
        \osu{1}{2}
        \dottedlineup{2}{3}
        \oszu{3}{5}
        \osd{5}{7}
        \dottedlineboth{7}{8}
        \spacelengthdown{1}{5}{$\lambda_L-1$}
        \lengtarrowuponeside{7}{5}{}
    \end{tikzpicture}
    \in 
    \connectedset^{k,j,m}_{n,d,g}
    ,
    \quad
    \begin{tikzpicture}[baseline=-0.5*\rd]
        \zboth{0}
        \osu{0}{2}
        \dottedlineup{2}{3}
        \oszu{3}{5}
        \osd{5}{7}
        \dottedlineboth{7}{8}
        \spacelengthdown{0}{5}{$\lambda_L$}
        \lengtarrowuponeside{7}{5}{}
    \end{tikzpicture}
    \in 
    \connectedset^{k, j-1, m-1}_{n,d,g}
$.

In the following, we consider the case of $\lambda_L=0$.
In this case, we also consider the contribution from the diagram in $Q_k^{j}$ with a list $\lambdabold_{1:-1}=\{0;\lambda_1-1,\ldots \}$, which is $
\begin{tikzpicture}[baseline=-0.5*\rd]
    \encloseup{0}{1}
    \osu{1}{2}
    \gap{0}{1}
    \zd{0}
    \lengtarrowdownoneside{2}{1}{}
    \dottedlineboth{2}{3}
\end{tikzpicture}
$ in~\eqref{eq:cancel-special} below.
According to the original definition of $A_i^{\sigma,\mu}$, a diagram with a list $\{0;\lambda_1-1,\ldots\}$ contributes to $A_1^{-,\sigma_i^R}$.
Nonetheless, we incorporate this contribution into $A_L^{-}$ rather than $A_1^{-,\sigma_i^R}$ for later convenience. 
This special treatment does not contradict the earlier arguments because this contribution was not considered previously, ensuring there is no double counting and maintaining overall consistency.

We consider the case in which $\lambda_L=0$, the $0$-th and $1$-th coast are adjacent, and $w>0$, such as
\begin{align}\label{eq:cancel-special}
    \dvx
    =
    \begin{tikzpicture}[baseline=-0.5*\rd]
        \osu{0}{2}
        \zd{0}
        \lengtarrowdownoneside{2}{0}{$\hspace{1.5em}\lambda_1\!+\!1$}
        \dottedlineboth{2}{3}
    \end{tikzpicture}
    .
\end{align}
In this case, we can see $\sigma^R_{0}=-$, and we have
\begin{align}
A_{0}^{-}\dvx=
    &
    (-1)^{n+m+g}
    \left\{
    -
    C^{j,m}_{n-1, d}\paren{\lambdabold_{L:+1}}
    \begin{tikzpicture}[baseline=-0.5*\rd]
        \encloseup{-1}{0}
        \osu{-1}{2}
        \Isd{-1}{0}
        \zd{0}
        \lengtarrowdownoneside{2}{0}{$\hspace{1.5em}\lambda_1\!+\!1$}
        \dottedlineboth{2}{3}
    \end{tikzpicture}
    -
    C^{j,m}_{n-1, d}\paren{\lambdabold_{L:+1}}
    \begin{tikzpicture}[baseline=-0.5*\rd]]
        \enclosedown{-1}{0}
        \belowarrowtikz{-1}{0}
        \osu{0}{2}
        \Isu{-1}{0}
        \oszd{-1}{0}
        \lengtarrowdownoneside{2}{0}{$\hspace{1.5em}\lambda_1\!+\!1$}
        \dottedlineboth{2}{3}
    \end{tikzpicture}
    \right.
    \nonumber\\
    &
    \hspace{7em}
    \left.
    +
    C^{j,m}_{n-1, d+1}\paren{\lambdabold_{1:-1}}
    \begin{tikzpicture}[baseline=-0.5*\rd]
        \encloseup{0}{1}
        \osu{1}{2}
        \gap{0}{1}
        \zd{0}
        \lengtarrowdownoneside{2}{1}{$\hspace{1.5em}\lambda_1$}
        \dottedlineboth{2}{3}
    \end{tikzpicture}
    +
    C^{j-1,m}_{n, d}\paren{\lambdabold_{\lambda_L\rightarrow 1+\lambda_1,\hat{1}}}
    \!\!\!\!\!\!
    \begin{tikzpicture}[baseline=-0.5*\rd]
        \osu{0}{2}
        \upverticalarrowtikz{0}
        \zu{0}
        \lengtarrowdownoneside{2}{0}{$\hspace{1.5em}\lambda_1\!+\!1$}
        \dottedlineboth{2}{3}
    \end{tikzpicture}
    \right\}
    \nonumber\\
    =
    &
    (-1)^{n+m+g}
    \left\{
    3C^{j,m}_{n-1, d}\paren{\lambdabold_{L:+1}}
    -
    C^{j,m}_{n-1, d+1}\paren{\lambdabold_{1:-1}}
    -
    C^{j-1,m}_{n, d}\paren{\lambdabold_{\lambda_L\rightarrow 1+\lambda_1,\hat{1}}}
    \right\}
    \dvx
    \nonumber\\
    =
    &
    (-1)^{n+m+g}
    \left\{
    C^{j,m}_{n-1, d}\paren{\lambdabold_{L:+1}}
    +
    C^{j,m}_{n-1, d+1}\paren{\lambdabold_{L:-1}}
    \right.
    \nonumber\\
    &
    \hspace{5em}
    \left.
    +
    \paren{
    C^{j,m}_{n-1, d+1}\paren{\lambdabold_{L:-1}}
    -
    C^{j,m}_{n-1, d+1}\paren{\lambdabold_{1:-1}}
    }
    -
    C^{j-1,m}_{n, d}\paren{\lambdabold_{\lambda_L\rightarrow 1+\lambda_1,\hat{1}}}
    \right\}
    \dvx
    \nonumber\\
    =
    &
    (-1)^{n+m+g}
    \left\{
    C^{j,m}_{n-1, d}\paren{\lambdabold_{L:+1}}
    +
    C^{j,m}_{n-1, d+1}\paren{\lambdabold_{L:-1}}
    \right.
    \nonumber\\
    &
    \hspace{5em}
    \left.
    +
    \paren{
    C^{j,m}_{n, d-1}\paren{\lambdabold_{i:+1}}
    -
    C^{j,m}_{n, d-1}\paren{\lambdabold_{L:+1}}
    }
    -
    C^{j-1,m}_{n, d}\paren{\lambdabold_{\lambda_L\rightarrow 1+\lambda_1,\hat{1}}}
    \right\}
    \dvx
    \nonumber\\
    =
    &
    (-1)^{n+m+g}
    \left\{
    C^{j,m}_{n-1, d}\paren{\lambdabold_{L:+1}}
    +
    C^{j,m}_{n-1, d+1}\paren{\lambdabold_{L:-1}}
    \right.
    -
    C^{j,m}_{n, d-1}\paren{\lambdabold_{L:+1}}
    \nonumber\\
    &
    \hspace{5em}
    \left.
    +
    \paren{
    C^{j,m}_{n, d-1}\paren{\lambdabold_{i:+1}}
    -
    C^{j-1,m}_{n, d}\paren{\lambdabold_{\lambda_L\rightarrow 1+\lambda_1,\hat{1}}}
    }
    \right\}
    \dvx
    \nonumber\\
    =
    &
    (-1)^{n+m+g}
    \left\{
    C^{j,m}_{n-1, d}\paren{\lambdabold_{L:+1}}
    +
    C^{j,m}_{n-1, d+1}\paren{\lambdabold_{L:-1}}
    -
    C^{j,m}_{n, d}\paren{\lambdabold_{L:-1}}
    +
    C^{j,m}_{n-1, d+1}\paren{\lambdabold_{L:-1}}
    \right\}
    \dvx
    \nonumber\\
    =
    &
    R
    \dvx
    ,
\end{align}
where $
    \begin{tikzpicture}[baseline=-0.5*\rd]
        \osu{1}{2}
        \gap{0}{1}
        \zd{0}
        \lengtarrowdownoneside{2}{1}{$\hspace{1.5em}\lambda_1$}
        \dottedlineboth{2}{3}
    \end{tikzpicture}
    \in 
    \connectedset^{k,j,m}_{n,d,g+1}
    ,
    \quad
    \begin{tikzpicture}[baseline=-0.5*\rd]
        \osu{0}{2}
        \zu{0}
        \lengtarrowdownoneside{2}{0}{$\hspace{1.5em}\lambda_1\!+\!1$}
        \dottedlineboth{2}{3}
    \end{tikzpicture}
    \in 
    \connectedset^{k,j-1,m}_{n,d,g}
$, and we used $
C^{j-1,m-1}_{n, d}\paren{{}_{0\rightarrow}(\lambdabold_{L:-1})}
=0
$
for $\lambda_L=0$, and used 
$
C^{j,m}_{n-1, d}\paren{\lambdabold_{L:+1}}
    =
    C^{j,m}_{n-1, d+1}\paren{\lambdabold_{L:-1}}
$ for $\lambda_L=0$ and, $
C^{j,m}_{n-1, d+1}\paren{\lambdabold_{L:-1}}
    -
    C^{j,m}_{n-1, d+1}\paren{\lambdabold_{1:-1}}
    =
    C^{j,m}_{n, d-1}\paren{\lambdabold_{i:+1}}
    -
    C^{j,m}_{n, d-1}\paren{\lambdabold_{L:+1}}
$ from~\eqref{eq:oxo}, and $
C^{j,m}_{n, d-1}\paren{\lambdabold_{i:+1}}
-
C^{j-1,m}_{n, d}\paren{\lambdabold_{\lambda_L\rightarrow 1+\lambda_1,\hat{1}}}
=
C^{j,m}_{n-1, d+1}\paren{\lambdabold_{L:-1}}
$ which can be derived from~\eqref{eq:betweenjLR}.

We consider the case in which $\lambda_L=0$, $d=0$ and $w=0\ (j=2m+1)$, such as 
\begin{align}
    \dvx
    =
    \begin{tikzpicture}[baseline=-0.5*\rd]
        \osu{0}{2}
        \zd{0}
        \lengtarrowdownoneside{2}{0}{$\hspace{1.5em}\lambda_R$}
        \dottedlineboth{2}{3}
    \end{tikzpicture}
    .
\end{align}
In this case, we can see $\sigma^R_{0}=-$, and we have
\begin{align}
A_{0}^{-}\dvx=
    &
    (-1)^{n+m+g}
    \left\{
    -
    C^{j,m}_{n-1, 0}\paren{\lambdabold_{L:+1}}
    \begin{tikzpicture}[baseline=-0.5*\rd]
        \encloseup{-1}{0}
        \osu{-1}{2}
        \Isd{-1}{0}
        \zd{0}
        \lengtarrowdownoneside{2}{0}{$\hspace{1.5em}\lambda_R$}
        \dottedlineboth{2}{3}
    \end{tikzpicture}
    -
    C^{j,m}_{n-1, 0}\paren{\lambdabold_{L:+1}}
    \begin{tikzpicture}[baseline=-0.5*\rd]]
        \enclosedown{-1}{0}
        \belowarrowtikz{-1}{0}
        \osu{0}{2}
        \Isu{-1}{0}
        \oszd{-1}{0}
        \lengtarrowdownoneside{2}{0}{$\hspace{1.5em}\lambda_R$}
        \dottedlineboth{2}{3}
    \end{tikzpicture}
    +
    C^{j,m}_{n-1, 1}\paren{\lambdabold_{R:-1}}
    \!\!\!
    \begin{tikzpicture}[baseline=-0.5*\rd]
        \encloseup{0}{1}
        \osu{1}{2}
        \gap{0}{1}
        \zd{0}
        \lengtarrowdownoneside{2}{1}{$\hspace{1.5em}\lambda_R-1$}
        \dottedlineboth{2}{3}
    \end{tikzpicture}
    \right\}
    \nonumber\\
    =
    &
    (-1)^{n+m+g}
    \left\{
    3C^{j,m}_{n-1,0}\paren{\lambdabold_{L:+1}}
    -
    C^{j,m}_{n-1, 1}\paren{\lambdabold_{R:-1}}
    \right\}
    \dvx
    \nonumber\\
    =
    &
    (-1)^{n+m+g}
    \left\{
    2
    C^{j,m}_{n-1, 1}\paren{\lambdabold_{L:-1}}
    +
    C^{j,m}_{n-1, 0}\paren{\lambdabold_{L:+1}}
    -
    C^{j,m}_{n, 0}\paren{\lambdabold_{L:-1}}
    \right\}
    \dvx
    \nonumber\\
    =
    &
    R
    \dvx
    ,
\end{align}
where we used 
$
C^{j-1,m-1}_{n, d}\paren{{}_{0\rightarrow}(\lambdabold_{L:-1})}
=0
$
for $\lambda_L=0$,
and we used 
$
C^{j,m}_{n-1, d}\paren{\lambdabold_{L:+1}}
    =
    C^{j,m}_{n-1, d+1}\paren{\lambdabold_{L:-1}}
$,
and
$
C^{j,m}_{n-1, 1}\paren{\lambdabold_{R:-1}}
    =
C^{j,m}_{n, 0}\paren{\lambdabold_{L:-1}}
$
for $\lambda_L=0$.

Thus, we have proved~\eqref{AL-}, $A_L^-=R$ for all the cases.
In the same way, we can prove~\eqref{AR-}.

This concludes all the proof in this section.

\newpage

\section{S6. Examples of higher-order local charges }
We present below three examples of the higher order local conserved quantity $Q_6, Q_7, Q_8$ that have not been found before.
In our diagram notation, $Q_k$ is represented more simply than the usual spin operator notation. Thus, we can read out the rule of the pattern of $Q_k$.
We write $Q_k$ as
\begin{align}
    Q_k
    =
    \sum_{j=0}^{j_f}
    U^j
    \sum_{n=0}^{\floor{\frac{k-1-j}{2}}}
    \sum_{d=0}^{\floor{\frac{k-1-j}{2}}-n}
    Q_k^j(k-j-2n-d,d)
    ,
\end{align}
where $Q_k^j(s,d)$ is the $(s,d)$ diagrams in $Q_k^j$ and $j_f=k-1(k-2)$ for even(odd) $k$.
$Q_k^j(s,d)$ is more explicitly written as 
\begin{align}
    Q_k^j(k-j-2n-d,d)
    =
    \sum_{m=0}^{\floor{\frac{j-1}{2}}}
    \sum_{g=0}^{d}
    \sum_{\dv\in \connectedset_{n,d,g}^{k,j,m}}
    (-1)^{n+m+g}
        C^{j,m}_{n,d}\paren{\boldsymbol{\lambda}_\dv}
        \dv
        .
\end{align}
All the expressions of $Q_k^j(s,d)$ below have $+\updownarrow$ in the RHS implicitly, and they are omitted.
\subsection{Explicit expressions for  $Q_6$}
\begin{figure}[H]
    \centering
    \includegraphics[width=0.8\linewidth]{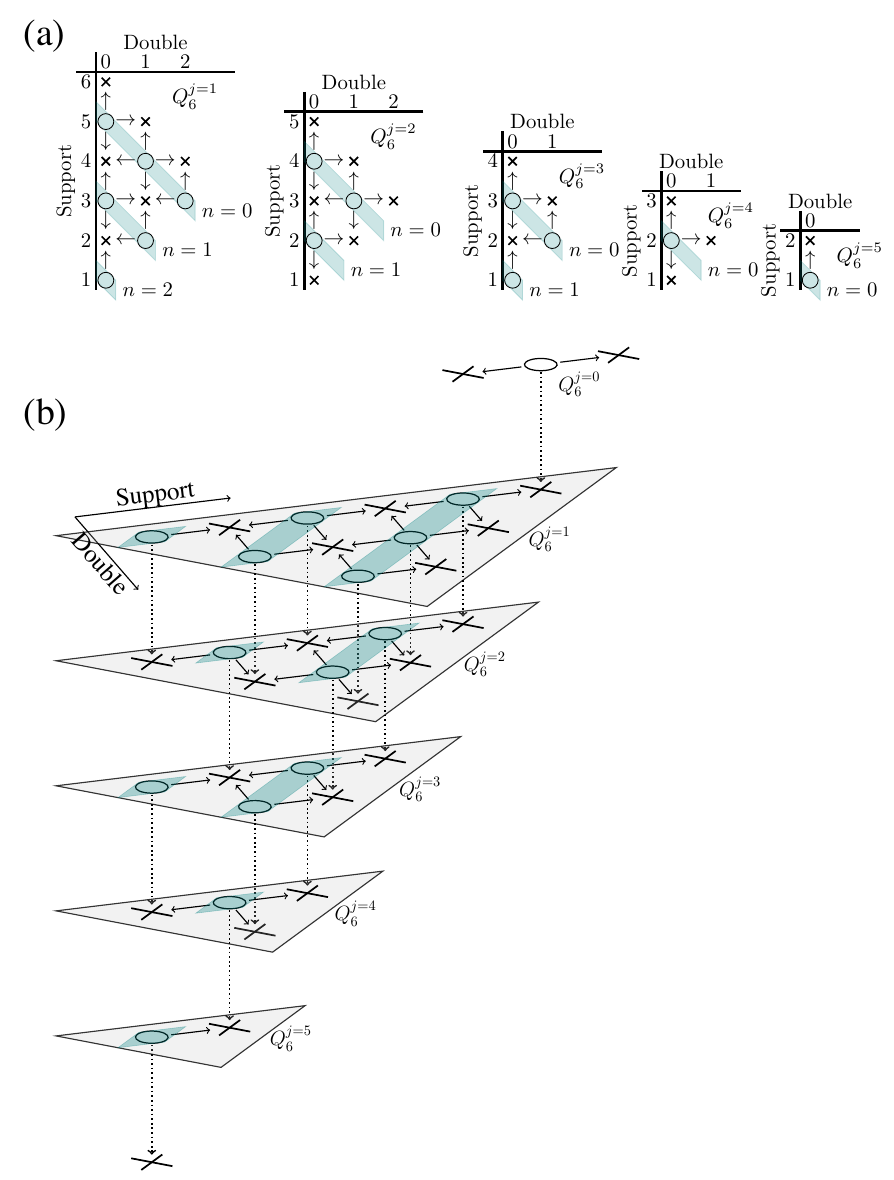}
    \caption{
    The structure of $Q_{6}^j$ for each $j$~(a) and all the structure of $Q_6$~(b).
    In (b), each plane represents the structure of $Q_{6}^j$ in (a),
    and the axis of support and double are omitted.
    The solid arrow in planes represents the commutator of diagrams with $H_\mathrm{0}$, and the vertical dotted arrow represents the commutator of diagrams with $H_\mathrm{int}$.
    }
    \label{Q6allfig}
\end{figure}

We show all the expressions for the components of $Q_6$.
We explain the structure of $Q_6^j$ for each $j$ in Fig.~\ref{Q6allfig}~(a) and all the structure of $Q_6$ in Fig.~\ref{Q6allfig}~(b). The circle at $(s,d)$ in Fig.~\ref{Q6allfig} represents $Q_k^j(s,d)$.
Each component of $Q_6$ is as follows:
\begin{align}
Q_{6}^{1}(5,0)=&
\begin{tikzpicture}[baseline=-0.5*\rd]
\oszu{0}{2}\oszd{2}{4}\Isd{0}{2}\Isu{2}{4}
\end{tikzpicture}+\begin{tikzpicture}[baseline=-0.5*\rd]
\oszu{1}{4}\oszd{0}{1}\Isu{0}{1}\Isd{1}{4}
\end{tikzpicture}+\begin{tikzpicture}[baseline=-0.5*\rd]
\oszu{0}{3}\oszd{3}{4}\Isd{0}{3}\Isu{3}{4}
\end{tikzpicture}+\begin{tikzpicture}[baseline=-0.5*\rd]
\oszu{0}{4}\zd{0}\Isd{0}{4}
\end{tikzpicture}+\begin{tikzpicture}[baseline=-0.5*\rd]
\oszu{0}{4}\zd{1}\Isd{0}{1}\Isd{1}{4}
\end{tikzpicture}+\begin{tikzpicture}[baseline=-0.5*\rd]
\oszu{0}{4}\zd{2}\Isd{0}{2}\Isd{2}{4}
\end{tikzpicture}+\begin{tikzpicture}[baseline=-0.5*\rd]
\oszu{0}{4}\zd{3}\Isd{0}{3}\Isd{3}{4}
\end{tikzpicture}+\begin{tikzpicture}[baseline=-0.5*\rd]
\oszu{0}{4}\zd{4}\Isd{0}{4}
\end{tikzpicture}\nonumber\\
Q_{6}^{1}(4,1)=&
-\begin{tikzpicture}[baseline=-0.5*\rd]
\oszu{0}{1}\oszd{2}{3}\Isd{0}{2}\Isu{1}{3}
\end{tikzpicture}-\begin{tikzpicture}[baseline=-0.5*\rd]
\oszu{1}{3}\zd{0}\Isu{0}{1}\Isd{0}{3}
\end{tikzpicture}-\begin{tikzpicture}[baseline=-0.5*\rd]
\oszu{0}{2}\zd{3}\Isd{0}{3}\Isu{2}{3}
\end{tikzpicture}+\begin{tikzpicture}[baseline=-0.5*\rd]
\oszu{0}{2}\oszd{1}{3}\Isd{0}{1}\Isu{2}{3}
\end{tikzpicture}+\begin{tikzpicture}[baseline=-0.5*\rd]
\oszu{0}{3}\oszd{0}{1}\Isd{1}{3}
\end{tikzpicture}+\begin{tikzpicture}[baseline=-0.5*\rd]
\oszu{0}{3}\oszd{1}{2}\Isd{0}{1}\Isd{2}{3}
\end{tikzpicture}+\begin{tikzpicture}[baseline=-0.5*\rd]
\oszu{0}{3}\oszd{2}{3}\Isd{0}{2}
\end{tikzpicture}\nonumber\\
Q_{6}^{1}(3,2)=&
\begin{tikzpicture}[baseline=-0.5*\rd]
\zu{0}\zd{2}\Isd{0}{2}\Isu{0}{2}
\end{tikzpicture}+\begin{tikzpicture}[baseline=-0.5*\rd]
\oszu{0}{2}\oszd{0}{2}
\end{tikzpicture}\nonumber\\
Q_{6}^{1}(3,0)=&
-\begin{tikzpicture}[baseline=-0.5*\rd]
\oszu{0}{1}\oszd{1}{2}\Isd{0}{1}\Isu{1}{2}
\end{tikzpicture}-\begin{tikzpicture}[baseline=-0.5*\rd]
\oszu{0}{2}\zd{0}\Isd{0}{2}
\end{tikzpicture}-\begin{tikzpicture}[baseline=-0.5*\rd]
\oszu{0}{2}\zd{1}\Isd{0}{1}\Isd{1}{2}
\end{tikzpicture}-\begin{tikzpicture}[baseline=-0.5*\rd]
\oszu{0}{2}\zd{2}\Isd{0}{2}
\end{tikzpicture}\nonumber\\
Q_{6}^{1}(2,1)=&
\begin{tikzpicture}[baseline=-0.5*\rd]
\zu{0}\zd{1}\Isd{0}{1}\Isu{0}{1}
\end{tikzpicture}-\begin{tikzpicture}[baseline=-0.5*\rd]
\oszu{0}{1}\oszd{0}{1}
\end{tikzpicture}\nonumber\\
Q_{6}^{1}(1,0)=&
\begin{tikzpicture}[baseline=-0.5*\rd]
\zu{0}\zd{0}
\end{tikzpicture}\nonumber\\
Q_{6}^{2}(4,0)=&
\begin{tikzpicture}[baseline=-0.5*\rd]
\osu{0}{2}\zd{0}\oszd{2}{3}\Isd{0}{2}\Isu{2}{3}
\end{tikzpicture}+\begin{tikzpicture}[baseline=-0.5*\rd]
\osu{1}{3}\oszd{0}{1}\zd{2}\Isd{1}{2}\Isu{0}{1}\Isd{2}{3}
\end{tikzpicture}+\begin{tikzpicture}[baseline=-0.5*\rd]
\osu{1}{3}\oszd{0}{1}\zd{3}\Isd{1}{3}\Isu{0}{1}
\end{tikzpicture}+\begin{tikzpicture}[baseline=-0.5*\rd]
\osu{0}{2}\zd{1}\oszd{2}{3}\Isd{1}{2}\Isd{0}{1}\Isu{2}{3}
\end{tikzpicture}+\begin{tikzpicture}[baseline=-0.5*\rd]
\zu{0}\oszu{1}{3}\Isu{0}{1}\osd{0}{1}\Isd{1}{3}
\end{tikzpicture}+\begin{tikzpicture}[baseline=-0.5*\rd]
\oszu{0}{1}\oszu{2}{3}\Isu{1}{2}\osd{1}{2}\Isd{0}{1}\Isd{2}{3}
\end{tikzpicture}+\begin{tikzpicture}[baseline=-0.5*\rd]
\oszu{0}{2}\zu{3}\Isu{2}{3}\osd{2}{3}\Isd{0}{2}
\end{tikzpicture}+\begin{tikzpicture}[baseline=-0.5*\rd]
\osu{0}{3}\zd{0}\zd{1}\Isd{0}{1}\Isd{1}{3}
\end{tikzpicture}+\begin{tikzpicture}[baseline=-0.5*\rd]
\osu{0}{3}\zd{0}\zd{2}\Isd{0}{2}\Isd{2}{3}
\end{tikzpicture}+\begin{tikzpicture}[baseline=-0.5*\rd]
\osu{0}{3}\zd{0}\zd{3}\Isd{0}{3}
\end{tikzpicture}+\begin{tikzpicture}[baseline=-0.5*\rd]
\osu{0}{3}\zd{1}\zd{2}\Isd{1}{2}\Isd{0}{1}\Isd{2}{3}
\end{tikzpicture}\nonumber\\&+\begin{tikzpicture}[baseline=-0.5*\rd]
\osu{0}{3}\zd{1}\zd{3}\Isd{1}{3}\Isd{0}{1}
\end{tikzpicture}+\begin{tikzpicture}[baseline=-0.5*\rd]
\osu{0}{3}\zd{2}\zd{3}\Isd{2}{3}\Isd{0}{2}
\end{tikzpicture}\nonumber\\
Q_{6}^{2}(3,1)=&
-\begin{tikzpicture}[baseline=-0.5*\rd]
\osu{1}{2}\zd{0}\zd{2}\Isd{0}{2}\Isu{0}{1}
\end{tikzpicture}-\begin{tikzpicture}[baseline=-0.5*\rd]
\osu{0}{1}\zd{0}\zd{2}\Isd{0}{2}\Isu{1}{2}
\end{tikzpicture}+\begin{tikzpicture}[baseline=-0.5*\rd]
\osu{0}{2}\zd{0}\oszd{1}{2}\Isd{0}{1}
\end{tikzpicture}+\begin{tikzpicture}[baseline=-0.5*\rd]
\osu{0}{2}\oszd{0}{1}\zd{2}\Isd{1}{2}
\end{tikzpicture}\nonumber\\
Q_{6}^{2}(2,0)=&
-\begin{tikzpicture}[baseline=-0.5*\rd]
\osu{0}{1}\zd{0}\zd{1}\Isd{0}{1}
\end{tikzpicture}\nonumber\\
Q_{6}^{3}(3,0)=&
\begin{tikzpicture}[baseline=-0.5*\rd]
\zu{0}\osu{1}{2}\Isu{0}{1}\osd{0}{1}\zd{2}\Isd{1}{2}
\end{tikzpicture}+\begin{tikzpicture}[baseline=-0.5*\rd]
\oszu{0}{2}\zd{0}\zd{1}\zd{2}\Isd{0}{1}\Isd{1}{2}
\end{tikzpicture}-\begin{tikzpicture}[baseline=-0.5*\rd]
\oszu{0}{1}\oszd{1}{2}\Isd{0}{1}\Isu{1}{2}
\end{tikzpicture}-\begin{tikzpicture}[baseline=-0.5*\rd]
\oszu{0}{2}\zd{0}\Isd{0}{2}
\end{tikzpicture}-\begin{tikzpicture}[baseline=-0.5*\rd]
\oszu{0}{2}\zd{1}\Isd{0}{1}\Isd{1}{2}
\end{tikzpicture}-\begin{tikzpicture}[baseline=-0.5*\rd]
\oszu{0}{2}\zd{2}\Isd{0}{2}
\end{tikzpicture}\nonumber\\
Q_{6}^{3}(2,1)=&
2\left(\begin{tikzpicture}[baseline=-0.5*\rd]
\zu{0}\zd{1}\Isd{0}{1}\Isu{0}{1}
\end{tikzpicture}-\begin{tikzpicture}[baseline=-0.5*\rd]
\oszu{0}{1}\oszd{0}{1}
\end{tikzpicture}\right)\nonumber\\
Q_{6}^{3}(1,0)=&
5\begin{tikzpicture}[baseline=-0.5*\rd]
\zu{0}\zd{0}
\end{tikzpicture}\nonumber\\
Q_{6}^{4}(2,0)=&
-2\begin{tikzpicture}[baseline=-0.5*\rd]
\osu{0}{1}\zd{0}\zd{1}\Isd{0}{1}
\end{tikzpicture}\nonumber\\
Q_{6}^{5}(1,0)=&
2\begin{tikzpicture}[baseline=-0.5*\rd]
\zu{0}\zd{0}
\end{tikzpicture}\nonumber
\end{align}

\newpage
\subsection{Explicit expressions for  $Q_7$}
\begin{figure}[H]
    \centering
    \includegraphics[width=0.8\linewidth]{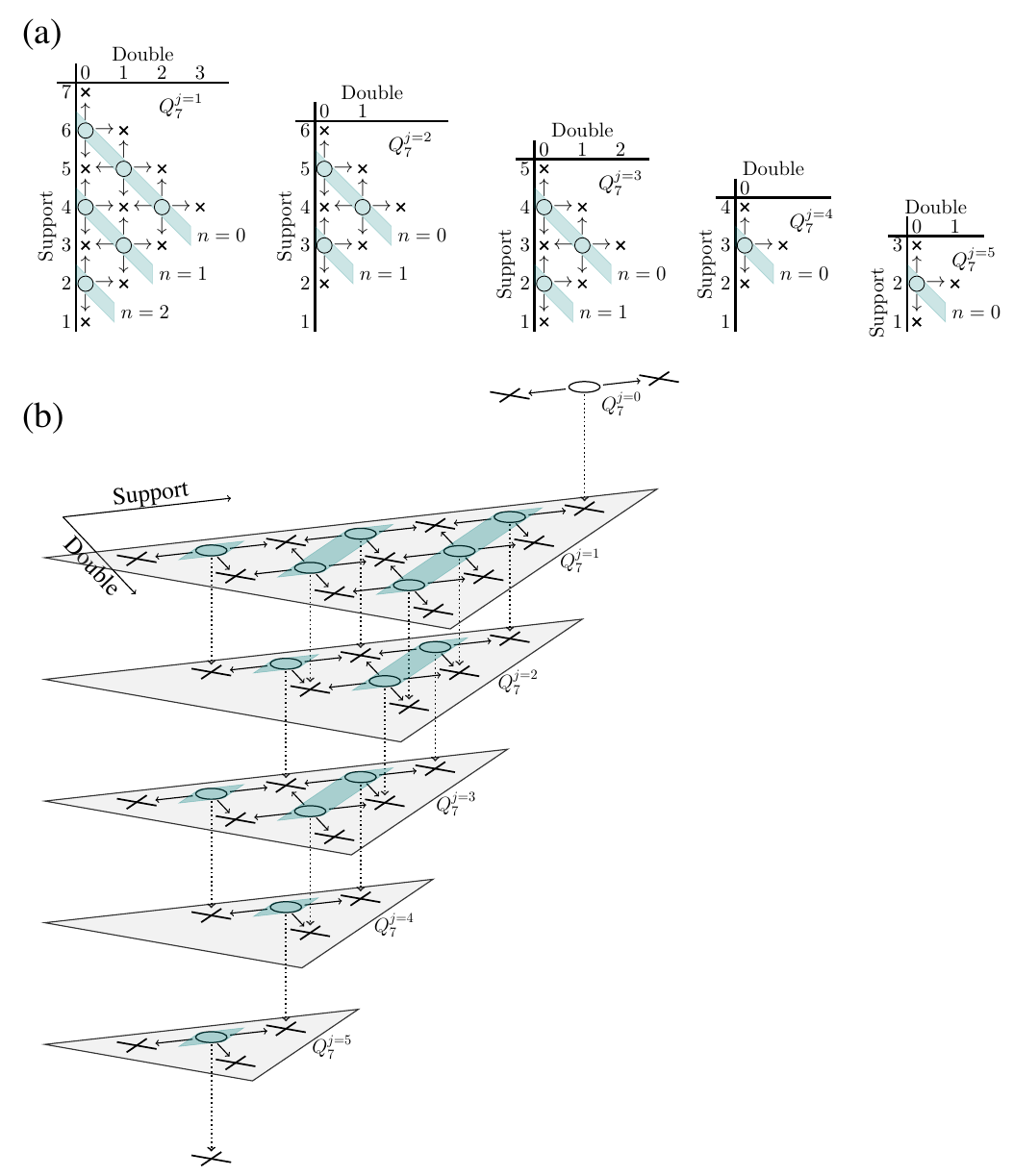}
    \caption{
    The structure of $Q_{7}^j$ for each $j$~(a) and all the structure of $Q_7$~(b).
    In (b), each plane represents the structure of $Q_{k}^j$ in (a),
    and the axis of support and double are omitted.
    The solid arrow in planes represents the commutator of diagrams with $H_\mathrm{0}$, and the vertical dotted arrow represents the commutator of diagrams with $H_\mathrm{int}$.
    }
    \label{Q7allfig}
\end{figure}

We show all the expressions for the components of $Q_7$.
We explain the structure of $Q_7^j$ for each $j$ in Fig.~\ref{Q7allfig}(a) and all the structure of $Q_7$ in Fig.~\ref{Q7allfig}(b). 
The circle at $(s,d)$ in Fig.~\ref{Q7allfig} represents $Q_7^j(s,d)$.

$Q_{2k+1}$ does not have $(s,s-1)$-connected diagrams because $(s,s-1)$-connected diagrams are even under space reflection and $Q_{2k+1}$ is odd under space reflection.
For $(s,s-1)$-connected diagrams, the unit number is always $2$, the types of the two units are both $-$ types, all columns in the diagram is overlap or gap, and the length of the two units are the same, such as
$
\begin{tikzpicture}[baseline=-0.5*\rd]
\oszu{0}{4}
\oszd{0}{4}
\end{tikzpicture}
$
and 
$
\begin{tikzpicture}[baseline=-0.5*\rd]
\gap{0}{3}
\zu{0}\zd{3}
\end{tikzpicture}
+
\begin{tikzpicture}[baseline=-0.5*\rd]
    \gap{0}{3}
    \zd{0}\zu{3}
    \end{tikzpicture}
$
, and these diagrams are even under space reflection.

Each component of $Q_7$ is as follows:
\begin{align}
Q_{7}^{1}(6,0)=&
\begin{tikzpicture}[baseline=-0.5*\rd]
\oszu{2}{5}\oszd{0}{2}\Isu{0}{2}\Isd{2}{5}
\end{tikzpicture}+\begin{tikzpicture}[baseline=-0.5*\rd]
\oszu{0}{3}\oszd{3}{5}\Isd{0}{3}\Isu{3}{5}
\end{tikzpicture}+\begin{tikzpicture}[baseline=-0.5*\rd]
\oszu{1}{5}\oszd{0}{1}\Isu{0}{1}\Isd{1}{5}
\end{tikzpicture}+\begin{tikzpicture}[baseline=-0.5*\rd]
\oszu{0}{4}\oszd{4}{5}\Isd{0}{4}\Isu{4}{5}
\end{tikzpicture}+\begin{tikzpicture}[baseline=-0.5*\rd]
\oszu{0}{5}\zd{0}\Isd{0}{5}
\end{tikzpicture}+\begin{tikzpicture}[baseline=-0.5*\rd]
\oszu{0}{5}\zd{1}\Isd{0}{1}\Isd{1}{5}
\end{tikzpicture}+\begin{tikzpicture}[baseline=-0.5*\rd]
\oszu{0}{5}\zd{2}\Isd{0}{2}\Isd{2}{5}
\end{tikzpicture}\nonumber\\&+\begin{tikzpicture}[baseline=-0.5*\rd]
\oszu{0}{5}\zd{3}\Isd{0}{3}\Isd{3}{5}
\end{tikzpicture}+\begin{tikzpicture}[baseline=-0.5*\rd]
\oszu{0}{5}\zd{4}\Isd{0}{4}\Isd{4}{5}
\end{tikzpicture}+\begin{tikzpicture}[baseline=-0.5*\rd]
\oszu{0}{5}\zd{5}\Isd{0}{5}
\end{tikzpicture}\nonumber\\
Q_{7}^{1}(5,1)=&
-\begin{tikzpicture}[baseline=-0.5*\rd]
\oszu{2}{4}\oszd{0}{1}\Isu{0}{2}\Isd{1}{4}
\end{tikzpicture}-\begin{tikzpicture}[baseline=-0.5*\rd]
\oszu{0}{2}\oszd{3}{4}\Isd{0}{3}\Isu{2}{4}
\end{tikzpicture}-\begin{tikzpicture}[baseline=-0.5*\rd]
\oszu{1}{4}\zd{0}\Isu{0}{1}\Isd{0}{4}
\end{tikzpicture}-\begin{tikzpicture}[baseline=-0.5*\rd]
\oszu{0}{3}\zd{4}\Isd{0}{4}\Isu{3}{4}
\end{tikzpicture}+\begin{tikzpicture}[baseline=-0.5*\rd]
\oszu{1}{4}\oszd{0}{2}\Isu{0}{1}\Isd{2}{4}
\end{tikzpicture}+\begin{tikzpicture}[baseline=-0.5*\rd]
\oszu{0}{3}\oszd{2}{4}\Isd{0}{2}\Isu{3}{4}
\end{tikzpicture}+\begin{tikzpicture}[baseline=-0.5*\rd]
\oszu{0}{4}\oszd{0}{1}\Isd{1}{4}
\end{tikzpicture}+\begin{tikzpicture}[baseline=-0.5*\rd]
\oszu{0}{4}\oszd{1}{2}\Isd{0}{1}\Isd{2}{4}
\end{tikzpicture}\nonumber\\&+\begin{tikzpicture}[baseline=-0.5*\rd]
\oszu{0}{4}\oszd{2}{3}\Isd{0}{2}\Isd{3}{4}
\end{tikzpicture}+\begin{tikzpicture}[baseline=-0.5*\rd]
\oszu{0}{4}\oszd{3}{4}\Isd{0}{3}
\end{tikzpicture}\nonumber\\
Q_{7}^{1}(4,2)=&
\begin{tikzpicture}[baseline=-0.5*\rd]
\oszu{2}{3}\zd{0}\Isu{0}{2}\Isd{0}{3}
\end{tikzpicture}+\begin{tikzpicture}[baseline=-0.5*\rd]
\oszu{0}{1}\zd{3}\Isd{0}{3}\Isu{1}{3}
\end{tikzpicture}+\begin{tikzpicture}[baseline=-0.5*\rd]
\oszu{0}{3}\oszd{0}{2}\Isd{2}{3}
\end{tikzpicture}+\begin{tikzpicture}[baseline=-0.5*\rd]
\oszu{0}{3}\oszd{1}{3}\Isd{0}{1}
\end{tikzpicture}\nonumber\\
Q_{7}^{1}(4,0)=&
-\begin{tikzpicture}[baseline=-0.5*\rd]
\oszu{1}{3}\oszd{0}{1}\Isu{0}{1}\Isd{1}{3}
\end{tikzpicture}-\begin{tikzpicture}[baseline=-0.5*\rd]
\oszu{0}{2}\oszd{2}{3}\Isd{0}{2}\Isu{2}{3}
\end{tikzpicture}-\begin{tikzpicture}[baseline=-0.5*\rd]
\oszu{0}{3}\zd{0}\Isd{0}{3}
\end{tikzpicture}-\begin{tikzpicture}[baseline=-0.5*\rd]
\oszu{0}{3}\zd{1}\Isd{0}{1}\Isd{1}{3}
\end{tikzpicture}-\begin{tikzpicture}[baseline=-0.5*\rd]
\oszu{0}{3}\zd{2}\Isd{0}{2}\Isd{2}{3}
\end{tikzpicture}-\begin{tikzpicture}[baseline=-0.5*\rd]
\oszu{0}{3}\zd{3}\Isd{0}{3}
\end{tikzpicture}\nonumber\\
Q_{7}^{1}(3,1)=&
\begin{tikzpicture}[baseline=-0.5*\rd]
\oszu{1}{2}\zd{0}\Isu{0}{1}\Isd{0}{2}
\end{tikzpicture}+\begin{tikzpicture}[baseline=-0.5*\rd]
\oszu{0}{1}\zd{2}\Isd{0}{2}\Isu{1}{2}
\end{tikzpicture}-\begin{tikzpicture}[baseline=-0.5*\rd]
\oszu{0}{2}\oszd{0}{1}\Isd{1}{2}
\end{tikzpicture}-\begin{tikzpicture}[baseline=-0.5*\rd]
\oszu{0}{2}\oszd{1}{2}\Isd{0}{1}
\end{tikzpicture}\nonumber\\
Q_{7}^{1}(2,0)=&
\begin{tikzpicture}[baseline=-0.5*\rd]
\oszu{0}{1}\zd{0}\Isd{0}{1}
\end{tikzpicture}+\begin{tikzpicture}[baseline=-0.5*\rd]
\oszu{0}{1}\zd{1}\Isd{0}{1}
\end{tikzpicture}\nonumber\\
Q_{7}^{2}(5,0)=&
\begin{tikzpicture}[baseline=-0.5*\rd]
\zu{1}\oszu{2}{4}\Isu{1}{2}\osd{0}{2}\Isu{0}{1}\Isd{2}{4}
\end{tikzpicture}+\begin{tikzpicture}[baseline=-0.5*\rd]
\zu{0}\oszu{2}{4}\Isu{0}{2}\osd{0}{2}\Isd{2}{4}
\end{tikzpicture}+\begin{tikzpicture}[baseline=-0.5*\rd]
\oszu{0}{1}\oszu{3}{4}\Isu{1}{3}\osd{1}{3}\Isd{0}{1}\Isd{3}{4}
\end{tikzpicture}+\begin{tikzpicture}[baseline=-0.5*\rd]
\oszu{0}{2}\zu{3}\Isu{2}{3}\osd{2}{4}\Isd{0}{2}\Isu{3}{4}
\end{tikzpicture}+\begin{tikzpicture}[baseline=-0.5*\rd]
\oszu{0}{2}\zu{4}\Isu{2}{4}\osd{2}{4}\Isd{0}{2}
\end{tikzpicture}+\begin{tikzpicture}[baseline=-0.5*\rd]
\osu{0}{3}\zd{0}\oszd{3}{4}\Isd{0}{3}\Isu{3}{4}
\end{tikzpicture}+\begin{tikzpicture}[baseline=-0.5*\rd]
\osu{1}{4}\oszd{0}{1}\zd{2}\Isd{1}{2}\Isu{0}{1}\Isd{2}{4}
\end{tikzpicture}+\begin{tikzpicture}[baseline=-0.5*\rd]
\osu{1}{4}\oszd{0}{1}\zd{3}\Isd{1}{3}\Isu{0}{1}\Isd{3}{4}
\end{tikzpicture}+\begin{tikzpicture}[baseline=-0.5*\rd]
\osu{1}{4}\oszd{0}{1}\zd{4}\Isd{1}{4}\Isu{0}{1}
\end{tikzpicture}\nonumber\\&+\begin{tikzpicture}[baseline=-0.5*\rd]
\osu{0}{3}\zd{1}\oszd{3}{4}\Isd{1}{3}\Isd{0}{1}\Isu{3}{4}
\end{tikzpicture}+\begin{tikzpicture}[baseline=-0.5*\rd]
\osu{0}{3}\zd{2}\oszd{3}{4}\Isd{2}{3}\Isd{0}{2}\Isu{3}{4}
\end{tikzpicture}+\begin{tikzpicture}[baseline=-0.5*\rd]
\zu{0}\oszu{1}{4}\Isu{0}{1}\osd{0}{1}\Isd{1}{4}
\end{tikzpicture}+\begin{tikzpicture}[baseline=-0.5*\rd]
\oszu{0}{1}\oszu{2}{4}\Isu{1}{2}\osd{1}{2}\Isd{0}{1}\Isd{2}{4}
\end{tikzpicture}+\begin{tikzpicture}[baseline=-0.5*\rd]
\oszu{0}{2}\oszu{3}{4}\Isu{2}{3}\osd{2}{3}\Isd{0}{2}\Isd{3}{4}
\end{tikzpicture}+\begin{tikzpicture}[baseline=-0.5*\rd]
\oszu{0}{3}\zu{4}\Isu{3}{4}\osd{3}{4}\Isd{0}{3}
\end{tikzpicture}+\begin{tikzpicture}[baseline=-0.5*\rd]
\osu{0}{4}\zd{0}\zd{1}\Isd{0}{1}\Isd{1}{4}
\end{tikzpicture}+\begin{tikzpicture}[baseline=-0.5*\rd]
\osu{0}{4}\zd{0}\zd{2}\Isd{0}{2}\Isd{2}{4}
\end{tikzpicture}+\begin{tikzpicture}[baseline=-0.5*\rd]
\osu{0}{4}\zd{0}\zd{3}\Isd{0}{3}\Isd{3}{4}
\end{tikzpicture}+\begin{tikzpicture}[baseline=-0.5*\rd]
\osu{0}{4}\zd{0}\zd{4}\Isd{0}{4}
\end{tikzpicture}\nonumber\\&+\begin{tikzpicture}[baseline=-0.5*\rd]
\osu{0}{4}\zd{1}\zd{2}\Isd{1}{2}\Isd{0}{1}\Isd{2}{4}
\end{tikzpicture}+\begin{tikzpicture}[baseline=-0.5*\rd]
\osu{0}{4}\zd{1}\zd{3}\Isd{1}{3}\Isd{0}{1}\Isd{3}{4}
\end{tikzpicture}+\begin{tikzpicture}[baseline=-0.5*\rd]
\osu{0}{4}\zd{1}\zd{4}\Isd{1}{4}\Isd{0}{1}
\end{tikzpicture}+\begin{tikzpicture}[baseline=-0.5*\rd]
\osu{0}{4}\zd{2}\zd{3}\Isd{2}{3}\Isd{0}{2}\Isd{3}{4}
\end{tikzpicture}+\begin{tikzpicture}[baseline=-0.5*\rd]
\osu{0}{4}\zd{2}\zd{4}\Isd{2}{4}\Isd{0}{2}
\end{tikzpicture}+\begin{tikzpicture}[baseline=-0.5*\rd]
\osu{0}{4}\zd{3}\zd{4}\Isd{3}{4}\Isd{0}{3}
\end{tikzpicture}\nonumber\\
Q_{7}^{2}(4,1)=&
-\begin{tikzpicture}[baseline=-0.5*\rd]
\zu{0}\oszu{2}{3}\Isu{0}{2}\osd{0}{1}\Isd{1}{3}
\end{tikzpicture}-\begin{tikzpicture}[baseline=-0.5*\rd]
\zu{0}\oszu{2}{3}\Isu{0}{2}\osd{1}{2}\Isd{0}{1}\Isd{2}{3}
\end{tikzpicture}-\begin{tikzpicture}[baseline=-0.5*\rd]
\oszu{0}{1}\zu{3}\Isu{1}{3}\osd{1}{2}\Isd{0}{1}\Isd{2}{3}
\end{tikzpicture}-\begin{tikzpicture}[baseline=-0.5*\rd]
\oszu{0}{1}\zu{3}\Isu{1}{3}\osd{2}{3}\Isd{0}{2}
\end{tikzpicture}-\begin{tikzpicture}[baseline=-0.5*\rd]
\osu{1}{3}\zd{0}\zd{2}\Isd{0}{2}\Isu{0}{1}\Isd{2}{3}
\end{tikzpicture}-\begin{tikzpicture}[baseline=-0.5*\rd]
\osu{1}{3}\zd{0}\zd{3}\Isd{0}{3}\Isu{0}{1}
\end{tikzpicture}-\begin{tikzpicture}[baseline=-0.5*\rd]
\osu{0}{2}\zd{0}\zd{3}\Isd{0}{3}\Isu{2}{3}
\end{tikzpicture}-\begin{tikzpicture}[baseline=-0.5*\rd]
\osu{0}{2}\zd{1}\zd{3}\Isd{1}{3}\Isd{0}{1}\Isu{2}{3}
\end{tikzpicture}+\begin{tikzpicture}[baseline=-0.5*\rd]
\zu{0}\oszu{1}{3}\Isu{0}{1}\osd{0}{2}\Isd{2}{3}
\end{tikzpicture}+\begin{tikzpicture}[baseline=-0.5*\rd]
\oszu{0}{1}\oszu{2}{3}\Isu{1}{2}\osd{0}{2}\Isd{2}{3}
\end{tikzpicture}+\begin{tikzpicture}[baseline=-0.5*\rd]
\oszu{0}{1}\oszu{2}{3}\Isu{1}{2}\osd{1}{3}\Isd{0}{1}
\end{tikzpicture}+\begin{tikzpicture}[baseline=-0.5*\rd]
\oszu{0}{2}\zu{3}\Isu{2}{3}\osd{1}{3}\Isd{0}{1}
\end{tikzpicture}\nonumber\\&+\begin{tikzpicture}[baseline=-0.5*\rd]
\osu{0}{3}\zd{0}\oszd{1}{2}\Isd{0}{1}\Isd{2}{3}
\end{tikzpicture}+\begin{tikzpicture}[baseline=-0.5*\rd]
\osu{0}{3}\zd{0}\oszd{2}{3}\Isd{0}{2}
\end{tikzpicture}+\begin{tikzpicture}[baseline=-0.5*\rd]
\osu{0}{3}\zd{1}\oszd{2}{3}\Isd{1}{2}\Isd{0}{1}
\end{tikzpicture}+\begin{tikzpicture}[baseline=-0.5*\rd]
\osu{0}{3}\oszd{0}{1}\zd{2}\Isd{1}{2}\Isd{2}{3}
\end{tikzpicture}+\begin{tikzpicture}[baseline=-0.5*\rd]
\osu{0}{3}\oszd{0}{1}\zd{3}\Isd{1}{3}
\end{tikzpicture}+\begin{tikzpicture}[baseline=-0.5*\rd]
\osu{0}{3}\oszd{1}{2}\zd{3}\Isd{2}{3}\Isd{0}{1}
\end{tikzpicture}\nonumber\\
Q_{7}^{2}(3,0)=&
-\begin{tikzpicture}[baseline=-0.5*\rd]
\zu{0}\oszu{1}{2}\Isu{0}{1}\osd{0}{1}\Isd{1}{2}
\end{tikzpicture}-\begin{tikzpicture}[baseline=-0.5*\rd]
\oszu{0}{1}\zu{2}\Isu{1}{2}\osd{1}{2}\Isd{0}{1}
\end{tikzpicture}-\begin{tikzpicture}[baseline=-0.5*\rd]
\osu{0}{2}\zd{0}\zd{1}\Isd{0}{1}\Isd{1}{2}
\end{tikzpicture}-\begin{tikzpicture}[baseline=-0.5*\rd]
\osu{0}{2}\zd{1}\zd{2}\Isd{1}{2}\Isd{0}{1}
\end{tikzpicture}-2\begin{tikzpicture}[baseline=-0.5*\rd]
\osu{0}{2}\zd{0}\zd{2}\Isd{0}{2}
\end{tikzpicture}\nonumber\\
Q_{7}^{3}(4,0)=&
\begin{tikzpicture}[baseline=-0.5*\rd]
\oszu{0}{2}\zd{0}\zd{1}\oszd{2}{3}\Isd{0}{1}\Isd{1}{2}\Isu{2}{3}
\end{tikzpicture}+\begin{tikzpicture}[baseline=-0.5*\rd]
\oszu{1}{3}\oszd{0}{1}\zd{2}\zd{3}\Isd{1}{2}\Isd{2}{3}\Isu{0}{1}
\end{tikzpicture}+\begin{tikzpicture}[baseline=-0.5*\rd]
\osu{0}{1}\oszu{2}{3}\Isu{1}{2}\zd{0}\osd{1}{2}\Isd{0}{1}\Isd{2}{3}
\end{tikzpicture}+\begin{tikzpicture}[baseline=-0.5*\rd]
\osu{0}{2}\zu{3}\Isu{2}{3}\zd{0}\osd{2}{3}\Isd{0}{2}
\end{tikzpicture}+\begin{tikzpicture}[baseline=-0.5*\rd]
\zu{0}\osu{1}{3}\Isu{0}{1}\osd{0}{1}\zd{2}\Isd{1}{2}\Isd{2}{3}
\end{tikzpicture}+\begin{tikzpicture}[baseline=-0.5*\rd]
\zu{0}\osu{1}{3}\Isu{0}{1}\osd{0}{1}\zd{3}\Isd{1}{3}
\end{tikzpicture}+\begin{tikzpicture}[baseline=-0.5*\rd]
\osu{0}{2}\zu{3}\Isu{2}{3}\zd{1}\osd{2}{3}\Isd{1}{2}\Isd{0}{1}
\end{tikzpicture}+\begin{tikzpicture}[baseline=-0.5*\rd]
\oszu{0}{1}\osu{2}{3}\Isu{1}{2}\osd{1}{2}\zd{3}\Isd{2}{3}\Isd{0}{1}
\end{tikzpicture}+\begin{tikzpicture}[baseline=-0.5*\rd]
\oszu{0}{3}\zd{0}\zd{1}\zd{2}\Isd{0}{1}\Isd{1}{2}\Isd{2}{3}
\end{tikzpicture}+\begin{tikzpicture}[baseline=-0.5*\rd]
\oszu{0}{3}\zd{0}\zd{1}\zd{3}\Isd{0}{1}\Isd{1}{3}
\end{tikzpicture}+\begin{tikzpicture}[baseline=-0.5*\rd]
\oszu{0}{3}\zd{0}\zd{2}\zd{3}\Isd{0}{2}\Isd{2}{3}
\end{tikzpicture}+\begin{tikzpicture}[baseline=-0.5*\rd]
\oszu{0}{3}\zd{1}\zd{2}\zd{3}\Isd{1}{2}\Isd{2}{3}\Isd{0}{1}
\end{tikzpicture}\nonumber\\&-\begin{tikzpicture}[baseline=-0.5*\rd]
\oszu{1}{3}\oszd{0}{1}\Isu{0}{1}\Isd{1}{3}
\end{tikzpicture}-\begin{tikzpicture}[baseline=-0.5*\rd]
\oszu{0}{2}\oszd{2}{3}\Isd{0}{2}\Isu{2}{3}
\end{tikzpicture}-\begin{tikzpicture}[baseline=-0.5*\rd]
\oszu{0}{3}\zd{0}\Isd{0}{3}
\end{tikzpicture}-\begin{tikzpicture}[baseline=-0.5*\rd]
\oszu{0}{3}\zd{1}\Isd{0}{1}\Isd{1}{3}
\end{tikzpicture}-\begin{tikzpicture}[baseline=-0.5*\rd]
\oszu{0}{3}\zd{2}\Isd{0}{2}\Isd{2}{3}
\end{tikzpicture}-\begin{tikzpicture}[baseline=-0.5*\rd]
\oszu{0}{3}\zd{3}\Isd{0}{3}
\end{tikzpicture}\nonumber\\
Q_{7}^{3}(3,1)=&
2\left(\begin{tikzpicture}[baseline=-0.5*\rd]
\oszu{1}{2}\zd{0}\Isu{0}{1}\Isd{0}{2}
\end{tikzpicture}+\begin{tikzpicture}[baseline=-0.5*\rd]
\oszu{0}{1}\zd{2}\Isd{0}{2}\Isu{1}{2}
\end{tikzpicture}-\begin{tikzpicture}[baseline=-0.5*\rd]
\oszu{0}{2}\oszd{0}{1}\Isd{1}{2}
\end{tikzpicture}-\begin{tikzpicture}[baseline=-0.5*\rd]
\oszu{0}{2}\oszd{1}{2}\Isd{0}{1}
\end{tikzpicture}\right)\nonumber\\
Q_{7}^{3}(2,0)=&
6\left(\begin{tikzpicture}[baseline=-0.5*\rd]
\oszu{0}{1}\zd{0}\Isd{0}{1}
\end{tikzpicture}+\begin{tikzpicture}[baseline=-0.5*\rd]
\oszu{0}{1}\zd{1}\Isd{0}{1}
\end{tikzpicture}\right)\nonumber\\
Q_{7}^{4}(3,0)=&
2\left(-\begin{tikzpicture}[baseline=-0.5*\rd]
\zu{0}\oszu{1}{2}\Isu{0}{1}\osd{0}{1}\Isd{1}{2}
\end{tikzpicture}-\begin{tikzpicture}[baseline=-0.5*\rd]
\oszu{0}{1}\zu{2}\Isu{1}{2}\osd{1}{2}\Isd{0}{1}
\end{tikzpicture}-\begin{tikzpicture}[baseline=-0.5*\rd]
\osu{0}{2}\zd{0}\zd{1}\Isd{0}{1}\Isd{1}{2}
\end{tikzpicture}-\begin{tikzpicture}[baseline=-0.5*\rd]
\osu{0}{2}\zd{0}\zd{2}\Isd{0}{2}
\end{tikzpicture}-\begin{tikzpicture}[baseline=-0.5*\rd]
\osu{0}{2}\zd{1}\zd{2}\Isd{1}{2}\Isd{0}{1}
\end{tikzpicture}\right)\nonumber\\
Q_{7}^{5}(2,0)=&
2\left(\begin{tikzpicture}[baseline=-0.5*\rd]
\oszu{0}{1}\zd{0}\Isd{0}{1}
\end{tikzpicture}+\begin{tikzpicture}[baseline=-0.5*\rd]
\oszu{0}{1}\zd{1}\Isd{0}{1}
\end{tikzpicture}\right)\nonumber
\end{align}

\newpage
\subsection{Explicit expressions for  $Q_8$}
\begin{figure}[H]
    \centering
    \includegraphics[width=0.84\linewidth]{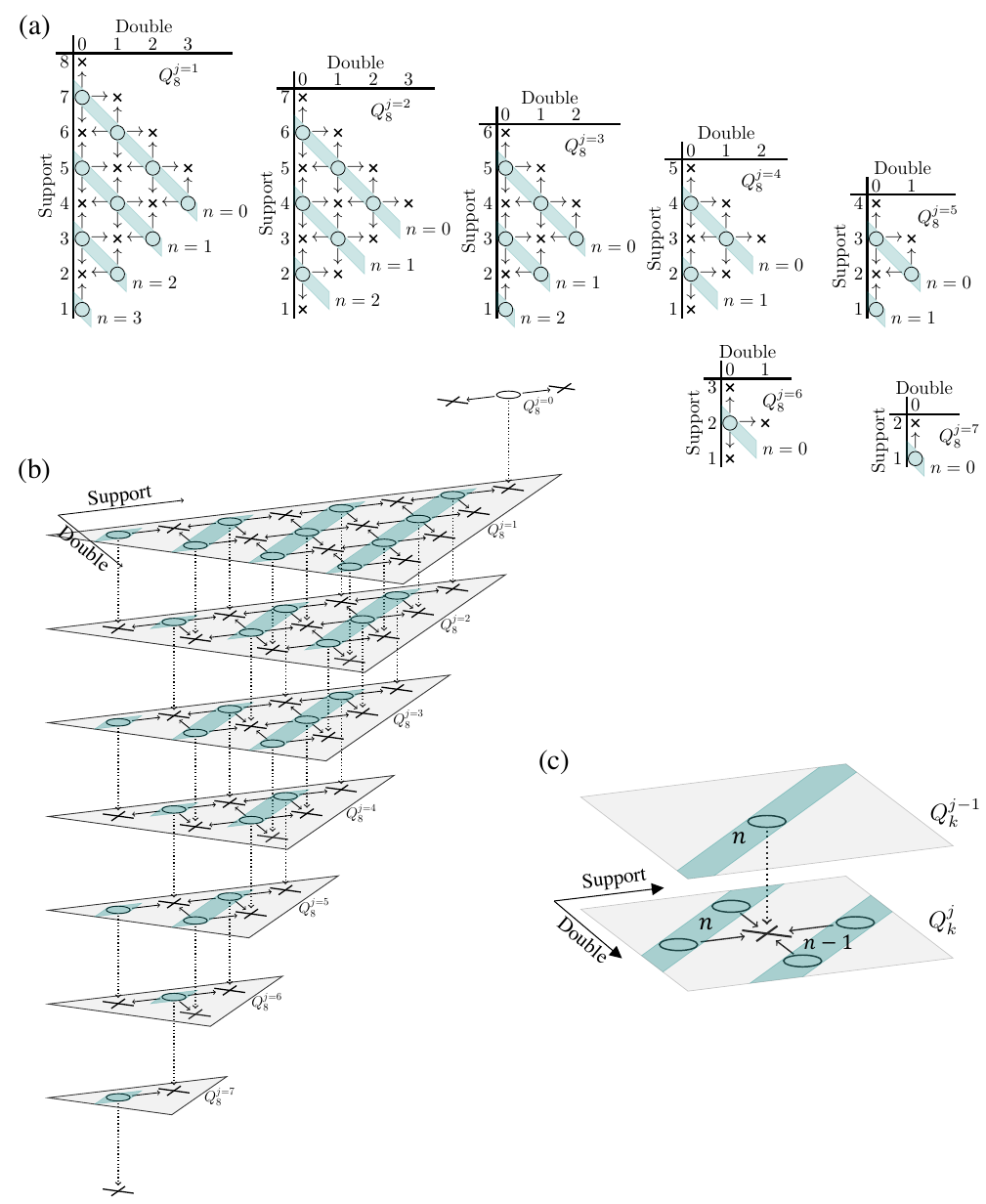}
    \caption{
    The structure of $Q_{8}^j$ for each $j$~(a), all the structure of $Q_8$~(b) and the basic structure of the cancellation of diagrams~(c). 
    In (b), each plane represents the structure of $Q_{8}^j$ in (a), and the axis of support and double are omitted.
    The solid arrow in planes represents the commutator of diagrams with $H_\mathrm{0}$, and the vertical dotted arrow represents the commutator of diagrams with $H_\mathrm{int}$.
    In (c), we only show the circle and arrows related to the cancellation at the crosses. 
    }
    \label{Q8allfig}
\end{figure}

We show all the expressions for the components of $Q_8$.
We explain the structure of $Q_8^j$ for each $j$ in Fig.~\ref{Q8allfig}~(a) and all the structure of $Q_8$ in Fig.~\ref{Q8allfig}~(b). The circle at $(s,d)$ in Fig.~\ref{Q8allfig} represents $Q_8^j(s,d)$.
From $Q_8$, the general structure of the cancellation of diagrams (Fig.~\ref{Q8allfig}~(c)) appears.
Each component of $Q_8$ is as follows:
\begin{align}
Q_{8}^{1}(7,0)=&
\nonumber
\end{align}

\newpage
\section{S7. Examples of coefficients $C_{n,d}^{j,m}\paren{\lambdabold}$}
We present several examples of the coefficients $C_{n,d}^{j,m}\paren{\lambdabold}$ in this section.
The coefficients presented here include enough information to construct up to $Q_{16}$.

We present the table for the value of $C_{n,d}^{j,m}\paren{\lambdabold}$ 
below for $m,n>0$.
The explicit expressions of $C_{n,d}^{j,m}\paren{\lambdabold}$ for $m=0$ or $n=0$ are presented in the main manuscript.
Using the invariance  of $C_{n,d}^{j,m}\paren{\lambdabold}$~\eqref{eq:LRswap} and~\eqref{eq:swap} and~\eqref{eq:minprop}, we restrict $\lambdabold=\{\lambda_L;\lambda_1,\ldots,\lambda_w;\lambda_R\}$~$(w=j-1-2m)$ to satisfy $\lambda_L\leq \lambda_R$ and $\lambda_1\leq \lambda_2\leq \cdots \leq \lambda_w$ and $0\leq \lambda_a\leq n$ without losing generality.
The columns indicate $(\lambda_L,\lambda_R)$.
The rows indicate the value of $d$ and $(\lambda_1,\lambda_2,\ldots,\lambda_w)$ which is represented by $\phi$ in the case of $w=0$, i.e., the case of $\lambdabold=\{\lambda_L;\lambda_R\}$.

\vspace{3em}

\begin{tabular}{cr"ccc}
&$C^{j=3,m=1}_{n=1,d}$&$(0,0)$&$(0,1)$&$(1,1)$\\\thickhline
&\multirow{1}{*}{d=0}\ \ $\phi$&$5$&$6$&$7$\\\hline
&\multirow{1}{*}{d=1}\ \ $\phi$&$10$&$11$&$12$\\\hline
&\multirow{1}{*}{d=2}\ \ $\phi$&$15$&$16$&$17$\\\hline
&\multirow{1}{*}{d=3}\ \ $\phi$&$20$&$21$&$22$\\\hline
&\multirow{1}{*}{d=4}\ \ $\phi$&$25$&$26$&$27$\\\hline
&\multirow{1}{*}{d=5}\ \ $\phi$&$30$&$31$&$32$\\\hline
\end{tabular}
\vspace{3em}

\begin{tabular}{cr"cccccc}
&$C^{j=3,m=1}_{n=2,d}$&$(0,0)$&$(0,1)$&$(0,2)$&$(1,1)$&$(1,2)$&$(2,2)$\\\thickhline
&\multirow{1}{*}{d=0}\ \ $\phi$&$16$&$20$&$21$&$24$&$25$&$26$\\\hline
&\multirow{1}{*}{d=1}\ \ $\phi$&$30$&$34$&$35$&$38$&$39$&$40$\\\hline
&\multirow{1}{*}{d=2}\ \ $\phi$&$44$&$48$&$49$&$52$&$53$&$54$\\\hline
&\multirow{1}{*}{d=3}\ \ $\phi$&$58$&$62$&$63$&$66$&$67$&$68$\\\hline
&\multirow{1}{*}{d=4}\ \ $\phi$&$72$&$76$&$77$&$80$&$81$&$82$\\\hline
\end{tabular}
\vspace{3em}

\begin{tabular}{cr"cccccccccc}
&$C^{j=3,m=1}_{n=3,d}$&$(0,0)$&$(0,1)$&$(0,2)$&$(0,3)$&$(1,1)$&$(1,2)$&$(1,3)$&$(2,2)$&$(2,3)$&$(3,3)$\\\thickhline
&\multirow{1}{*}{d=0}\ \ $\phi$&$40$&$50$&$54$&$55$&$60$&$64$&$65$&$68$&$69$&$70$\\\hline
&\multirow{1}{*}{d=1}\ \ $\phi$&$70$&$80$&$84$&$85$&$90$&$94$&$95$&$98$&$99$&$100$\\\hline
&\multirow{1}{*}{d=2}\ \ $\phi$&$100$&$110$&$114$&$115$&$120$&$124$&$125$&$128$&$129$&$130$\\\hline
&\multirow{1}{*}{d=3}\ \ $\phi$&$130$&$140$&$144$&$145$&$150$&$154$&$155$&$158$&$159$&$160$\\\hline
\end{tabular}
\vspace{3em}

\begin{tabular}{cr"ccccccccccccccc}
&$C^{j=3,m=1}_{n=4,d}$&$(0,0)$&$(0,1)$&$(0,2)$&$(0,3)$&$(0,4)$&$(1,1)$&$(1,2)$&$(1,3)$&$(1,4)$&$(2,2)$&$(2,3)$&$(2,4)$&$(3,3)$&$(3,4)$&$(4,4)$\\\thickhline
&\multirow{1}{*}{d=0}\ \ $\phi$&$85$&$105$&$115$&$119$&$120$&$125$&$135$&$139$&$140$&$145$&$149$&$150$&$153$&$154$&$155$\\\hline
&\multirow{1}{*}{d=1}\ \ $\phi$&$140$&$160$&$170$&$174$&$175$&$180$&$190$&$194$&$195$&$200$&$204$&$205$&$208$&$209$&$210$\\\hline
&\multirow{1}{*}{d=2}\ \ $\phi$&$195$&$215$&$225$&$229$&$230$&$235$&$245$&$249$&$250$&$255$&$259$&$260$&$263$&$264$&$265$\\\hline
\end{tabular}
\vspace{3em}

\resizebox{50em}{!}{
\begin{tabular}{cr"ccccccccccccccccccccc}
&$C^{j=3,m=1}_{n=5,d}$&$(0,0)$&$(0,1)$&$(0,2)$&$(0,3)$&$(0,4)$&$(0,5)$&$(1,1)$&$(1,2)$&$(1,3)$&$(1,4)$&$(1,5)$&$(2,2)$&$(2,3)$&$(2,4)$&$(2,5)$&$(3,3)$&$(3,4)$&$(3,5)$&$(4,4)$&$(4,5)$&$(5,5)$\\\thickhline
&\multirow{1}{*}{d=0}\ \ $\phi$&$161$&$196$&$216$&$226$&$230$&$231$&$231$&$251$&$261$&$265$&$266$&$271$&$281$&$285$&$286$&$291$&$295$&$296$&$299$&$300$&$301$\\\hline
&\multirow{1}{*}{d=1}\ \ $\phi$&$252$&$287$&$307$&$317$&$321$&$322$&$322$&$342$&$352$&$356$&$357$&$362$&$372$&$376$&$377$&$382$&$386$&$387$&$390$&$391$&$392$\\\hline
\end{tabular}
}
\vspace{3em}

\resizebox{50em}{!}{
\begin{tabular}{cr"cccccccccccccccccccccccccccc}
&$C^{j=3,m=1}_{n=6,d}$&$(0,0)$&$(0,1)$&$(0,2)$&$(0,3)$&$(0,4)$&$(0,5)$&$(0,6)$&$(1,1)$&$(1,2)$&$(1,3)$&$(1,4)$&$(1,5)$&$(1,6)$&$(2,2)$&$(2,3)$&$(2,4)$&$(2,5)$&$(2,6)$&$(3,3)$&$(3,4)$&$(3,5)$&$(3,6)$&$(4,4)$&$(4,5)$&$(4,6)$&$(5,5)$&$(5,6)$&$(6,6)$\\\thickhline
&\multirow{1}{*}{d=0}\ \ $\phi$&$280$&$336$&$371$&$391$&$401$&$405$&$406$&$392$&$427$&$447$&$457$&$461$&$462$&$462$&$482$&$492$&$496$&$497$&$502$&$512$&$516$&$517$&$522$&$526$&$527$&$530$&$531$&$532$\\\hline
\end{tabular}
}
\vspace{3em}

\begin{tabular}{cr"ccc}
&$C^{j=4,m=1}_{n=1,d}$&$(0,0)$&$(0,1)$&$(1,1)$\\\thickhline
&\multirow{2}{*}{d=0}\ \ $(0)$&$10$&$11$&$12$\\
&$(1)$&$14$&$15$&$16$\\\hline
&\multirow{2}{*}{d=1}\ \ $(0)$&$15$&$16$&$17$\\
&$(1)$&$20$&$21$&$22$\\\hline
&\multirow{2}{*}{d=2}\ \ $(0)$&$20$&$21$&$22$\\
&$(1)$&$26$&$27$&$28$\\\hline
&\multirow{2}{*}{d=3}\ \ $(0)$&$25$&$26$&$27$\\
&$(1)$&$32$&$33$&$34$\\\hline
&\multirow{2}{*}{d=4}\ \ $(0)$&$30$&$31$&$32$\\
&$(1)$&$38$&$39$&$40$\\\hline
\end{tabular}
\vspace{3em}

\begin{tabular}{cr"cccccc}
&$C^{j=4,m=1}_{n=2,d}$&$(0,0)$&$(0,1)$&$(0,2)$&$(1,1)$&$(1,2)$&$(2,2)$\\\thickhline
&\multirow{3}{*}{d=0}\ \ $(0)$&$30$&$34$&$35$&$38$&$39$&$40$\\
&$(1)$&$50$&$55$&$56$&$60$&$61$&$62$\\
&$(2)$&$56$&$61$&$62$&$66$&$67$&$68$\\\hline
&\multirow{3}{*}{d=1}\ \ $(0)$&$44$&$48$&$49$&$52$&$53$&$54$\\
&$(1)$&$69$&$74$&$75$&$79$&$80$&$81$\\
&$(2)$&$76$&$81$&$82$&$86$&$87$&$88$\\\hline
&\multirow{3}{*}{d=2}\ \ $(0)$&$58$&$62$&$63$&$66$&$67$&$68$\\
&$(1)$&$88$&$93$&$94$&$98$&$99$&$100$\\
&$(2)$&$96$&$101$&$102$&$106$&$107$&$108$\\\hline
&\multirow{3}{*}{d=3}\ \ $(0)$&$72$&$76$&$77$&$80$&$81$&$82$\\
&$(1)$&$107$&$112$&$113$&$117$&$118$&$119$\\
&$(2)$&$116$&$121$&$122$&$126$&$127$&$128$\\\hline
\end{tabular}
\vspace{3em}

\begin{tabular}{cr"cccccccccc}
&$C^{j=4,m=1}_{n=3,d}$&$(0,0)$&$(0,1)$&$(0,2)$&$(0,3)$&$(1,1)$&$(1,2)$&$(1,3)$&$(2,2)$&$(2,3)$&$(3,3)$\\\thickhline
&\multirow{4}{*}{d=0}\ \ $(0)$&$70$&$80$&$84$&$85$&$90$&$94$&$95$&$98$&$99$&$100$\\
&$(1)$&$128$&$142$&$147$&$148$&$156$&$161$&$162$&$166$&$167$&$168$\\
&$(2)$&$158$&$173$&$178$&$179$&$188$&$193$&$194$&$198$&$199$&$200$\\
&$(3)$&$166$&$181$&$186$&$187$&$196$&$201$&$202$&$206$&$207$&$208$\\\hline
&\multirow{4}{*}{d=1}\ \ $(0)$&$100$&$110$&$114$&$115$&$120$&$124$&$125$&$128$&$129$&$130$\\
&$(1)$&$172$&$186$&$191$&$192$&$200$&$205$&$206$&$210$&$211$&$212$\\
&$(2)$&$207$&$222$&$227$&$228$&$237$&$242$&$243$&$247$&$248$&$249$\\
&$(3)$&$216$&$231$&$236$&$237$&$246$&$251$&$252$&$256$&$257$&$258$\\\hline
&\multirow{4}{*}{d=2}\ \ $(0)$&$130$&$140$&$144$&$145$&$150$&$154$&$155$&$158$&$159$&$160$\\
&$(1)$&$216$&$230$&$235$&$236$&$244$&$249$&$250$&$254$&$255$&$256$\\
&$(2)$&$256$&$271$&$276$&$277$&$286$&$291$&$292$&$296$&$297$&$298$\\
&$(3)$&$266$&$281$&$286$&$287$&$296$&$301$&$302$&$306$&$307$&$308$\\\hline
\end{tabular}
\vspace{3em}

\begin{tabular}{cr"ccccccccccccccc}
&$C^{j=4,m=1}_{n=4,d}$&$(0,0)$&$(0,1)$&$(0,2)$&$(0,3)$&$(0,4)$&$(1,1)$&$(1,2)$&$(1,3)$&$(1,4)$&$(2,2)$&$(2,3)$&$(2,4)$&$(3,3)$&$(3,4)$&$(4,4)$\\\thickhline
&\multirow{5}{*}{d=0}\ \ $(0)$&$140$&$160$&$170$&$174$&$175$&$180$&$190$&$194$&$195$&$200$&$204$&$205$&$208$&$209$&$210$\\
&$(1)$&$270$&$300$&$314$&$319$&$320$&$330$&$344$&$349$&$350$&$358$&$363$&$364$&$368$&$369$&$370$\\
&$(2)$&$356$&$390$&$405$&$410$&$411$&$424$&$439$&$444$&$445$&$454$&$459$&$460$&$464$&$465$&$466$\\
&$(3)$&$396$&$431$&$446$&$451$&$452$&$466$&$481$&$486$&$487$&$496$&$501$&$502$&$506$&$507$&$508$\\
&$(4)$&$406$&$441$&$456$&$461$&$462$&$476$&$491$&$496$&$497$&$506$&$511$&$512$&$516$&$517$&$518$\\\hline
&\multirow{5}{*}{d=1}\ \ $(0)$&$195$&$215$&$225$&$229$&$230$&$235$&$245$&$249$&$250$&$255$&$259$&$260$&$263$&$264$&$265$\\
&$(1)$&$355$&$385$&$399$&$404$&$405$&$415$&$429$&$434$&$435$&$443$&$448$&$449$&$453$&$454$&$455$\\
&$(2)$&$455$&$489$&$504$&$509$&$510$&$523$&$538$&$543$&$544$&$553$&$558$&$559$&$563$&$564$&$565$\\
&$(3)$&$500$&$535$&$550$&$555$&$556$&$570$&$585$&$590$&$591$&$600$&$605$&$606$&$610$&$611$&$612$\\
&$(4)$&$511$&$546$&$561$&$566$&$567$&$581$&$596$&$601$&$602$&$611$&$616$&$617$&$621$&$622$&$623$\\\hline
\end{tabular}

\newpage

\resizebox{50em}{!}{
\begin{tabular}{cr"ccccccccccccccccccccc}
&$C^{j=4,m=1}_{n=5,d}$&$(0,0)$&$(0,1)$&$(0,2)$&$(0,3)$&$(0,4)$&$(0,5)$&$(1,1)$&$(1,2)$&$(1,3)$&$(1,4)$&$(1,5)$&$(2,2)$&$(2,3)$&$(2,4)$&$(2,5)$&$(3,3)$&$(3,4)$&$(3,5)$&$(4,4)$&$(4,5)$&$(5,5)$\\\thickhline
&\multirow{6}{*}{d=0}\ \ $(0)$&$252$&$287$&$307$&$317$&$321$&$322$&$322$&$342$&$352$&$356$&$357$&$362$&$372$&$376$&$377$&$382$&$386$&$387$&$390$&$391$&$392$\\
&$(1)$&$502$&$557$&$587$&$601$&$606$&$607$&$612$&$642$&$656$&$661$&$662$&$672$&$686$&$691$&$692$&$700$&$705$&$706$&$710$&$711$&$712$\\
&$(2)$&$692$&$757$&$791$&$806$&$811$&$812$&$822$&$856$&$871$&$876$&$877$&$890$&$905$&$910$&$911$&$920$&$925$&$926$&$930$&$931$&$932$\\
&$(3)$&$806$&$875$&$910$&$925$&$930$&$931$&$944$&$979$&$994$&$999$&$1000$&$1014$&$1029$&$1034$&$1035$&$1044$&$1049$&$1050$&$1054$&$1055$&$1056$\\
&$(4)$&$856$&$926$&$961$&$976$&$981$&$982$&$996$&$1031$&$1046$&$1051$&$1052$&$1066$&$1081$&$1086$&$1087$&$1096$&$1101$&$1102$&$1106$&$1107$&$1108$\\
&$(5)$&$868$&$938$&$973$&$988$&$993$&$994$&$1008$&$1043$&$1058$&$1063$&$1064$&$1078$&$1093$&$1098$&$1099$&$1108$&$1113$&$1114$&$1118$&$1119$&$1120$\\\hline
\end{tabular}
}

\vspace{3em}

\begin{tabular}{cr"ccc}
&$C^{j=5,m=1}_{n=1,d}$&$(0,0)$&$(0,1)$&$(1,1)$\\\thickhline
&\multirow{3}{*}{d=0}\ \ $(0,0)$&$15$&$16$&$17$\\
&$(0,1)$&$20$&$21$&$22$\\
&$(1,1)$&$25$&$26$&$27$\\\hline
&\multirow{3}{*}{d=1}\ \ $(0,0)$&$20$&$21$&$22$\\
&$(0,1)$&$26$&$27$&$28$\\
&$(1,1)$&$32$&$33$&$34$\\\hline
&\multirow{3}{*}{d=2}\ \ $(0,0)$&$25$&$26$&$27$\\
&$(0,1)$&$32$&$33$&$34$\\
&$(1,1)$&$39$&$40$&$41$\\\hline
&\multirow{3}{*}{d=3}\ \ $(0,0)$&$30$&$31$&$32$\\
&$(0,1)$&$38$&$39$&$40$\\
&$(1,1)$&$46$&$47$&$48$\\\hline
&\multirow{3}{*}{d=4}\ \ $(0,0)$&$35$&$36$&$37$\\
&$(0,1)$&$44$&$45$&$46$\\
&$(1,1)$&$53$&$54$&$55$\\\hline
\end{tabular}
\vspace{3em}

\begin{tabular}{cr"cccccc}
&$C^{j=5,m=1}_{n=2,d}$&$(0,0)$&$(0,1)$&$(0,2)$&$(1,1)$&$(1,2)$&$(2,2)$\\\thickhline
&\multirow{6}{*}{d=0}\ \ $(0,0)$&$44$&$48$&$49$&$52$&$53$&$54$\\
&$(0,1)$&$69$&$74$&$75$&$79$&$80$&$81$\\
&$(0,2)$&$76$&$81$&$82$&$86$&$87$&$88$\\
&$(1,1)$&$101$&$107$&$108$&$113$&$114$&$115$\\
&$(1,2)$&$108$&$114$&$115$&$120$&$121$&$122$\\
&$(2,2)$&$115$&$121$&$122$&$127$&$128$&$129$\\\hline
&\multirow{6}{*}{d=1}\ \ $(0,0)$&$58$&$62$&$63$&$66$&$67$&$68$\\
&$(0,1)$&$88$&$93$&$94$&$98$&$99$&$100$\\
&$(0,2)$&$96$&$101$&$102$&$106$&$107$&$108$\\
&$(1,1)$&$126$&$132$&$133$&$138$&$139$&$140$\\
&$(1,2)$&$134$&$140$&$141$&$146$&$147$&$148$\\
&$(2,2)$&$142$&$148$&$149$&$154$&$155$&$156$\\\hline
&\multirow{6}{*}{d=2}\ \ $(0,0)$&$72$&$76$&$77$&$80$&$81$&$82$\\
&$(0,1)$&$107$&$112$&$113$&$117$&$118$&$119$\\
&$(0,2)$&$116$&$121$&$122$&$126$&$127$&$128$\\
&$(1,1)$&$151$&$157$&$158$&$163$&$164$&$165$\\
&$(1,2)$&$160$&$166$&$167$&$172$&$173$&$174$\\
&$(2,2)$&$169$&$175$&$176$&$181$&$182$&$183$\\\hline
&\multirow{6}{*}{d=3}\ \ $(0,0)$&$86$&$90$&$91$&$94$&$95$&$96$\\
&$(0,1)$&$126$&$131$&$132$&$136$&$137$&$138$\\
&$(0,2)$&$136$&$141$&$142$&$146$&$147$&$148$\\
&$(1,1)$&$176$&$182$&$183$&$188$&$189$&$190$\\
&$(1,2)$&$186$&$192$&$193$&$198$&$199$&$200$\\
&$(2,2)$&$196$&$202$&$203$&$208$&$209$&$210$\\\hline
\end{tabular}
\vspace{3em}

\newpage

\begin{tabular}{cr"cccccccccc}
&$C^{j=5,m=1}_{n=3,d}$&$(0,0)$&$(0,1)$&$(0,2)$&$(0,3)$&$(1,1)$&$(1,2)$&$(1,3)$&$(2,2)$&$(2,3)$&$(3,3)$\\\thickhline
&\multirow{10}{*}{d=0}\ \ $(0,0)$&$100$&$110$&$114$&$115$&$120$&$124$&$125$&$128$&$129$&$130$\\
&$(0,1)$&$172$&$186$&$191$&$192$&$200$&$205$&$206$&$210$&$211$&$212$\\
&$(0,2)$&$207$&$222$&$227$&$228$&$237$&$242$&$243$&$247$&$248$&$249$\\
&$(0,3)$&$216$&$231$&$236$&$237$&$246$&$251$&$252$&$256$&$257$&$258$\\
&$(1,1)$&$279$&$298$&$304$&$305$&$317$&$323$&$324$&$329$&$330$&$331$\\
&$(1,2)$&$323$&$343$&$349$&$350$&$363$&$369$&$370$&$375$&$376$&$377$\\
&$(1,3)$&$332$&$352$&$358$&$359$&$372$&$378$&$379$&$384$&$385$&$386$\\
\hline
\end{tabular}
\vspace{3em}

\begin{tabular}{cr"ccc}
&$C^{j=5,m=2}_{n=1,d}$&$(0,0)$&$(0,1)$&$(1,1)$\\\thickhline
&\multirow{1}{*}{d=0}\ \ $\phi$&$21$&$25$&$29$\\\hline
&\multirow{1}{*}{d=1}\ \ $\phi$&$52$&$57$&$62$\\\hline
&\multirow{1}{*}{d=2}\ \ $\phi$&$92$&$98$&$104$\\\hline
&\multirow{1}{*}{d=3}\ \ $\phi$&$141$&$148$&$155$\\\hline
&\multirow{1}{*}{d=4}\ \ $\phi$&$199$&$207$&$215$\\\hline
\end{tabular}
\vspace{3em}

\begin{tabular}{cr"cccccc}
&$C^{j=5,m=2}_{n=2,d}$&$(0,0)$&$(0,1)$&$(0,2)$&$(1,1)$&$(1,2)$&$(2,2)$\\\thickhline
&\multirow{1}{*}{d=0}\ \ $\phi$&$124$&$159$&$165$&$195$&$201$&$207$\\\hline
&\multirow{1}{*}{d=1}\ \ $\phi$&$294$&$337$&$344$&$381$&$388$&$395$\\\hline
&\multirow{1}{*}{d=2}\ \ $\phi$&$507$&$558$&$566$&$610$&$618$&$626$\\\hline
&\multirow{1}{*}{d=3}\ \ $\phi$&$763$&$822$&$831$&$882$&$891$&$900$\\\hline
\end{tabular}
\vspace{3em}

\begin{tabular}{cr"cccccccccc}
&$C^{j=5,m=2}_{n=3,d}$&$(0,0)$&$(0,1)$&$(0,2)$&$(0,3)$&$(1,1)$&$(1,2)$&$(1,3)$&$(2,2)$&$(2,3)$&$(3,3)$\\\thickhline
&\multirow{1}{*}{d=0}\ \ $\phi$&$532$&$701$&$752$&$760$&$878$&$930$&$938$&$982$&$990$&$998$\\\hline
&\multirow{1}{*}{d=1}\ \ $\phi$&$1193$&$1398$&$1457$&$1466$&$1610$&$1670$&$1679$&$1730$&$1739$&$1748$\\\hline
&\multirow{1}{*}{d=2}\ \ $\phi$&$2002$&$2242$&$2309$&$2319$&$2489$&$2557$&$2567$&$2625$&$2635$&$2645$\\\hline
\end{tabular}
\vspace{3em}

\begin{tabular}{cr"ccccccccccccccc}
&$C^{j=5,m=2}_{n=4,d}$&$(0,0)$&$(0,1)$&$(0,2)$&$(0,3)$&$(0,4)$&$(1,1)$&$(1,2)$&$(1,3)$&$(1,4)$&$(2,2)$&$(2,3)$&$(2,4)$&$(3,3)$&$(3,4)$&$(4,4)$\\\thickhline
&\multirow{1}{*}{d=0}\ \ $\phi$&$1837$&$2434$&$2673$&$2740$&$2750$&$3065$&$3312$&$3380$&$3390$&$3560$&$3628$&$3638$&$3696$&$3706$&$3716$\\\hline
&\multirow{1}{*}{d=1}\ \ $\phi$&$3891$&$4607$&$4882$&$4957$&$4968$&$5351$&$5633$&$5709$&$5720$&$5916$&$5992$&$6003$&$6068$&$6079$&$6090$\\\hline
\end{tabular}
\vspace{3em}

\newpage
\resizebox{50em}{!}{
\begin{tabular}{cr"ccccccccccccccccccccc}
&$C^{j=5,m=2}_{n=5,d}$&$(0,0)$&$(0,1)$&$(0,2)$&$(0,3)$&$(0,4)$&$(0,5)$&$(1,1)$&$(1,2)$&$(1,3)$&$(1,4)$&$(1,5)$&$(2,2)$&$(2,3)$&$(2,4)$&$(2,5)$&$(3,3)$&$(3,4)$&$(3,5)$&$(4,4)$&$(4,5)$&$(5,5)$\\\thickhline
&\multirow{1}{*}{d=0}\ \ $\phi$&$5403$&$7127$&$7948$&$8257$&$8340$&$8352$&$8956$&$9811$&$10128$&$10212$&$10224$&$10674$&$10992$&$11076$&$11088$&$11310$&$11394$&$11406$&$11478$&$11490$&$11502$\\\hline
\end{tabular}
}

\vspace{3em}
\begin{tabular}{cr"ccc}
&$C^{j=6,m=1}_{n=1,d}$&$(0,0)$&$(0,1)$&$(1,1)$\\\thickhline
&\multirow{4}{*}{d=0}\ \ $(0,0,0)$&$20$&$21$&$22$\\
&$(0,0,1)$&$26$&$27$&$28$\\
&$(0,1,1)$&$32$&$33$&$34$\\
&$(1,1,1)$&$38$&$39$&$40$\\\hline
&\multirow{4}{*}{d=1}\ \ $(0,0,0)$&$25$&$26$&$27$\\
&$(0,0,1)$&$32$&$33$&$34$\\
&$(0,1,1)$&$39$&$40$&$41$\\
&$(1,1,1)$&$46$&$47$&$48$\\\hline
&\multirow{4}{*}{d=2}\ \ $(0,0,0)$&$30$&$31$&$32$\\
&$(0,0,1)$&$38$&$39$&$40$\\
&$(0,1,1)$&$46$&$47$&$48$\\
&$(1,1,1)$&$54$&$55$&$56$\\\hline
&\multirow{4}{*}{d=3}\ \ $(0,0,0)$&$35$&$36$&$37$\\
&$(0,0,1)$&$44$&$45$&$46$\\
&$(0,1,1)$&$53$&$54$&$55$\\
&$(1,1,1)$&$62$&$63$&$64$\\\hline
\end{tabular}
\vspace{3em}

\begin{tabular}{cr"cccccc}
&$C^{j=6,m=1}_{n=2,d}$&$(0,0)$&$(0,1)$&$(0,2)$&$(1,1)$&$(1,2)$&$(2,2)$\\\thickhline
&\multirow{10}{*}{d=0}\ \ $(0,0,0)$&$58$&$62$&$63$&$66$&$67$&$68$\\
&$(0,0,1)$&$88$&$93$&$94$&$98$&$99$&$100$\\
&$(0,0,2)$&$96$&$101$&$102$&$106$&$107$&$108$\\
&$(0,1,1)$&$126$&$132$&$133$&$138$&$139$&$140$\\
&$(0,1,2)$&$134$&$140$&$141$&$146$&$147$&$148$\\
&$(0,2,2)$&$142$&$148$&$149$&$154$&$155$&$156$\\
&$(1,1,1)$&$172$&$179$&$180$&$186$&$187$&$188$\\
&$(1,1,2)$&$180$&$187$&$188$&$194$&$195$&$196$\\
&$(1,2,2)$&$188$&$195$&$196$&$202$&$203$&$204$\\
&$(2,2,2)$&$196$&$203$&$204$&$210$&$211$&$212$\\\hline
&\multirow{10}{*}{d=1}\ \ $(0,0,0)$&$72$&$76$&$77$&$80$&$81$&$82$\\
&$(0,0,1)$&$107$&$112$&$113$&$117$&$118$&$119$\\
&$(0,0,2)$&$116$&$121$&$122$&$126$&$127$&$128$\\
&$(0,1,1)$&$151$&$157$&$158$&$163$&$164$&$165$\\
&$(0,1,2)$&$160$&$166$&$167$&$172$&$173$&$174$\\
&$(0,2,2)$&$169$&$175$&$176$&$181$&$182$&$183$\\
&$(1,1,1)$&$204$&$211$&$212$&$218$&$219$&$220$\\
&$(1,1,2)$&$213$&$220$&$221$&$227$&$228$&$229$\\
&$(1,2,2)$&$222$&$229$&$230$&$236$&$237$&$238$\\
&$(2,2,2)$&$231$&$238$&$239$&$245$&$246$&$247$\\\hline
&\multirow{10}{*}{d=2}\ \ $(0,0,0)$&$86$&$90$&$91$&$94$&$95$&$96$\\
&$(0,0,1)$&$126$&$131$&$132$&$136$&$137$&$138$\\
&$(0,0,2)$&$136$&$141$&$142$&$146$&$147$&$148$\\
&$(0,1,1)$&$176$&$182$&$183$&$188$&$189$&$190$\\
&$(0,1,2)$&$186$&$192$&$193$&$198$&$199$&$200$\\
&$(0,2,2)$&$196$&$202$&$203$&$208$&$209$&$210$\\
&$(1,1,1)$&$236$&$243$&$244$&$250$&$251$&$252$\\
&$(1,1,2)$&$246$&$253$&$254$&$260$&$261$&$262$\\
&$(1,2,2)$&$256$&$263$&$264$&$270$&$271$&$272$\\
&$(2,2,2)$&$266$&$273$&$274$&$280$&$281$&$282$\\\hline
\end{tabular}

\newpage

\begin{tabular}{cr"ccc}
&$C^{j=6,m=2}_{n=1,d}$&$(0,0)$&$(0,1)$&$(1,1)$\\\thickhline
&\multirow{2}{*}{d=0}\ \ $(0)$&$52$&$57$&$62$\\
&$(1)$&$66$&$71$&$76$\\\hline
&\multirow{2}{*}{d=1}\ \ $(0)$&$92$&$98$&$104$\\
&$(1)$&$112$&$118$&$124$\\\hline
&\multirow{2}{*}{d=2}\ \ $(0)$&$141$&$148$&$155$\\
&$(1)$&$168$&$175$&$182$\\\hline
&\multirow{2}{*}{d=3}\ \ $(0)$&$199$&$207$&$215$\\
&$(1)$&$234$&$242$&$250$\\\hline
\end{tabular}
\vspace{3em}

\begin{tabular}{cr"cccccc}
&$C^{j=6,m=2}_{n=2,d}$&$(0,0)$&$(0,1)$&$(0,2)$&$(1,1)$&$(1,2)$&$(2,2)$\\\thickhline
&\multirow{3}{*}{d=0}\ \ $(0)$&$294$&$337$&$344$&$381$&$388$&$395$\\
&$(1)$&$435$&$485$&$492$&$536$&$543$&$550$\\
&$(2)$&$462$&$512$&$519$&$563$&$570$&$577$\\\hline
&\multirow{3}{*}{d=1}\ \ $(0)$&$507$&$558$&$566$&$610$&$618$&$626$\\
&$(1)$&$706$&$765$&$773$&$825$&$833$&$841$\\
&$(2)$&$741$&$800$&$808$&$860$&$868$&$876$\\\hline
&\multirow{3}{*}{d=2}\ \ $(0)$&$763$&$822$&$831$&$882$&$891$&$900$\\
&$(1)$&$1029$&$1097$&$1106$&$1166$&$1175$&$1184$\\
&$(2)$&$1073$&$1141$&$1150$&$1210$&$1219$&$1228$\\\hline
\end{tabular}
\vspace{3em}

\begin{tabular}{cr"cccccccccc}
&$C^{j=6,m=2}_{n=3,d}$&$(0,0)$&$(0,1)$&$(0,2)$&$(0,3)$&$(1,1)$&$(1,2)$&$(1,3)$&$(2,2)$&$(2,3)$&$(3,3)$\\\thickhline
&\multirow{4}{*}{d=0}\ \ $(0)$&$1193$&$1398$&$1457$&$1466$&$1610$&$1670$&$1679$&$1730$&$1739$&$1748$\\
&$(1)$&$1956$&$2220$&$2288$&$2297$&$2492$&$2561$&$2570$&$2630$&$2639$&$2648$\\
&$(2)$&$2222$&$2495$&$2563$&$2572$&$2776$&$2845$&$2854$&$2914$&$2923$&$2932$\\
&$(3)$&$2266$&$2539$&$2607$&$2616$&$2820$&$2889$&$2898$&$2958$&$2967$&$2976$\\\hline
&\multirow{4}{*}{d=1}\ \ $(0)$&$2002$&$2242$&$2309$&$2319$&$2489$&$2557$&$2567$&$2625$&$2635$&$2645$\\
&$(1)$&$3064$&$3371$&$3448$&$3458$&$3686$&$3764$&$3774$&$3842$&$3852$&$3862$\\
&$(2)$&$3406$&$3723$&$3800$&$3810$&$4048$&$4126$&$4136$&$4204$&$4214$&$4224$\\
&$(3)$&$3460$&$3777$&$3854$&$3864$&$4102$&$4180$&$4190$&$4258$&$4268$&$4278$\\\hline
\end{tabular}
\vspace{3em}

\begin{tabular}{cr"ccccccccccccccc}
&$C^{j=6,m=2}_{n=4,d}$&$(0,0)$&$(0,1)$&$(0,2)$&$(0,3)$&$(0,4)$&$(1,1)$&$(1,2)$&$(1,3)$&$(1,4)$&$(2,2)$&$(2,3)$&$(2,4)$&$(3,3)$&$(3,4)$&$(4,4)$\\\thickhline
&\multirow{5}{*}{d=0}\ \ $(0)$&$3891$&$4607$&$4882$&$4957$&$4968$&$5351$&$5633$&$5709$&$5720$&$5916$&$5992$&$6003$&$6068$&$6079$&$6090$\\
&$(1)$&$6849$&$7840$&$8190$&$8276$&$8287$&$8866$&$9224$&$9311$&$9322$&$9583$&$9670$&$9681$&$9757$&$9768$&$9779$\\
&$(2)$&$8253$&$9319$&$9680$&$9766$&$9777$&$10421$&$10790$&$10877$&$10888$&$11160$&$11247$&$11258$&$11334$&$11345$&$11356$\\
&$(3)$&$8680$&$9757$&$10118$&$10204$&$10215$&$10870$&$11239$&$11326$&$11337$&$11609$&$11696$&$11707$&$11783$&$11794$&$11805$\\
&$(4)$&$8745$&$9822$&$10183$&$10269$&$10280$&$10935$&$11304$&$11391$&$11402$&$11674$&$11761$&$11772$&$11848$&$11859$&$11870$\\\hline
\end{tabular}
\vspace{3em}

\newpage
\begin{tabular}{cr"ccc}
&$C^{j=7,m=2}_{n=1,d}$&$(0,0)$&$(0,1)$&$(1,1)$\\\thickhline
&\multirow{3}{*}{d=0}\ \ $(0,0)$&$92$&$98$&$104$\\
&$(0,1)$&$112$&$118$&$124$\\
&$(1,1)$&$132$&$138$&$144$\\\hline
&\multirow{3}{*}{d=1}\ \ $(0,0)$&$141$&$148$&$155$\\
&$(0,1)$&$168$&$175$&$182$\\
&$(1,1)$&$195$&$202$&$209$\\\hline
&\multirow{3}{*}{d=2}\ \ $(0,0)$&$199$&$207$&$215$\\
&$(0,1)$&$234$&$242$&$250$\\
&$(1,1)$&$269$&$277$&$285$\\\hline
&\multirow{3}{*}{d=3}\ \ $(0,0)$&$266$&$275$&$284$\\
&$(0,1)$&$310$&$319$&$328$\\
&$(1,1)$&$354$&$363$&$372$\\\hline
\end{tabular}
\vspace{3em}

\begin{tabular}{cr"cccccc}
&$C^{j=7,m=2}_{n=2,d}$&$(0,0)$&$(0,1)$&$(0,2)$&$(1,1)$&$(1,2)$&$(2,2)$\\\thickhline
&\multirow{6}{*}{d=0}\ \ $(0,0)$&$507$&$558$&$566$&$610$&$618$&$626$\\
&$(0,1)$&$706$&$765$&$773$&$825$&$833$&$841$\\
&$(0,2)$&$741$&$800$&$808$&$860$&$868$&$876$\\
&$(1,1)$&$940$&$1007$&$1015$&$1075$&$1083$&$1091$\\
&$(1,2)$&$975$&$1042$&$1050$&$1110$&$1118$&$1126$\\
&$(2,2)$&$1010$&$1077$&$1085$&$1145$&$1153$&$1161$\\\hline
&\multirow{6}{*}{d=1}\ \ $(0,0)$&$763$&$822$&$831$&$882$&$891$&$900$\\
&$(0,1)$&$1029$&$1097$&$1106$&$1166$&$1175$&$1184$\\
&$(0,2)$&$1073$&$1141$&$1150$&$1210$&$1219$&$1228$\\
&$(1,1)$&$1339$&$1416$&$1425$&$1494$&$1503$&$1512$\\
&$(1,2)$&$1383$&$1460$&$1469$&$1538$&$1547$&$1556$\\
&$(2,2)$&$1427$&$1504$&$1513$&$1582$&$1591$&$1600$\\\hline
&\multirow{6}{*}{d=2}\ \ $(0,0)$&$1062$&$1129$&$1139$&$1197$&$1207$&$1217$\\
&$(0,1)$&$1404$&$1481$&$1491$&$1559$&$1569$&$1579$\\
&$(0,2)$&$1458$&$1535$&$1545$&$1613$&$1623$&$1633$\\
&$(1,1)$&$1800$&$1887$&$1897$&$1975$&$1985$&$1995$\\
&$(1,2)$&$1854$&$1941$&$1951$&$2029$&$2039$&$2049$\\
&$(2,2)$&$1908$&$1995$&$2005$&$2083$&$2093$&$2103$\\\hline
\end{tabular}
\vspace{3em}

\begin{tabular}{cr"ccc}
&$C^{j=7,m=3}_{n=1,d}$&$(0,0)$&$(0,1)$&$(1,1)$\\\thickhline
&\multirow{1}{*}{d=0}\ \ $\phi$&$84$&$98$&$112$\\\hline
&\multirow{1}{*}{d=1}\ \ $\phi$&$232$&$252$&$272$\\\hline
&\multirow{1}{*}{d=2}\ \ $\phi$&$453$&$480$&$507$\\\hline
&\multirow{1}{*}{d=3}\ \ $\phi$&$760$&$795$&$830$\\\hline
\end{tabular}
\vspace{3em}

\begin{tabular}{cr"cccccc}
&$C^{j=7,m=3}_{n=2,d}$&$(0,0)$&$(0,1)$&$(0,2)$&$(1,1)$&$(1,2)$&$(2,2)$\\\thickhline
&\multirow{1}{*}{d=0}\ \ $\phi$&$752$&$953$&$980$&$1161$&$1188$&$1215$\\\hline
&\multirow{1}{*}{d=1}\ \ $\phi$&$2010$&$2290$&$2325$&$2578$&$2613$&$2648$\\\hline
&\multirow{1}{*}{d=2}\ \ $\phi$&$3820$&$4191$&$4235$&$4571$&$4615$&$4659$\\\hline
\end{tabular}
\vspace{3em}

\begin{tabular}{cr"cccccccccc}
&$C^{j=7,m=3}_{n=3,d}$&$(0,0)$&$(0,1)$&$(0,2)$&$(0,3)$&$(1,1)$&$(1,2)$&$(1,3)$&$(2,2)$&$(2,3)$&$(3,3)$\\\thickhline
&\multirow{1}{*}{d=0}\ \ $\phi$&$4720$&$6242$&$6613$&$6657$&$7855$&$8235$&$8279$&$8615$&$8659$&$8703$\\\hline
&\multirow{1}{*}{d=1}\ \ $\phi$&$12088$&$14167$&$14641$&$14695$&$16340$&$16824$&$16878$&$17308$&$17362$&$17416$\\\hline
\end{tabular}
\vspace{3em}

\begin{tabular}{cr"ccccccccccccccc}
&$C^{j=7,m=3}_{n=4,d}$&$(0,0)$&$(0,1)$&$(0,2)$&$(0,3)$&$(0,4)$&$(1,1)$&$(1,2)$&$(1,3)$&$(1,4)$&$(2,2)$&$(2,3)$&$(2,4)$&$(3,3)$&$(3,4)$&$(4,4)$\\\thickhline
&\multirow{1}{*}{d=0}\ \ $\phi$&$23203$&$31302$&$33996$&$34585$&$34650$&$40016$&$42825$&$43425$&$43490$&$45645$&$46245$&$46310$&$46845$&$46910$&$46975$\\\hline
\end{tabular}
\vspace{3em}

\newpage
\begin{tabular}{cr"ccc}
&$C^{j=8,m=2}_{n=1,d}$&$(0,0)$&$(0,1)$&$(1,1)$\\\thickhline
&\multirow{4}{*}{d=0}\ \ $(0,0,0)$&$141$&$148$&$155$\\
&$(0,0,1)$&$168$&$175$&$182$\\
&$(0,1,1)$&$195$&$202$&$209$\\
&$(1,1,1)$&$222$&$229$&$236$\\\hline
&\multirow{4}{*}{d=1}\ \ $(0,0,0)$&$199$&$207$&$215$\\
&$(0,0,1)$&$234$&$242$&$250$\\
&$(0,1,1)$&$269$&$277$&$285$\\
&$(1,1,1)$&$304$&$312$&$320$\\\hline
&\multirow{4}{*}{d=2}\ \ $(0,0,0)$&$266$&$275$&$284$\\
&$(0,0,1)$&$310$&$319$&$328$\\
&$(0,1,1)$&$354$&$363$&$372$\\
&$(1,1,1)$&$398$&$407$&$416$\\\hline
\end{tabular}
\vspace{3em}

\begin{tabular}{cr"ccc}
&$C^{j=8,m=3}_{n=1,d}$&$(0,0)$&$(0,1)$&$(1,1)$\\\thickhline
&\multirow{2}{*}{d=0}\ \ $(0)$&$232$&$252$&$272$\\
&$(1)$&$280$&$300$&$320$\\\hline
&\multirow{2}{*}{d=1}\ \ $(0)$&$453$&$480$&$507$\\
&$(1)$&$528$&$555$&$582$\\\hline
&\multirow{2}{*}{d=2}\ \ $(0)$&$760$&$795$&$830$\\
&$(1)$&$870$&$905$&$940$\\\hline
\end{tabular}
\vspace{3em}

\begin{tabular}{cr"cccccc}
&$C^{j=8,m=3}_{n=2,d}$&$(0,0)$&$(0,1)$&$(0,2)$&$(1,1)$&$(1,2)$&$(2,2)$\\\thickhline
&\multirow{3}{*}{d=0}\ \ $(0)$&$2010$&$2290$&$2325$&$2578$&$2613$&$2648$\\
&$(1)$&$2770$&$3085$&$3120$&$3408$&$3443$&$3478$\\
&$(2)$&$2880$&$3195$&$3230$&$3518$&$3553$&$3588$\\\hline
&\multirow{3}{*}{d=1}\ \ $(0)$&$3820$&$4191$&$4235$&$4571$&$4615$&$4659$\\
&$(1)$&$4986$&$5401$&$5445$&$5825$&$5869$&$5913$\\
&$(2)$&$5140$&$5555$&$5599$&$5979$&$6023$&$6067$\\\hline
\end{tabular}
\vspace{3em}

\begin{tabular}{cr"cccccccccc}
&$C^{j=8,m=3}_{n=3,d}$&$(0,0)$&$(0,1)$&$(0,2)$&$(0,3)$&$(1,1)$&$(1,2)$&$(1,3)$&$(2,2)$&$(2,3)$&$(3,3)$\\\thickhline
&\multirow{4}{*}{d=0}\ \ $(0)$&$12088$&$14167$&$14641$&$14695$&$16340$&$16824$&$16878$&$17308$&$17362$&$17416$\\
&$(1)$&$18358$&$20911$&$21439$&$21493$&$23568$&$24106$&$24160$&$24644$&$24698$&$24752$\\
&$(2)$&$20042$&$22649$&$23177$&$23231$&$25360$&$25898$&$25952$&$26436$&$26490$&$26544$\\
&$(3)$&$20250$&$22857$&$23385$&$23439$&$25568$&$26106$&$26160$&$26644$&$26698$&$26752$\\\hline
\end{tabular}
\vspace{3em}

\begin{tabular}{cr"ccc}
&$C^{j=9,m=3}_{n=1,d}$&$(0,0)$&$(0,1)$&$(1,1)$\\\thickhline
&\multirow{3}{*}{d=0}\ \ $(0,0)$&$453$&$480$&$507$\\
&$(0,1)$&$528$&$555$&$582$\\
&$(1,1)$&$603$&$630$&$657$\\\hline
&\multirow{3}{*}{d=1}\ \ $(0,0)$&$760$&$795$&$830$\\
&$(0,1)$&$870$&$905$&$940$\\
&$(1,1)$&$980$&$1015$&$1050$\\\hline
&\multirow{3}{*}{d=2}\ \ $(0,0)$&$1166$&$1210$&$1254$\\
&$(0,1)$&$1320$&$1364$&$1408$\\
&$(1,1)$&$1474$&$1518$&$1562$\\\hline
\end{tabular}
\vspace{3em}

\begin{tabular}{cr"ccc}
&$C^{j=9,m=4}_{n=1,d}$&$(0,0)$&$(0,1)$&$(1,1)$\\\thickhline
&\multirow{1}{*}{d=0}\ \ $\phi$&$330$&$378$&$426$\\\hline
&\multirow{1}{*}{d=1}\ \ $\phi$&$975$&$1050$&$1125$\\\hline
&\multirow{1}{*}{d=2}\ \ $\phi$&$2035$&$2145$&$2255$\\\hline
\end{tabular}

\newpage

\begin{tabular}{cr"cccccc}
&$C^{j=9,m=4}_{n=2,d}$&$(0,0)$&$(0,1)$&$(0,2)$&$(1,1)$&$(1,2)$&$(2,2)$\\\thickhline
&\multirow{1}{*}{d=0}\ \ $\phi$&$4035$&$5020$&$5130$&$6040$&$6150$&$6260$\\\hline
&\multirow{1}{*}{d=1}\ \ $\phi$&$11605$&$13101$&$13255$&$14641$&$14795$&$14949$\\\hline
\end{tabular}
\vspace{3em}

\begin{tabular}{cr"cccccccccc}
&$C^{j=9,m=4}_{n=3,d}$&$(0,0)$&$(0,1)$&$(0,2)$&$(0,3)$&$(1,1)$&$(1,2)$&$(1,3)$&$(2,2)$&$(2,3)$&$(3,3)$\\\thickhline
&\multirow{1}{*}{d=0}\ \ $\phi$&$33913$&$44341$&$46487$&$46695$&$55419$&$57619$&$57827$&$59819$&$60027$&$60235$\\\hline
\end{tabular}
\vspace{3em}

\begin{tabular}{cr"ccc}
&$C^{j=10,m=4}_{n=1,d}$&$(0,0)$&$(0,1)$&$(1,1)$\\\thickhline
&\multirow{2}{*}{d=0}\ \ $(0)$&$975$&$1050$&$1125$\\
&$(1)$&$1140$&$1215$&$1290$\\\hline
&\multirow{2}{*}{d=1}\ \ $(0)$&$2035$&$2145$&$2255$\\
&$(1)$&$2310$&$2420$&$2530$\\\hline
\end{tabular}
\vspace{3em}

\begin{tabular}{cr"cccccc}
&$C^{j=10,m=4}_{n=2,d}$&$(0,0)$&$(0,1)$&$(0,2)$&$(1,1)$&$(1,2)$&$(2,2)$\\\thickhline
&\multirow{3}{*}{d=0}\ \ $(0)$&$11605$&$13101$&$13255$&$14641$&$14795$&$14949$\\
&$(1)$&$15246$&$16896$&$17050$&$18590$&$18744$&$18898$\\
&$(2)$&$15675$&$17325$&$17479$&$19019$&$19173$&$19327$\\\hline
\end{tabular}
\vspace{3em}

\begin{tabular}{cr"ccc}
&$C^{j=11,m=5}_{n=1,d}$&$(0,0)$&$(0,1)$&$(1,1)$\\\thickhline
&\multirow{1}{*}{d=0}\ \ $\phi$&$1287$&$1452$&$1617$\\\hline
&\multirow{1}{*}{d=1}\ \ $\phi$&$3982$&$4257$&$4532$\\\hline
\end{tabular}
\vspace{3em}

\begin{tabular}{cr"cccccc}
&$C^{j=11,m=5}_{n=2,d}$&$(0,0)$&$(0,1)$&$(0,2)$&$(1,1)$&$(1,2)$&$(2,2)$\\\thickhline
&\multirow{1}{*}{d=0}\ \ $\phi$&$20152$&$24618$&$25047$&$29238$&$29667$&$30096$\\\hline
\end{tabular}
\vspace{3em}

\begin{tabular}{cr"ccc}
&$C^{j=12,m=5}_{n=1,d}$&$(0,0)$&$(0,1)$&$(1,1)$\\\thickhline
&\multirow{2}{*}{d=0}\ \ $(0)$&$3982$&$4257$&$4532$\\
&$(1)$&$4554$&$4829$&$5104$\\\hline
\end{tabular}
\vspace{3em}

\begin{tabular}{cr"ccc}
&$C^{j=13,m=6}_{n=1,d}$&$(0,0)$&$(0,1)$&$(1,1)$\\\thickhline
&\multirow{1}{*}{d=0}\ \ $\phi$&$5005$&$5577$&$6149$\\\hline
\end{tabular}
\vspace{3em}


\begin{thebibliography}{65}
\bibitem [{wad()}]{wadachinote}%
  \BibitemOpen
  \href@noop {} {\bibinfo {title} {{Though the conserved quantities for the
  Hubbard model are studied by the expansion of the transfer
  matrix~\cite{Olmedilla1988}, it is not an easy task to obtain our local $Q_k$
  from the expansion transfer matrix, as noted
  in~\cite{wadati1998}}}}\BibitemShut {NoStop}%
\bibitem [{boo()}]{boostnote}%
  \BibitemOpen
  \href@noop {} {\bibinfo {title} {{ Though the generalization of boost
  operator for the Hubbard model was studied in~\cite{PhysRevLett.86.5096}, it
  seems difficult to obtain our $Q_k$ using it because it has a differential
  operator and needs qualitatively different treatment from the usual boost
  operator, as noted in~\cite{10.21468/SciPostPhys.9.3.040}}}}\BibitemShut
  {NoStop}%
\bibitem [{com()}]{commutator_note}%
  \BibitemOpen
  \href@noop {} {\bibinfo {title} {{Here, the commutator is denoted as $[A,B]
  \equiv \frac{1}{2}(AB-BA)$.}}}\BibitemShut {Stop}%
\end{thebibliography}
\end{document}